\documentclass[twocolumn,aps,prb,showpacs,raggedbottom,nobalancelastpage,amssymb,superscriptaddress]{revtex4}

\usepackage{amsmath}
\usepackage{amsfonts}
\usepackage{graphicx}
\usepackage{appendix}
\usepackage{dsfont}
\usepackage{amssymb}
\usepackage{bm}
\usepackage{color}
\usepackage{soul}
\usepackage{natbib}
\setstcolor{red}
\usepackage{nccmath}
\usepackage{array}
\newcounter{rowcount}
\newcommand{\mean}[1]{\langle {#1}\rangle}

\newcommand{\ket}[1]{\left| #1\right\rangle}
\newcommand{\bra}[1]{\left\langle #1\right|}
\newcommand{\ta}{\mathcal{A}^{(q)}}
\newcommand{\tb}{\mathcal{B}^{(q)}}

\newcommand{\td}{\mathcal{D}^{(q)}}
\newcommand{\tad}{\mathcal{A}^{(q)\dag}}
\newcommand{\tbd}{\mathcal{B}^{(q)\dag}}
\newcommand{\tcd}{\mathcal{C}^{(q)\dag}}
\newcommand{\tdd}{\mathcal{D}^{(q)\dag}}

\begin{document}
\title{Hanbury Brown and Twiss Correlations in Quantum Hall Systems}
\author{Gabriele Campagnano}
\altaffiliation{Contributed equally to this work.}
\affiliation{Dipartimento di Fisica, Universit\`{a} di Napoli Federico II, Via Cintia,80126 Napoli, Italy}
\affiliation{CNR-SPIN,  Via Cintia, 80126, Napoli,  Italy}
\affiliation{Department of Condensed Matter Physics, The Weizmann Institute of Science, Rehovot 76100, Israel}
\author{Oded Zilberberg}
\altaffiliation{Contributed equally to this work.}
\affiliation{Institute for Theoretical Physics, ETH Zurich, 8093 Z{\"u}rich, Switzerland}
\affiliation{Department of Condensed Matter Physics, The Weizmann Institute of Science, Rehovot 76100, Israel}
\author{Igor V. Gornyi}
\affiliation{Institut f\"{u}r Nanotechnologie, Karlsruhe Institute of Technology, 76021 Karlsruhe, Germany}
\affiliation{A.~F. Ioffe Physico-Technical Institute, 194021 St. Petersburg, Russia}
\author{Yuval Gefen}
\affiliation{Department of Condensed Matter Physics, The Weizmann Institute of Science,  Rehovot 76100, Israel}

\begin{abstract}
We study a  Hanbury Brown and Twiss (HBT) interferometer formed with chiral edge channels of a quantum Hall system. HBT cross correlations are calculated for a device operating both in the integer and fractional quantum Hall regimes, the latter at Laughlin filling fractions.  We find that in both cases, when the current is dominated by electron tunneling,  current-current correlations show  antibunching, characteristic of fermionic correlations. When the current-current correlations are dominated by  quasiparticle tunneling, the correlations reveal bunching, characteristic of bosons. For electron tunneling, we use the Keldysh technique, and show that  the result for fractional filling factors can be obtained in a simple way from the results  of the integer case. It is shown that quasiparticle-dominated cross-current correlations can be  analyzed by means of a quantum master equation approach.  We present here a detailed derivation of the results [G. Campaganano \textit{et al.}, Phys. Rev. Lett. 107, 106802 (2012)] and generalize them to all Laughlin fractions.
\end{abstract}
\pacs{73.43.Cd}
\maketitle
\section{Introduction}
A single-particle interference may be observed  through the measurement of  light intensity on a screen,  or electric current that reaches a certain drain. Such interference is a manifestation of the wave nature of particles. By contrast, the seminal  experiments of Hanbury Brown and Twiss~\cite{hanburybrowntwiss,hanburybrowntwiss2} (HBT) have introduced the notion of two-particle interference. The latter addresses two other, arguably less trivial, aspects of quantum mechanics, which are manifestations of non-local nature: entanglement and quantum statistics. In the original HBT experiment, the two-particle observables were correlations of light, originating from two uncorrelated and spatially separated sources, and collected at two detectors. This kind of experiment was performed either with photons traveling astronomical distances or in table-top size experiments, (see, e.g., Ref.~[\onlinecite{Lahini:2010}] and references therein). The possibility of observing correlations between uncorrelated sources can be understood as the interference of two two-particle amplitudes~\cite{Fano61}.

The observation of HBT interference with electrons, rather than photons, has been facilitated only recently, due to advances in fabrication and measurement techniques of low-temperature  nano-scale semiconducting  devices. The theoretical predictions  of Ref.~[\onlinecite{samuelsson04}] have been eventually confirmed in experiment~\cite{neder2007}, which employed  chiral edge channels of a quantum Hall system at an integer filling factor. Edge channels of quantum Hall systems have provided an ideal playground to realize and study electronic interference vis-a-vis Mach-Zehnder and Fabry-Perot interferometers~\cite{Wees:1989,Yang:2003,Camino:2005}. In the case of an integer filling factor, electrons may propagate along the edge channels only in one direction, with backscattering due to  impurities or random potential at the edge  suppressed.  It follows that a segment of an edge channel may be thought of as an ideal electronic equivalent of an optical wave guide. Gate-modulated constrictions, so-called quantum point contacts  (QPCs), function as tunable beam splitters,  where impinging particles may  either be deflected from one edge channel to another,  or continue to propagate along the same edge. Although the  highly reduced probability of backscattering makes interferometers realized with edge channels almost ideal
systems, electron-electron interaction and environment-induced dephasing may still play an important role, reducing the visibility of the interference patterns~\cite{Yang:2003}, or modifying their expected behavior~\cite{neder2006}. Such effects are still the subject of ongoing research~\cite{chalker,chalker2009,chalker2010,blanter2007,halperin2011,Bagrets:2012,Bagrets:2013}.

Most interestingly, electronic interferometers may also be considered in the fractional quantum Hall regime. In this case, elementary excitations that can tunnel between edges are not necessarily electrons. In the case of weak inter-edge tunneling, the tunneling amplitude for emergent quasiparticles, a.k.a.~anyons which carry fractional charge, are more relevant (in the renormalization group sense) than electron tunneling amplitudes\cite{KaneFisher1992,glazmanFisher}. Anyons are predicted to obey fractional statistics~\cite{Arovas84}. For Laughlin filling factors [i.e.~, $\nu= 1/(2n+1)$ with integer $n$], quasiparticles are Abelian anyons (the focus of this work). The exchange of two identical anyons  introduces a statistical phase factor to the many-body wave function, $e^{i\theta}$, with the  statistical angle $\theta= \pm\pi/(2n + 1)$. The sign ambiguity of the statistical factor implies that the outcome of anyon exchange depends on details of the trajectory employed to realize this exchange. It follows that  without further assumptions one can not  develop second quantized formalism  for single anyons  (i.e., one can not write a field operator representing the creation or annihilation of an anyon). Instead, one may resort to generalized Klein factors, or follow the kinetics of individual anyons that carry statistical flux quanta with them~\cite{kane03,feldman}. Other observed filling factors may lead to a complex edge channel structure and the emergence of exotic quasi-particles~\cite{Polchinski:1994,Wang:to_be}; for instance, of non-Abelian anyons~\cite{nayak2008} at $\nu=5/2$. Several proposals have been made concerning  current and shot noise measurement in Fabry-Perot and Mach-Zehnder interferometers as diagnostics  of fractional statistics~\cite{chamon1997,stern2006,feldman2006,feldman,feldman2007,ponomarenko2007,law2008,Sassetti:2012,Levkivskyi:2012}. We note that notwithstanding  certain published results~\cite{Willet:2009,Kang:2011}, there are to date no undisputed claims of experimental observation of anyonic interferometry.

%%%%%%%%%%%%%%%%%%%%%%%%%%%%%%%%%%%%%%%%%%%%%%%%%%%%%%%%%%%%%%
\begin{figure}[!ht] \begin{center}
\includegraphics[width=\columnwidth]{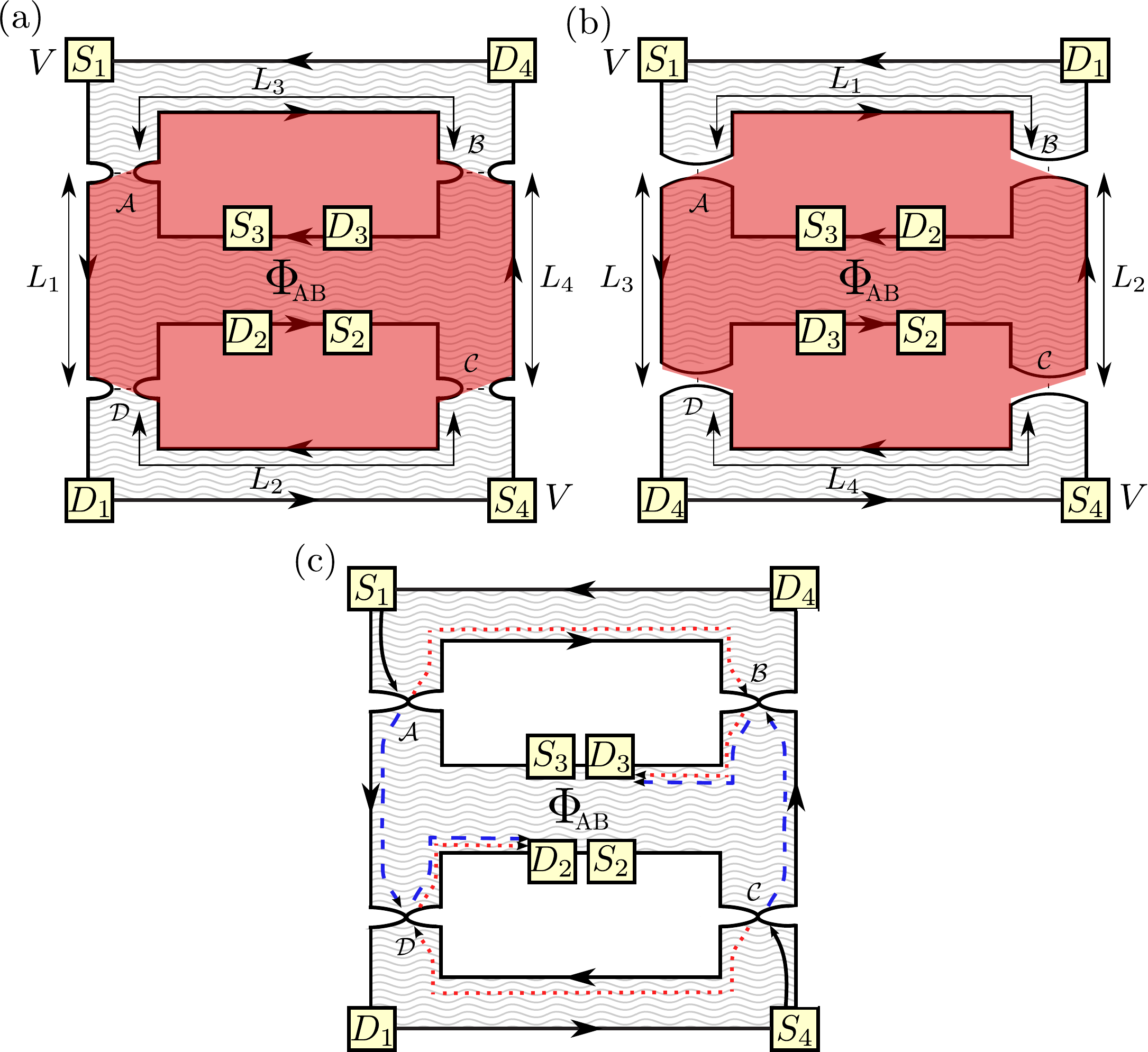}
\end{center}
\caption{Schematic representation of an electronic HBT interferometer. The electron liquid confined to the plane  is subject to a strong transverse magnetic field. For an integer filling factor, as well as for certain  fractions, the bulk of the electronic system forms an incompressible liquid (depicted in yellow), whose edge excitations have dissipation-less chiral propagation (solid lines with arrows). The edges forming the HBT interferometer are the lines $\overline{S_iD_i}$ with i=1,2,3,4. The external edges, $\overline{S_1D_1}$ and  $\overline{S_4D_4}$,  are kept at potential $V$, and the internal ones, $\overline{S_2D_2}$ and  $\overline{S_3D_3}$, are grounded ($V=0$). Inter-edge tunneling (dashed lines) takes place at the four QPCs, $\mathcal{A}$, $\mathcal{B}$, $\mathcal{C}$, $\mathcal{D}$. The distance, $L_i$, is between two consecutive QPCs along edges $\overline{S_iD_i}$. The enclosed area of the HBT (marked by the red overlaying layer) is threaded by a flux, $\Phi_{\rm AB}$, of the applied magnetic field. (a) Edge channels configuration for  almost open QPCs, which corresponds to Fig.~\ref{pinching}(a). For filling factor $\nu=1$, tunneling of electrons occurs at the four QPCs. For filling factor $\nu=1/3$, tunneling of electrons and Laughlin quasiparticles is possible. (b) Edge channels configuration for pinched QPCs, which corresponds to Fig.~\ref{pinching}(b). Here, both for filling factor $\nu=1$ and filling factor $\nu=1/3$, only tunneling of electrons is possible at the four QPCs. (c) Illustration of a flux-sensitive two-particle process. Here, two quasiparticles/electrons are transferred from edges 1 and 4 to edges 2 and 3, the process is AB-sensitive due to the interference between two amplitudes $A_1$ and $A_2$. In $A_1$ a quasiparticle/electron tunnels from edge 1 to edge 3 and a second quasiparticle/electron tunnels from edge 4 to edge 2 (red dotted line). In $A_2$ a quasiparticle/electron tunnels from edge 1 to edge 2 and a second quasiparticle/electron tunnels from edge 4 to edge 3 (blue dashed line).} \label{scheme}
\end{figure}
%%%%%%%%%%%%%%%%%%%%%%%%%%%%%%%%%%%%%%%%%%%%%%%%%%%%%%%%%%%%%%

In this paper, we consider a HBT interferometer realized with edge channels. Our basic setup is depicted in Fig.~\ref{scheme}. The device is made of a two-dimensional electron gas, subject to a strong perpendicular magnetic field and geometrically constrained by gate voltages. The system is assumed to be in the quantum Hall regime, either integer or fractional. By tuning the strength of the applied magnetic field and/or the density of the two-dimensional electron gas, one can control the filling factor at which the system operates. For $\nu = 1$, the low-energy excitations are electrons, while for $\nu= 1/(2n+1)$, the elementary excitations are Laughlin quasiparticles with charge $q = e/(2n+1)$ and Abelian fractional statistics~\cite{Arovas84} $\theta= \pm\pi/(2n+1)$. The setup consists of four active edge channels labeled by $i$ ($i = 1; 2; 3; 4$), each connecting a source $S_i$ to a drain $D_i$. Tunneling between edges takes place at four QPCs which we label as $\mathcal{A}$, $\mathcal{B}$, $\mathcal{C}$, and $\mathcal{D}$. The setup is topologically equivalent to the one addressed in Ref.~[\onlinecite{samuelsson04}], which had been used  to study HBT interference  in  integer filling fractions. In the weak tunneling regime considered throughout most of this paper, the chemical potentials of each source also set the chemical potential of the respective connected edges. Nonequilibrium effects due to current leakage from a voltage-biased edge to a non-biased edge~\cite{Gutman:2009,Gutman:2010} are neglected. We assume that sources $S_1$ and $S_4$ are kept at potential $eV$, while  sources $S_2$ and $S_3$ are kept at potential $eV=0$. Note that our setup studies true HBT interference of particles emitted from two uncorrelated sources and detected at two drains, in contrast to setups where the particles are emitted from a single source~\cite{safi}.

\begin{figure}[!ht] \begin{center}
\includegraphics[width=8cm]{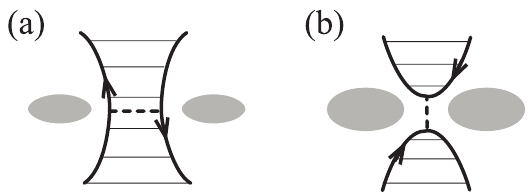}
\end{center}
\caption{A schematic representation of the gates (in gray) controlling a quantum point contact. Thin lines mark the incompressible electron puddle;  bold lines represent the edges; dashed lines represent a tunneling bridge between the two edges. (a) The quantum point contact  is almost open and tunneling between the edges controlled by anyon tunneling is weak. (b) The quantum point contact is pinched off.  Consequently the edges are deformed. Only (weak) tunneling of electrons between the two newly formed puddles (weak link) is allowed~\cite{kane1997}.
} \label{pinching}
\end{figure}

The voltage bias, the  strength of the magnetic field,  or the bulk electronic densities are not
the only control parameters available. By varying the applied gate voltage at each quantum point contact one can control the geometry of the edge channels. The two limiting cases of inter-edge tunneling bridges are depicted in Fig.~\ref{pinching}.  Figure \ref{pinching}(a) shows an almost open QPC: for filling fraction $\nu= 1$, electrons may tunnel between the two edges; for filling fraction  $\nu= 1/(2n+1)$, both tunneling of quasiparticles  and electrons is possible. The respective tunneling operators are relevant/irrelevant in the renormalization group sense. This implies that as one lowers the applied voltage/temperature (and as long as the tunneling amplitude is small), tunneling is dominated by quasi-particle processes.  Figure \ref{pinching}(b) depicts  the case of a considerably pinched off QPC. In this case, for either  $\nu= 1$ or  $1/(2n+1)$ filling fractions, only electrons may tunnel between the two newly formed edges~\cite{kane1997} (see, however, Ref.~[\onlinecite{Shopen:2005}]). % Below we consider various hybrids of HBT devices with electron/quasi-particle tunneling bridges.

We present here a systematic study of current-current correlations (HBT correlations) measured at different drains. This is done for an interferometer consisting of the edges of a bulk $\nu=1$ device,  which is then compared with an interferometer consisting of the edges of a bulk $\nu=1/3$ device, and is finally generalized to other Laughlin fractions. We begin by considering the $\nu=1$ case. Samuelsson \textit{et al.}~\cite{samuelsson04} employed the Landauer-B\"{u}ttiker scattering approach to analyze HBT correlations. Here we repeat the derivation employing a Keldysh technique applied to a bosonic model of the interferometer. We present this calculation having two goals in mind: first we develop
a framework which allows one, in the future, to explicitly include electron-electron interactions in the integer case.  Second, we later employ this approach to study interferometers operating in fractional (Laughlin) filling fractions, where the current is dominated by electron (rather than quasi-particle) tunneling  [see, e.g., the setup of Fig.~\ref{scheme}(b)].

We then move on to consider HBT interferometers operating in the fractional regime (Laughlin fractions), where the current is dominated by quasi-particle (anyon) tunneling  [cf. e.g., the setting of Fig.~\ref{scheme}(a)]. In this case, application of the Keldysh technique for general interferometer parameters and general voltage-temperature regimes is highly complicated. Even when we treat the tunneling amplitudes of the interferometer  perturbatively, we show below that the treatment of general voltage-temperature regimes is virtually impossible, as it involves (for  $\nu= 1/3$) $12^{\rm th}$-order perturbation theory analysis. Here, we focus on the ``high-temperature regime”, where the thermal length defined by $L_T=\hbar \beta v$ with $\beta=1/(k_B T)$ is smaller than the length of each of the interferometer's arms [see Appendix \ref{twevlth_order_break} below, e.g., Eq.~\eqref{highT-condition}]. This simplifies the Keldysh diagrams considerably, as discussed in Sec.~\ref{fractionalqp}. Another complication we need to tackle is how to account for fractional quantum statistics of the anyons. As was noted above, one can not associate statistical Klein factors (see, e.g.~, Ref.~[\onlinecite{vondelft}]) (at least in a simple minded way) to anyonic operators~\cite{remarkYuval}. There are three ways to overcome this hurdle: (i) One may associate a Klein factor with a bilinear  form of anyonic operators (e.g., an inter-edge tunneling quasiparticle tunneling operator~\cite{kane03,feldman}); (ii) One may associate a statistical flux tube with each anyon and follow the trajectories of the interfering anyons, the latter represented by charge-flux composites; (iii) One may embed the various chiral edges within a single closed chiral ``super edge''~\cite{Oshikawa:2006,Altland:to_be}. Following such a  procedure,  the exchange of two anyons does  not any more lead to ambiguity (for example, it is always done clockwise). The resulting statistical phase is thus well defined. In the present analysis, we adopt the second approach. Our accounting of the anyon interfering trajectories and inclusion of the statistical fluxes will be translated to a master-equation approach. The building blocks of that master equation are calculated quantum mechanically.

%%%%%%%%%%%%%%%%%%%%%%%%%%%%%%%%%%%%%%%%%%%%%%%%%%%%%%%%%%%%%%
%%%%%%%%%%%%%%%%%%%%%%%%%%%%%%%%%%%%%%%%%%%%%%%%%%%%%%%%%%%%%%

The present paper is structured as follows. Section \ref{main_results} serves as a guide for the reader. We point out the main issues to be analyzed and discussed (making reference to the main equations and figures), and distinguish between facts that were known previously, and the major new results of the present analysis. %We also discuss briefly the amenability of our predictions to experimental verification.
Section \ref{integerfermionic} is a study of the $\nu=1$ case. This case has been studied  earlier (cf.~Ref.~[\onlinecite{samuelsson04}]). In the absence of electron-electron interactions, it  may  be solved exactly introducing the scattering matrix of the HBT interferometer. Here, we employ a non-equilibrium fermionic Keldysh approach, for  the case of weak tunneling at the four QPCs
[see Fig.~\ref{scheme}(a)]. The case  of strongly pinched-off  QPCs is readily analyzed as well, given  the self-duality with the former case [see Fig.~\ref{scheme}(b)]. In Sec.~\ref{integerbosonic}, we repeat our analysis of the $\nu=1$ case, this time employing  a bosonized  representation of our model. The advantage of the analysis presented in Secs.~\ref{integerfermionic} and \ref{integerbosonic}, as compared with previous analysis based on single-particle scattering matrix approach, is that our present study allows the inclusion of electron-electron interactions on the edge. Section \ref{fractionalmodel} sets the stage for the analysis of interference at fractional filling factors $\nu=1/(2n+1)$.  Using a bosonized picture of the edge, we define the various tunneling operators, including the statistical flux. Currrent-current correlations are calculated for  electron-tunneling-dominated dynamics in Sec.~\ref{sec_elec_over_frac}, employing the Keldysh technique. In Sec.~\ref{fractionalqp}, we address quasiparticle tunneling, resorting to a master-equation technique. We first outline the analysis of the  time evolution of statistical flux trapped inside the interferometer and the ensuing current-current correlations, and then present a quantum-mechanical calculation of the rates to be incorporated in the master equation. % The analysis of the various electron-quasiparticle hybrid setups is presented in Section \ref{hybridsec}.
To put our analysis  in the right context, and to provide important details of our analysis, we have included here a few appendices. Appendix \ref{app_scatter} repeats the scattering matrix analysis of Ref.~[\onlinecite{samuelsson04}] of HBT correlations for non-interacting electrons. Detailed derivation of certain equations in the main text is presented in Appenidx \ref{app_calcs23} [Eqs.~(\ref{s0}) and \eqref{sphi}], Appendix \ref{app_integer_bosonic} [Eq.~(\ref{electrons_in_integer})],  and Appendix \ref{app_equiv_bos_ferm} [Eq.~\eqref{derivatives1}]. The complexity of tackling quasiparticle correlations that depend on Aharonov-Bohm flux, employing the Keldysh technique, is outlined in Appendix \ref{twevlth_order_break}. Evaluation of two-quasiparticle rates that affect the kinematics of the statistical flux  state of the interferometer is presented in Appendices \ref{app_rate2} and \ref{app_rate1}.

\section{Main results and discussion}
\label{main_results}

We begin by presenting \textit{a guide to the main results of our analysis}. We first consider an electronic HBT interferometer operating in the $\nu=1$ regime, where we ignore electron-electron interactions. We derive (Section \ref{integerfermionic}) an expression for cross-current correlations [Eq.~\eqref{noisegeneral1}], which includes both flux-independent [Eq.~\eqref{s0}] and flux-dependent [Eq.~\eqref{sphi}] terms. Our results are identical to those of Ref.~[\onlinecite{samuelsson04}], which were derived by employing a Landauer-B\"{u}ttiker scattering approach, valid for non-interacting electrons. Our approach here, which utilizes the Keldysh technique, is advantageous as it allows for the incorporation of electron-electron interaction. This, however, is left for future analysis.

We next (Sec.~\ref{integerbosonic}) repeat the calculation ($\nu=1$) within the framework of a bosonic theory of the edge modes. The formal expression we obtain for the flux-dependent correlations is new [Eq.~(\ref{electrons_in_integer})], but leads to the same result as in the previous section. The merit of employing a bosonized version of our action is two-fold: it would allow the inclusion of electron-electron interaction (as is the case with the fermionic Keldysh approach described in Sec.~\ref{integerfermionic}). More importantly, it can be readily generalized to allow the study of anyonic HBT correlations.

We generalize the bosonic description of the HBT to the fractional quantum Hall regime (Sec.~\ref{fractionalmodel}), focusing on Laughlin filling factors. 
Our new results for the flux-dependent current-current correlator, for a generalized Laughlin fraction $\nu$, are given for the case where the particles tunneling at the ``beam splitters'' are electrons [Sec.~\ref{sec_elec_over_frac}, Eq.~\eqref{derivatives1}], or Laughlin anyons [Sec.~\ref{fractionalqp}, Eq.~\eqref{mainres1gen}]. Our analysis is carried out employing a quantum master-equation analysis. Central to this analysis is the observation that the interferometer is characterized by the number of flux tubes trapped inside; hence we define a number of flux states for the interferometer. The kinematics whereby transitions among flux states take place is governed by one- and two-particle rates, classified in Table \ref{table1} (cf. Ref.~[\onlinecite{campagnano2012}]). The rates are explicitly calculated [Eqs. \eqref{single-qp-transf3} and \eqref{2-phi-rate-bis}]. Our analysis is mostly carried out for zero-frequency correlations. We show that the flux-dependent correlations have non-vanishing contributions only from ``auto-terms'', depicted in Fig.~\ref{autoterm}. We further find here that this statement about non-vanishing contributions coming only from auto-terms holds in the finite-frequency regime too, cf.~Eq.~\eqref{finfreq}.

\textit{Flux periodicity and tunneling strength}. In the three scenarios considered here, $\nu=1$ (electron tunneling), $\nu=1/(2n+1)$ (electron tunneling), and $\nu=1/(2n+1)$ (quasiparticle tunneling), the current-current correlations are periodic in the flux, with period of the Dirac flux quantum $\Phi_0=hc/e$. This is not surprising in view of general gauge invariance arguments, and has far reaching implications on the dependence of the current-current correlations on the tunneling amplitudes, \{$\Gamma_i$\}. Consider, for example, the $\nu=1/3$ case. The lowest order (in $\Gamma$; for the moment we assume that all $\Gamma_i$'s are of the same order, $\sim\Gamma$) contribution involves interfering paths made up of four tunneling events. This contribution is therefore $\sim\Gamma^4$. When $1/3$ anyons are concerned, such a $\Gamma^4$ contribution will lead to a term whose flux periodicity is $3\Phi_0$, in marked contradiction to the gauge invariance requirement. Indeed, we have checked that this contribution vanishes. The leading, non-vanishing, contribution, giving rise to a $\Phi_0$ periodicity, involves three pairs of quasiparticles, and naively would scale as $\sim\Gamma^{12}$. As we discuss in Section \ref{kinetic} and in Appendix \ref{twevlth_order_break}, single particle processes renormalize this to $\sim\Gamma^8$. We extend this analysis to a general $\nu=1/(2n+1)$. In short, gauge invariance dictates $\Phi_0$ periodicity, and implies that the smaller $\nu$ the higher is the order in $\Gamma$.

\textit{What can one observe?} The AB-dependent contribution to the correlations we calculate will be negative/positive depending on whether they are dominated by electron tunneling/quasiparticle tunneling. The latter corresponds to partial-bunching akin to bosons, indicating a substantial   statistical transmutation from the underlying electronic degrees of freedom. We also find qualitative differences between the anyonic signal and the corresponding bosonic or fermionic signals, indicating that anyons can not be simply thought as intermediate between bosons and fermions. %One

\section{Integer filling fraction: fermionic framework}\label{integerfermionic}
As a pedagogical introduction, we consider here the HBT interferometer at integer filling factor, and calculate current-current correlations measured at two different drains.
Such a quantity can be obtained in a straightforward manner using a  Landauer-B\"{u}ttiker  scattering approach \cite{samuelsson04} (see Appendix \ref{app_scatter}), but it is nevertheless interesting  to reproduce here known
results using a Keldysh non-equilibrium formalism. In this section we use a standard fermionic representation for the electron operators (in the next section the same result is obtained using a bosonic representation of the fermionic operators.)

We first consider the case of almost open QPCs [cf. Fig.~\ref{scheme}(a)]. The model Hamiltonian can be written as $H=H_0+H_T$, where $H_0$ is the bare Hamiltonian describing the four edges, and $H_T$ describes the tunneling
of electrons at the four QPCs. For an edge $j$ ($j=\{1,2,3,4\}$) of length $\mathcal{L}_j$, the electron field operator at point $x$ is given by
\begin{equation}
\hat{\psi}_j(x)=\frac{1}{\sqrt{\mathcal{L}_j}}\sum_{k_j}e^{ik_j x}\hat{c}_{k_j,j} \, ,
\label{electron_field_operator}
\end{equation}
with periodic boundary conditions, i.e. $k_j=2\pi n/\mathcal{L}_j$ and  $n$ integer. We assume $k_j \mathcal{L}_j \gg 1$ in which case the particular boundary conditions employed are of little consequence. Here, $c_{k_j,j}$ is the electron annihilation operator for a state of wave vector $k_j$ on edge $j$. Creation and annihilation operators  satisfy standard fermionic anti-commutations rules,
$\{\hat{c}_{k_j,j}^\dag,\hat{c}_{k_{j'}',j'}\}=\delta_{k_j,k_{j'}'}\delta_{j,j'}$ and $\{\hat{c}_{k_j,j},\hat{c}_{k_{j'}',j'}\}=0$. We assume that, in absence of tunneling, each edge $j$ is at equilibrium with a reservoir at chemical potential $\mu_j$. Hence,
\begin{align}
H_0=-i \hbar v\sum_{j=1}^4\int dx :\hat{\psi}_j^\dag(x) \partial_x \hat{\psi}_j(x): \, ,
\end{align}
where normal ordering is defined with respect to the state filled up to the lowest chemical potential in the problem,
$\prod_{j=1,..,4; \epsilon(k_j)\leq \min\{\mu_j\}} \hat{c}_{k_j,j}^\dag |0\rangle$. Notice that we take into account only the four edges $1,...,4$ defining the HBT (cf. Fig.~\ref{scheme}).

The tunneling Hamiltonian $H_T$ is given by
\begin{align}
H_T=(\mathcal{A}+\mathcal{B}+\mathcal{C}+\mathcal{D})+h.c \, ,
\end{align}
with
\begin{align}\label{tunneling-operators}
\mathcal{A}=&e^{2\pi i \Phi_{\rm AB}/\Phi_0} \Gamma_\mathcal{A} \,  \hat{\psi}_1^\dag(0)\hat{\psi}_3(0)\, ,\nonumber\\
\mathcal{B}=&\Gamma_\mathcal{B} \, \hat{\psi}_4^\dag(L_4)\hat{\psi}_3(L_3)\, ,\nonumber\\
\mathcal{C}=& \Gamma_\mathcal{C} \, \hat{\psi}_4^\dag(0)\hat{\psi}_2(0) \, , \nonumber\\
 \mathcal{D}=&\Gamma_\mathcal{D} \, \hat{\psi}_1^\dag(L_1)\hat{\psi}_2(L_2) \,,
\end{align}
where on each edge $j$ the point $x=0$ is chosen at the first QPC in the direction of chirality and the distance to the second QPC in the direction of chirality defines the lengths of the interferometer arms, $L_j$ (not to be confused with the length of the full edge $\mathcal{L}_j$). We choose a gauge such that the flux of the magnetic field threading the interferometer, $\Phi_{\rm AB}$, is ascribed to the tunneling operator of QPC $\mathcal{A}$ [see Fig.~\ref{scheme}(a)].
For future reference, it is convenient to introduce the shorthand notation $H_{T_\mathcal{A}}=\mathcal{A}+\mathcal{A}^\dag$, and similarly for  tunneling operators at the other three QPCs.
%\[
%H_T=H_{T_A}+H_{T_B}+H_{T_C}+H_{T_D}
%\]=(e/h c)\oint d \vec{l} \cdot \vec{A}

\begin{figure}
\includegraphics[width=0.6\columnwidth]{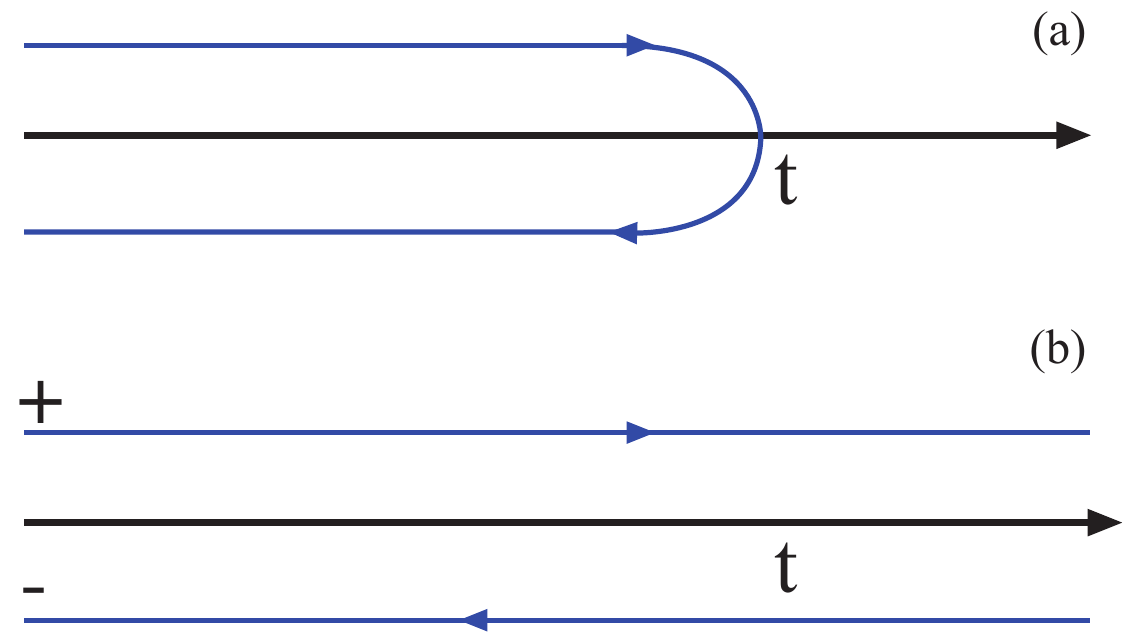}
\caption{Keldysh contour, the direction of the arrows indicates the ordering of times along
the Keldysh contour. The original contour in (a) can be extended to the one shown in (b),
here $+$ and $-$ indicate the upper and the lower branches and will be used to define the four components
of the Keldysh Green's function.}\label{fig:keldysh}
\end{figure}

In order to study the stationary non-equilibrium problem, we use the Keldysh technique \cite{keldysh}. We assume that a time $t=-\infty$ the tunneling Hamiltonian $H_T$ is zero and then adiabatically turned on. The expectation value of a generic operator $\hat{O}$ at time $t$ can be expressed  as
\begin{equation}\label{keldysh1}
\langle\hat{O}^{H}(t)\rangle=\langle  T_K \hat{O}(t) \mathcal{S}_K\rangle \, .
\end{equation}
Here and in the following, operators $\hat{O}^{H}$ are in the Heisenberg representation, whereas operators $\hat{O}$ are in the interaction representation with respect to $H_0$. We have introduced the Keldysh action
\begin{equation}\label{keldysh_action}
\mathcal{S}_K=T_K \exp\{-\frac{i}{\hbar}\int_K H_T(\tau)d\tau \}\,,
\end{equation}
where $T_K$ is the Keldysh time ordering  operator and the Keldysh contour $K$  is represented in Fig.~\ref{fig:keldysh}. A point $\tau$ on the Keldysh contour is specified by the time coordinate $t$ along the real axis and the  branch index $\eta=\{+1,-1\}$. In the following for such a point we will use the notation $\tau \equiv  t_\eta$. Equation~(\ref{keldysh1}) can be generalized to the expectation value of $n$ operators,
\begin{multline}\label{keldysh2}
\langle T_K  \hat{O}^H_1(t_{1,\eta_1})...\hat{O}^H_n(t_{n,\eta_n})       \rangle=\\
\langle  T_K   \hat{O}_1(t_{1,\eta_1})...\hat{O}_n(t_{n,\eta_n})   \mathcal{S}_K\rangle \, .
\end{multline}

%%%%%%%%%%%%%%%%%%%%%%%%%%%%%%%%%%%%%
%%%%%%%%%%%%%%%%%%%%%%%%%%%%%%%%%%%

For each edge channel $j$ the Keldysh Green's function is defined as
$G_j^{\eta_1 \eta_2}(x_1,t_{1};x_2,t_{2})\equiv G_j(x_1,t_{1,\eta_1};x_2,t_{2,\eta_2})=-i\langle T_K \hat{\psi}_j(x_1,t_{1,\eta_1})\hat{\psi}_j^\dag (x_2,t_{2,\eta_2})\rangle$.  The four
components in  the $2\times 2$ Keldysh space, corresponding to the four different choices of the branch indices, are
\begin{align}
 &G_j^{\eta_1 \eta_2}(x_1,t_1;x_2,t_2)
 =\\
&=\left(
\begin{array}{cc}
 G^{++}_j(x_1,t_1;x_2,t_2) &  G^{+-}_j(x_1,t_1;x_2,t_2) \\
  G^{-+}_j(x_1,t_1;x_2,t_2) &  G^{--}_j(x_1,t_1;x_2,t_2)
\end{array} \right)
 \nonumber \\
&=\resizebox{.9\hsize}{!}{$\displaystyle  \left(
\begin{array}{cc}
-i\langle T \hat{\psi}_j(x_1,t_1)\hat{\psi}_j^\dag (x_2,t_2)\rangle & i\langle \hat{\psi}_j^\dag (x_2,t_2)   \hat{\psi}_j(x_1,t_1)  \rangle \\
 -i\langle   \hat{\psi}_j(x_1,t_1) \hat{\psi}_j^\dag (x_2,t_2)  \rangle & -i\langle \tilde{T} \hat{\psi}_j(x_1,t_1)\hat{\psi}_j^\dag (x_2,t_2)\rangle
\end{array}
\right) \, ,$}\nonumber
\end{align}
where $T$ and $\tilde{T}$ are the time and anti-time ordering operators, respectively (not to be confused with Keldysh time ordering $T_K$). Notice that the two components $G_j^{+-}(x_1,t_1;x_2,t_2)$ and $G_j^{-+}(x_1,t_1;x_2,t_2)$  correspond to the standard \emph{lesser}, $G_j^{<}(x_1,t_1;x_2,t_2)$, and \emph{greater}, $G_j^{>}(x_1,t_1;x_2,t_2)$, Green's functions, respectively.

We assume that our system is translational invariant in space (far from boundaries) and in time (steady state). Hence, we can write the Green's function in terms of two parameters, i.e. $G_j^{\eta_1 \eta_2}(x_1,t_1;x_2,t_2)=G_j^{\eta_1 \eta_2}(\Delta x,\Delta t)$ with $\Delta x=x_1-x_2$ and $\Delta t = t_1-t_2$. It is convenient to express the four components of the Keldysh Green's function in mixed energy-space representation defining: $G_{j}^{\eta_1 \eta_2}(\Delta x,\omega)=\int dt \exp(i \omega  \Delta t) G_{j}^{\eta_1 \eta_2}(\Delta x,\Delta t)$. In this representation the four components of the Keldysh Green's functions are
\begin{align}\label{FT1}
G_j^{++}(\Delta x,\omega)=&\frac{i}{v}e^{i \omega \Delta x/ v}[f(\hbar \omega-\mu_i)-\Theta(\Delta x)]  \, , \\ \label{FT2}
G_j^{+-}(\Delta x,\omega)=&\frac{i}{v}e^{i \omega \Delta x/v} f(\hbar \omega-\mu_i) \, , \\ \label{FT3}
G_j^{-+}(\Delta x,\omega)=&-\frac{i}{v}e^{i \omega \Delta x/ v} [1-f(\hbar \omega-\mu_i)]  \, , \\ \label{FT4}
G_j^{--}(\Delta x,\omega)=&\frac{i}{v}e^{i \omega \Delta x/ v}[f(\hbar \omega-\mu_i)-\Theta(-\Delta x)]  \, ,
\end{align}
where $f(\omega)=[1+\exp(\beta \hbar \omega)]^{-1}$   [$\beta=1/(k_B T)$] is the Fermi-Dirac distribution function, and $\Theta(x)$  the Heaviside step function. In the following we will evaluate non-local Green's functions, $\Delta x\ne 0$.

The current operator at point $x$ on edge $j$ can be written as $\hat{I}_j(x)=e v :\hat{\psi}^\dag_j(x) \hat{\psi}_j(x):$. Let us define the zero-frequency (cross) current-current correlation function measured at two drains $D_i$ and $D_j$ ($i\ne  j$):
\begin{multline}\label{noisegeneral1}
S_{ij}(0)=\frac{1}{2}\int_{-\infty}^{+\infty} dt_1 \langle \langle  \hat{I}^H_i(x_i,t_1) \hat{I}^H_j(x_j,t_2) \\ +\hat{I}^H_j(x_j,t_2)\hat{I}^H_i(x_i,t_1)  \rangle \rangle \, ,
\end{multline}
where $\langle \langle \hat{O}_i \hat{O}_j\rangle \rangle \equiv \langle  \hat{O}_i \hat{O}_j\rangle-\langle  \hat{O}_i \rangle \langle \hat{O}_j\rangle$.
The point $x_i$ ($x_j$) is chosen to be after the last QPC on the respective edge (in the direction of chirality), i.e. $x_i>L_i$ ($x_j>L_j$), see Fig.~\ref{scheme}.
Owing to the chiral propagation, the above condition is sufficient to express
the current-current correlations measured at the drains $D_i$ and $D_j$.

Plugging the expression of the current operators, we can rewrite Eq.~\eqref{noisegeneral1} on the Keldysh contour,
\begin{multline}\label{noisegeneral2}
S_{ij}(0)= \\
 \frac{e^2 v^2}{2} \sum_{\eta=\pm 1}\int_{-\infty}^{+\infty} dt_1 \langle \langle  T_K
: (\hat{\psi}^H_i)^\dag(x_i,t_{1,\eta} + \eta \, 0^+) \hat{\psi}_i^H (x_i,t_{1,\eta}):  \\ \times
 : (\hat{\psi}^H_j)^\dag(x_j,t_{2,-\eta}- \eta \, 0^+) \hat{\psi}_j^H (x_j,t_{2,-\eta}):
\rangle \rangle\,.
\end{multline}
Notice that we have introduced time splitting, using the infinitesimal positive number $0^+$, in order to preserve the correct ordering of the operators under the action of $T_K$. The operators are still in the Heisenberg representation. Making use of Eq.~(\ref{keldysh2}), we write Eq.~\eqref{noisegeneral2} in the interaction representation,
\begin{multline}\label{noisegeneral3}
S_{ij}(0)= \\
\frac{e^2 v^2}{2} \sum_{\eta=\pm1}\int_{-\infty}^{+\infty} dt_1 \langle \langle  T_K
: \hat{\psi}^\dag_i(x_i,t_{1,\eta}+ \eta \, 0^+) \hat{\psi}_i(x_i,t_{1,\eta}):  \\ \times
 : \hat{\psi}^\dag_j(x_j,t_{2,-\eta}- \eta \, 0^+) \hat{\psi}_j(x_j,t_{2,-\eta}):  \mathcal{S}_K
\rangle \rangle \, .
\end{multline}

Equation (\ref{noisegeneral3}) is exact. Next, we assume that tunneling at the four QPCs is weak and expand $\mathcal{S}_K$ in powers of $H_T$. The lowest non-vanishing contribution to cross-current correlations, $i\ne j$, is fourth order in $H_T$,
%\begin{widetext}
\begin{multline}\label{generalsij}
\lefteqn{S_{ij}(0)=\frac{e^2 v^2}{2}\frac{(-i)^4}{4!\hbar^4} \sum_{\eta= \pm1 }
\int_{-\infty}^{+\infty} dt_1 \int_K d\tau_1 d\tau_2 d\tau_3 d \tau_4}
\\
 \langle \langle  T_K
: \hat{\psi}^\dag_i(x_i,t_{1,\eta}+ \eta \, 0^+) \hat{\psi}_i(x_i,t_{1,\eta}):  \\
 : \hat{\psi}^\dag_j(x_j,t_{2,-\eta}- \eta \, 0^+) \hat{\psi}_j(x_j,t_{2,-\eta}):   \\    H_T(\tau_1)H_T(\tau_2)H_T(\tau_3)H_T(\tau_4)
\rangle \rangle\,.
\end{multline}
%\end{widetext}

In the following, we consider explicitly the cross-current correlations at drains $D_2$ and $D_3$, $S_{23}$. In Appendix \ref{app_calcs23}, we derive its outcome in detail. Here, we highlight the results. In essence, from Eq.~(\ref{generalsij}) we obtain four  contributions, proportional to $|\Gamma_\mathcal{A}|^2|\Gamma_\mathcal{D}|^2$, $|\Gamma_\mathcal{B}|^2|\Gamma_\mathcal{C}|^2$, $\Gamma_\mathcal{A} \Gamma_\mathcal{B}^* \Gamma_\mathcal{C} \Gamma_\mathcal{D}^*$, and $\Gamma_\mathcal{A}^* \Gamma_\mathcal{B} \Gamma_\mathcal{C}^* \Gamma_\mathcal{D}^*$. We assume that edges 1 and 4 are kept  at chemical potential $\mu_1=\mu_4=eV$, and edges 2 and 3 are grounded, $\mu_2=\mu_3=0$. With this symmetric choice, we can write $S_{23}$ as
\begin{multline}
\label{s23_divide_contrib}
S_{23}=\frac{1}{\hbar^4 v^4}\Big[(|\Gamma_\mathcal{A}|^2|\Gamma_\mathcal{D}|^2+|\Gamma_\mathcal{B}|^2|\Gamma_\mathcal{C}|^2)S_0 \\+(
e^{i \Phi_{\rm AB}/\Phi_0}\Gamma_\mathcal{A} \Gamma_\mathcal{B}^* \Gamma_\mathcal{C} \Gamma_\mathcal{D}^*S_{\Phi}+c.c. )\Big] \, ,
\end{multline}
where $S_0$ is a contribution to $S_{23}$ due to processes that involve tunneling across two QPCs, i.e. $\mathcal{A}$ and $\mathcal{D}$, or $\mathcal{C}$ and $\mathcal{B}$, and $S_{\Phi}$ is a contribution due to tunneling across all four QPCs. The prefactor of the latter contribution is modulated by the flux of the magnetic field threading the interferometer, $\Phi_{\rm AB}$. These contributions are
\begin{align}
S_0=&-\frac{e^2}{2\pi}\int d\omega \left[ f(\hbar \omega)-f(\hbar \omega-eV)\right]^2 \label{s0}\\
 =&\frac{e^2}{\hbar\beta}\frac{1- \pi \alpha \coth[\pi \alpha]}{\pi }\, ,\nonumber\\
S_{\Phi}=&-\frac{e^2}{2\pi }\int d\omega \left[ f(\hbar \omega)-f(\hbar \omega-eV)\right]^2\label{sphi}
\\ &\times
 \exp \left\{ i \frac{\omega}{v}(L_1+L_4-L_2-L_3)\right\}\nonumber\\
=&\frac{e^2}{\hbar\beta} \frac{\left(\Delta \tilde{L} \cos\left[\alpha\Delta \tilde{L}\right]-\pi\coth\left[\pi\alpha\right]\sin\left[\alpha\Delta \tilde{L}\right] \right)}{\pi\sinh(\Delta \tilde{L})}\nonumber\\
&\times e^{i\alpha\Delta \tilde{L}} \, ,\nonumber
\end{align}
where we defined the unitless parameters $\alpha=eV \beta/(2\pi)$ and $\Delta \tilde{L}=\pi (L_1+L_4-L_2-L_3)/(\hbar \beta v)$, with $\beta=1/(K_B T)$.

Our result, Eqs.~(\ref{s0}) and (\ref{sphi}), coincides
with the results of Ref.~[\onlinecite{samuelsson04}], where the problem was treated using a Landauer-B\"{u}ttiker scattering
approach. For completeness, we provide in Appendix \ref{app_scatter} the scattering matrix treatment of the problem and present the
equivalence with the method shown here.

We have presented the calculation for the geometry of Fig.~\ref{scheme}(a). This corresponds to small voltages applied to the gates of the QPCs. For pinched QPCs [cf. Fig.~\ref{scheme}(b)], at filling factor $\nu=1$, Eqs.~(\ref{s0}) and (\ref{sphi}) are the same expressions for $S_{23}$ under re-labeling $L_1 \leftrightarrow L_2$, $L_3 \leftrightarrow L_4$ and $\Gamma_{\mathcal{A}} \leftrightarrow  \Gamma_{\mathcal{C}}$.

%%%%%%%%%%%%%%%%%%%%%%%%%%%%%%%%%%%%%%%%%%%%%%%%%%%%%%%%%%%
\section{Integer filling fraction: calculation in terms of bosonic operators}\label{integerbosonic}

\begin{figure}[ht] \begin{center}
\includegraphics[width=6cm]{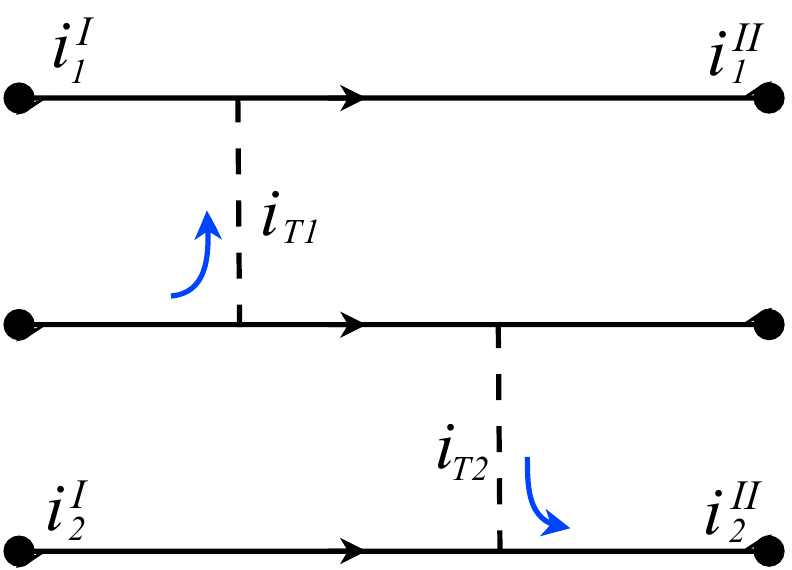}
\end{center}
\caption{Correlation of tunneling currents does not always correspond to correlation of currents measured at the chiral edges (or at the drains).
In this figure, we give an example to illustrate such a difference. The three lines represent edge states with chiral propagation. The dashed lines represent tunneling between edges. $i_1^I$ and $i_2^I$ ($i_1^{II}$ and $i_2^{II}$) are currents measured before (after) the tunneling points. $i_{T1}$ and $i_{T2}$ are tunneling currents. Assume that current fluctuations are slow on the time scale of propagation along the edges and consider the correlator $\langle \langle i_1^{II}  i_2^{II}\rangle\rangle$. Current conservation dictates $i_j^{II}=i_j^{I}+i_{Tj}$ ($j=1,2$), which yields $\langle \langle i_1^{II}  i_2^{II}\rangle\rangle$= $\langle \langle i_1^{I}  i_{T2}\rangle\rangle$+ $\langle \langle i_{T1} i_{T2}  \rangle\rangle$. We have used the fact that  $\langle \langle i_1^{I}  i_2^{I}\rangle\rangle=  \langle \langle  i_{T1}  i_2^{I} \rangle\rangle=0$. The correlator $\langle \langle i_1^{I}  i_{T2}\rangle\rangle$ is not vanishing at a finite temperature.
 Hence, correlations between tunneling currents and correlations between currents measured at  drains are not equal at a finite temperature.  } \label{threedges}
\end{figure}

In this section, we address again current-current correlations when the system is set to integer filling factor, $\nu=1$. We carry on this analysis from a different point of view: (i) we use a bosonic representation of the Fermi fields (to be defined in the following), and (ii) following common practice in the literature on electron interferometry \cite{kane03,vish,safi,feldman}, we introduce tunneling currents and not currents defined as densities.
We define tunneling currents towards edge 2 and edge 3 as the time derivative of the total charge on these edges.
 With this choice, it is sufficient to expand $\mathcal{S}_K$ to second order rather than to fourth order as we did in the previous section. Surely, there is a price to pay for such a simplification: naively, one could think that both definitions yield the same current-current correlation. This is actually not the case; only the magnetic flux modulated part of $S_{23}$, i.e., $S_{\Phi}$, is the same for both approaches. In Fig.~\ref{threedges}, we consider a simplified geometry to give a heuristic motivation why correlations of tunneling currents are not equivalent to correlations of currents measured at the drains.

Here, we follow closely the approach of Ref.~[\onlinecite{chalker}]. We briefly repeat the standard steps of operator bosonization~\cite{vondelft,Affleck:2013}, applying them to the present geometry. We can express $H_0$ in terms of plasmonic modes. Boson creation operators  are defined as
\begin{align}
b^\dag_{q_j,j}=\frac{i}{\sqrt{n_{q_j}}}\sum_{k_j} c_{k_j+q_j,j}^\dag c_{k_j,j}\,,
\end{align}
for $q_j>0$. These boson operators satisfy the canonical commutations relation $[b_{q_i,i},b^\dag_{q_j,j}]=\delta_{q_i q_j}\delta_{ij}$.
We can now introduce the bosonic field operators
\begin{align}
\phi_{j}(x)=-\sum_{q_j>0} \frac{1}{\sqrt{n_{q_j}}} \:
\left(e^{-iq_j x}b_{q_j, j}^{\dagger}+ e^{iq_j x}b_{q_j, j}\right)e^{-l_c q_j/2}\,,
\end{align}
where $l_c$ is an ultraviolet cut-off.
This new fields obeys the commutation relation
\begin{align}
[\phi_{i}(x), \partial_y\phi_{j}(y) ] = -2 \pi i
\delta_{i j} \delta(x-y)\,.
\end{align}
Fermion and boson field operators are related by [cf. Eq.~\eqref{electron_field_operator}]
\begin{align}
\hat{\psi}_{j}(x)=\frac{F_{j}}{ \sqrt{2\pi l_c}}  e^{i 2\pi {\hat N}_{j}
x/\mathcal{L}_j} e^{-i\phi_{j}(x)}\,,
\end{align}
where $F_{j}$ are  Klein operators that satisfy the
anticommutation relation $\{ F^{\dagger}_{i}, F^{\phantom{\dagger}}_{j} \}
= 2 \delta_{i j}$ and ${\hat N}_{j}{=} \int\!   dx \:\rho_{j}(x)$ is the total charge operator for edge ${j}$.
The Hamiltonian $H_0$ can be expressed in terms of the new fields as
\begin{align}
H_0= \sum_{j=1}^4\left[ \frac{\pi\hbar  v}{\mathcal{L}_j}\hat N_{j}(\hat N_{j}+1)+\frac{\hbar v}{4\pi}\int dx (\partial_x\phi_j)^2\right]\, ,
\end{align}
where $v$ is the velocity of edge excitations. The average value  $\langle \hat N_j \rangle$ is related to
 the chemical potential  $\mu_j$ by $\mu_j=2\pi \hbar v \langle \hat N_j \rangle/\mathcal{L}_j$; linearizing the dependence of the Hamiltonian on $\hat N_j$ we have
\begin{align}
H_0= \sum_{j=1}^4\left[ \mu_j{\hat N_{j}}+\frac{\hbar v}{4\pi}\int dx (\partial_x\phi_j)^2\right]\, ,
\end{align}
The tunneling Hamiltonian remains to be described by Eq.~(\ref{tunneling-operators}).

%%%%%%%%%%%%%%%%%%%%%%%%%%%%%%%%%%%%%%%%
%%%%%%%%%%%%%%%%%%%%%%%%%%%%%%%%%%%%%%%%

The Keldysh Green's function with the above representation of the Fermi fields can be written as a multiplication of two expectation values,
\begin{multline}
G_{j}^{\eta_1 \eta_2}(x_1,t_{1};x_2,t_{2})=\\ -\frac{i}{2\pi l_c}
\langle  T_k  F_j(t_{1,\eta_1}) e^{i 2\pi N_j x_1/\mathcal{L}_j}    e^{-i 2\pi N_j x_2/\mathcal{L}_j}
F_j^\dag (t_{2,\eta_2})  \rangle \\\times
\langle T_K           e^{-i\phi_j(x_1,t_{1,\eta_1})} e^{i\phi_j(x_2,t_{2,\eta_2})}        \rangle\,.
\end{multline}
These expectation values are straightforward to evaluate.
The one involving Klein factors yields
\begin{multline}
\langle  T_k  F_j(t_{1,\eta_1}) e^{i 2\pi N_j x_1/\mathcal{L}_j}    e^{-i 2\pi N_j x_2/\mathcal{L}_j}
F_j^\dag (t_{2,\eta_2})  \rangle = \\
e^{i\mu_j(x_1/v-t_1)/\hbar} e^{-i\mu_j(x_2/v-t_2)/\hbar} \chi_{\eta_1 \eta_2}(t_1-t_2)\,,
\end{multline}
where $$\chi_{\eta_1 \eta_2}(t_1-t_2)=(\eta_2-\eta_1)/2+\text{sign}(t_1-t_2)(\eta_1+\eta_2)/2.$$
The  bosonic part of the two-point correlation function reads as~\cite{vondelft,giamarchi}
%\begin{align}
%\label{eq:GtoS}
%&\frac{1}{2 \pi l_c} \langle T_K \;
%e^{-i\phi_j(x_1,t_{1,\eta_1})}e^{i\phi_j(x_2,t_{2,\eta_2})}\rangle =\\
%&\frac{1}{2\pi}
%\frac{1}{\frac{\hbar \beta  v}{\pi}\sinh\left\{\frac{\pi}{\hbar \beta v}\left([(t_1-t_2)v-(x_1-x_2)]\chi_{\eta_1,\eta_2}(t_1-t_2)+i l_c\right)\right\}}
%\,.\nonumber
%\end{align}
\begin{align}
& \langle T_K \;
e^{-i\phi_j(x_1,t_{1,\eta_1})}e^{i\phi_j(x_2,t_{2,\eta_2})}\rangle =
\frac{\pi l_c}{\hbar \beta v}\label{eq:GtoS} \\
&\times \frac{1}{\sin\left\{\frac{\pi}{\hbar \beta}\left(i\left[t_1-t_2-\frac{x_1-x_2}{v}\right]\chi_{\eta_1,\eta_2}(t_1-t_2)+\frac{l_c}{v}\right)\right\}}
\,.\nonumber
\end{align}

Let us introduce the tunneling current  operators
${\hat I_{T2} ,\hat I_{T3}}$, representing the tunneling currents from  edges $1$ and $4$ to edges $2$ and $3$.
They can be obtained from the
time evolution of the total charge operator on each edge, as
\begin{equation}\label{tunncurr1}
{\hat I_{T2}} =e \frac{d}{dt}{\hat N_2}=\frac{i e}{\hbar}[H,\hat N_2]= -\frac{ie}{\hbar}\left(\mathcal{C}+\mathcal{D}-h.c.\right)\,,
\end{equation}
\begin{equation}
\label{tunncurr2}
{\hat I_{T3}}= e \frac{d}{dt}{\hat N_3}=\frac{i e}{\hbar}[H,\hat N_3]= -\frac{ie}{\hbar}\left(\mathcal{A}+\mathcal{B}-h.c.\right)\,.
\end{equation}

We are now ready to address the zero-frequency tunneling current-current correlation. Once more, we can
introduce Keldysh time ordering and write
\begin{align}\label{stunnel}
S^{(T)}_{23}(0)=\frac{1}{2}\int_{-\infty}^{\infty}dt\; \sum_{\eta=\pm1} \langle \langle T_K \hat I_{T2}(0_\eta)\hat I_{T3}(t_{-\eta})\rangle \rangle \,.
\end{align}
Notice that we named the correlator $S^{(T)}_{23}$ to stress that it measures the correlation between tunneling currents.
Rewriting the operators in the interaction representation, we have
\begin{align}
 \resizebox{.85\hsize}{!}{$\displaystyle S^{(T)}_{23}(0)=\frac{1}{2}\int_{-\infty}^{\infty}dt\; \sum_{\eta=\pm1} \langle \langle T_K I_{T2}(0_\eta) I_{T3}(t_{-\eta}) \mathcal{S}_K\rangle \rangle \,.$}
\end{align}
In this case it is sufficient to expand $\mathcal{S}_K$ to second order in $H_T$ and find
\begin{multline}\label{s23t}
S_{23}^{(T)}(0)=\frac{(-i)^2}{4\hbar^2}\sum_{\eta=\pm1}  \int_{-\infty}^{\infty} dt \int_K d\tau_1 d\tau_2
\times \\  \langle \langle T_K  I_{T2}(0_\eta) I_{T3}(t_{-\eta})H_T(\tau_1)H_T(\tau_2)
\rangle \rangle \,.
\end{multline}
Collecting terms proportional to $\Gamma_A \Gamma_B^* \Gamma_C\Gamma_D^*$ from Eq.~(\ref{s23t}) we obtain a contribution
\begin{multline}\label{s23t-a} \frac{e^{i\Phi_{\rm AB}/\Phi_0}}{\hbar^4 v^4}
\Gamma_A \Gamma_B^* \Gamma_C\Gamma_D^* S_{\Phi}=\frac{e^2}{4\hbar^4}
 \sum_{\eta=\pm1} \int_{-\infty}^{\infty} dt \int_K d\tau_1 d\tau_2  \\
\langle T_K
\Big\{
\mathcal{C}(0_\eta)\mathcal{A}(t_{-\eta})\left[\mathcal{B}^\dagger(\tau_1)\mathcal{D}^\dagger(\tau_2)+\mathcal{D}^\dagger(\tau_1)\mathcal{B}^\dagger(\tau_2)\right]  \\
 -\mathcal{C}(0_\eta)\mathcal{B}^\dagger(t_{-\eta})\left[\mathcal{A}(\tau_1)\mathcal{D}^\dagger(\tau_2)+\mathcal{D}^\dagger(\tau_1)\mathcal{A}(\tau_2)\right]    \\
 -\mathcal{D}^\dagger(0_\eta)\mathcal{A}(t_{-\eta})\left[\mathcal{B}^\dagger(\tau_1)\mathcal{C}(\tau_2)+\mathcal{C}(\tau_1)\mathcal{B}^\dagger(\tau_2)\right] \\
+ \mathcal{D}^\dagger(0_\eta)\mathcal{B}^\dagger(t_{-\eta})\left[\mathcal{A}(\tau_1)\mathcal{C}(\tau_2)+\mathcal{C}(\tau_1)\mathcal{A}(\tau_2)\right] \Big\} \rangle\,.
\end{multline}

In Appendix \ref{app_integer_bosonic}, we obtain
\begin{widetext}
%\begin{small}
\begin{multline}\label{electrons_in_integer}
S_{\Phi}=\frac{e^2}{(2\pi)^4 }\sum_{ \eta_1,\eta_2} \eta_1 \eta_2
\int dt dt_1 dt_2
 e^{i\mu(L_1/v+L_4/v+t_1+t_2-t)/\hbar}
 \\ \times
\frac{1}{\frac{\hbar\beta }{\pi}\sinh[\frac{\pi}{\hbar\beta v}(-v\, t_1-L_1-i \eta_1 l_c    )]}
\frac{1}{\frac{\hbar\beta }{\pi}\sinh[\frac{\pi}{\hbar\beta v}(v\, t_2+L_2+i\eta_2 l_c )]}
\\ \times
\frac{1}{\frac{\hbar\beta }{\pi}\sinh[\frac{\pi}{\hbar\beta v}(v\, (t_1-t)+L_3+i \eta_1 l_c  )]}
\frac{1}{\frac{\hbar\beta }{\pi}\sinh[\frac{\pi}{\hbar\beta v}(v\, (t-t_2)-L_4-i \eta_2 l_c  )]}\,.
\end{multline}
%\end{small}
\end{widetext}
In order to perform the above integrals it is convenient to use the following relation
 \begin{equation}
 f(\pm \omega)=\pm\frac{ i}{2\hbar\beta}\int_{-\infty}^{\infty} dt \frac{1}{\sinh\left[\frac{\pi}{\hbar\beta v}(v t\pm il_c)\right]}e^{i\omega t}\,,
 \end{equation}
where $f(\omega)$ is the Fermi-Dirac function. As in Appendix \ref{app_calcs23}, we can represent  the correlation functions in terms of their Fourier transforms and perform the straightforward integrals and the Keldysh sums. Thus, the outcome of Eq.~(\ref{electrons_in_integer}) is equal to that of Eq.~(\ref{sphi}).

 It may seem that the effort of performing this calculation in bosonic representation has little merit, as we have obtained the same result as the fermionic analysis. Nevertheless, the expression in Eq.~\eqref{electrons_in_integer} would prove useful in Sec.~\ref{sec_elec_over_frac} where we analyze electron tunneling between edges with fractional filling factors.

%%%%%%%%%%%%%%%%%%%%%%%%%%%%%%%%%%%%%%%%%%%%%%%%%%%%%%%%%%%%%%
\section{Model of the interferometer at fractional filling factors}\label{fractionalmodel}
In the second part of this paper, we consider the HBT interferometer operating in the FQHE regime, specifically at a simple Laughlin filling factors $\nu=1/(2n+1)$. For the sake of being specific, we will discuss the $\nu=1/3$ case. In this case no straightforward  Landauer-B\"{u}ttiker analysis is readily available, and one is forced to study the system with the non-equilibrium methods introduced in the previous sections. We  consider  two possible scenarios: (i) the QPCs are almost open [cf. Fig.~\ref{scheme}(a)], and (ii) the QPCs  are  pinched [cf. Fig.~\ref{scheme}(b)]. As explained before, for case (ii)  only tunneling of  electrons
has to be taken into account. For case  (i),  instead, even at relatively low energy, the relevant tunneling operator is that of quasiparticles, which leads to a profoundly different behavior of the system, as will be shown in the following. Here, we first introduce the Hamiltonian describing the system in the absence of tunneling; then, with the introduction of the tunneling operators, we specialize our analysis to either case (i) or (ii).

\subsection{Fractional edge model}
In the fractional quantum Hall effect (FQHE) regime, the spectrum of the bulk excitations of the electronic liquid is gapped. As we saw in the integer filling factor case, the low-energy physics is confined to gapless excitations at the edges of the system.
For the Laughlin filling factors considered here, the dynamics of the edge channels is described by a chiral Luttinger liquid where the interaction constant $g$ is set equal to the filling fraction~\cite{wenbook} $\nu$.
Our geometry is depicted in Fig.~\ref{scheme}. The Hamiltonian of the  four edge channels reads as
\begin{equation}\label{fractionalhamiltonian}
H_0=\frac{\hbar v}{4\pi}\sum_{l=1}^4\int dx (\partial_x\phi_l)^2,
\end{equation}
where $v$ is the (renormalized) chiral velocity of the edge excitations. The four Bose fields $\{ \phi_l \}$ satisfy the
following commutation rules:
\begin{equation}\label{commutationrules}
\left[  \phi_l(x,t=0),\phi_k(x',t=0) \right]=i \pi \delta_{lk}\mbox{sgn}(x-x')\,.
\end{equation}
Note that we have chosen a convention where the commutation relations do \textit{not} contain $\nu$.

On each edge channel, the charge density operator is related to the bosonic field by
\begin{equation}\label{fractionaldensity}
\rho_k=-\frac{\sqrt{\nu }e}{2\pi}\partial_x\phi_k\,.
\end{equation}
Making use of the commutation rules in Eq.~(\ref{commutationrules}), one can verify that the operators $\mbox{exp}(i \phi_l/\sqrt{\nu} )$ and $\mbox{exp}(i \sqrt{\nu} \phi_l )$ are, respectively, proportional to the electron and the quasiparticle creation operator on the edge $l$.

\subsection{Model for electron tunneling}
Although electronic operators do anticommute on the same edge, in order to enforce anticommutation
between operators on different edges one has to introduce a set of Klein factors as was previously done in Sec.~\ref{integerbosonic}.  In most cases, i.e., in a two-edge geometry with a single tunnel barrier \cite{KaneFisher1992,martin-klein},  it is not necessary to introduce such Klein factors; in the case our geometry [cf.~Fig.~\ref{scheme}(b)], though, failing to do so would lead to a wrong sign in the expression for the current-current correlation.

The electron tunneling operators at the four QPCs read as
\begin{align}\label{tunneling-op-frac-e}
\mathcal{A}^{(e)}=&\frac{\Gamma_\mathcal{A}}{l_c} \, F_1^\dag F_3 \, e^{2\pi i \Phi_{\rm AB}/\Phi_0 } e^{i(\phi_1(0)-\phi_3(0))/\sqrt{\nu}}
\,,\nonumber\\
\mathcal{B}^{(e)}=&\frac{\Gamma_\mathcal{B}}{l_c}\, F_4^\dag F_3 \,e^{i (\phi_4(L_4)-\phi_3(L_3))/\sqrt{\nu}}\,,\nonumber \\
\mathcal{C}^{(e)}=&\frac{\Gamma_\mathcal{C}}{l_c}\, F_4^\dag F_2\, e^{i (\phi_4(0)-\phi_2(0))/\sqrt{\nu}}\,,\nonumber\\
\mathcal{D}^{(e)}=&\frac{\Gamma_\mathcal{D}}{l_c}\, F_1^\dag F_2\, e^{i(\phi_1(L_1)-\phi_2(L_2))/\sqrt{\nu}}\, .
\end{align}
where the tunneling Hamiltonian in this case reads as
\begin{equation}
H_T=\mathcal{A}^{(e)}+\mathcal{B}^{(e)}+\mathcal{C}^{(e)}+\mathcal{D}^{(e)}+h.c.
\end{equation}
Notice that we have reintroduced the Aharonov-Bohm phase due to the magnetic flux $\Phi_{\rm AB}$.  Electrons moving along the edges  acquire a phase given by $(e/hc)\oint d\vec{l} \cdot \vec{A}$.  We choose again a gauge where the total AB-phase, $\mbox{exp}(  2\pi i \Phi_{\rm AB} / \Phi_0)$, acquired by going around the loop $\mathcal{A}\rightarrow \mathcal{B} \rightarrow \mathcal{C} \rightarrow \mathcal{D}\rightarrow \mathcal{A}$, is assigned to the tunneling operator at QPC $\mathcal{A}$.
Once again, we assume that the external edge channels $1$ and $4$, connecting $S_1$ to $D_1$ and $S_4$ to $D_4$ respectively,
are kept at chemical potential $eV$. Edge channels $2$ and $3$, connecting $S_2$ to $D_2$ and $D_3$ to $S_3$ respectively, are grounded.

%%%%%%%%%%%%%%%%%%%%%%%%%%%%%%%%%%%%%%%%%%%%%%%%%%%%%%%%%%%%%%%%%%
\subsection{Model for quasiparticle tunneling}\label{modelquasiparticle}
%%%%%%%%%%%%%%%%%%%%%%%%%%%%%%%%%%%%%%%%%%%%%%%%%%%%%%%%%%%%%%%%%%
\begin{figure}
\includegraphics[width=8cm]{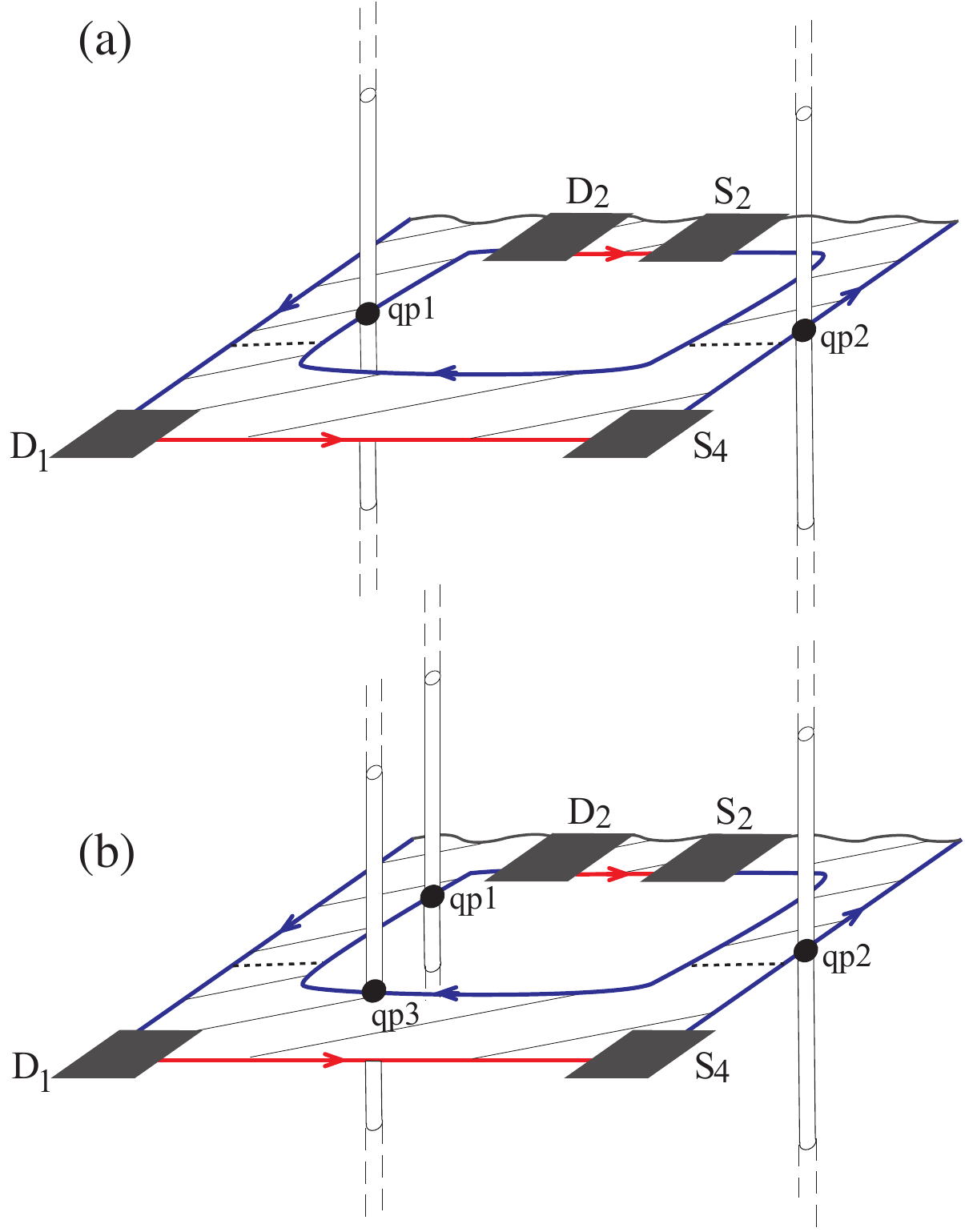}
\caption{A segment of the two-particle interferometer in two different topological states. In (a) there is one quasiparticle (qp1) trapped inside the interferometer, in this case $\Phi_{stat}=\Phi_0$ (assuming no quasiparticles in the other section of the interferometer). In (b) a second quasiparticle (qp3) has tunneled towards the inner edge of the interferometer; this corresponds to $\Phi_{stat}=2\Phi_0$. Notice that the quasiparticle moving on the external edge (qp2) does not contribute to the statistical flux. }
\label{trappedflux}
\end{figure}
%%%%%%%%%%%%%%%%%%%%%%%%%%%%%%%%%%%%%%%%%%%%%%%%%%%%%%%%%%%%%%%%%%

We now turn to the case of almost open QPCs. This situation corresponds to the edge geometry depicted in Fig.~\ref{scheme}(a); for such a geometry, tunneling of quasiparticles \emph{and} electrons can take place at the four QPCs between the edge channels. At Laughlin filling factor $\nu=1/(2n+1)$  quasiparticle tunneling  is more relevant, in a renormalization group sense, than electron tunneling. We focus on a range of temperatures and voltage that render
electron tunneling negligible as compared with quasiparticle tunneling. The low-energy physics of the edges is still described by the Hamiltonian in Eq.~(\ref{fractionalhamiltonian}), but we need to express the tunneling operators in terms of quasiparticle operators $\exp(i\sqrt{\nu}\phi_k)$. Notice that in this case, the
%tunneling amplitudes in the interaction representation will acquire a phase
%\begin{equation}
%\Gamma_\mathcal{A},\Gamma_\mathcal{B},\Gamma_\mathcal{C},\Gamma_\mathcal{D}\sim \exp(-i \nu eVt/\hbar) \, ,
%\end{equation} and the
Aharonov-Bohm phase reads as $(\nu e/hc)\oint d\vec{l} \cdot \vec{A}$.
The quasiparticle tunneling operators at the four QPCs are selected to be:
\begin{align}\label{tunneling-op-frac}
\mathcal{A}^{(q)}&=\frac{\Gamma_\mathcal{A}}{l_c} e^{2\pi i \nu \Phi_{\rm AB}/\Phi_0 } e^{i\sqrt{\nu}(\phi_1(0)-\phi_3(0))} \,,\nonumber \\
\mathcal{B}^{(q)}&=\frac{\Gamma_\mathcal{B}}{l_c}   e^{i\sqrt{\nu}(\phi_4(L_4)-\phi_3(L_3))} \,, \nonumber\\
\mathcal{C}^{(q)}&=\frac{\Gamma_\mathcal{C}}{l_c}  e^{i\sqrt{\nu}(\phi_4(0)-\phi_2(0))}\,,\nonumber\\
\mathcal{D}^{(q)}&=\frac{\Gamma_\mathcal{D}}{l_c}    e^{i\sqrt{\nu}(\phi_1(L_1)-\phi_2(L_2))}\,.
\end{align}
The tunneling Hamiltonian is then
\begin{equation}
H_T=\mathcal{A}^{(q)}+\mathcal{B}^{(q)}+\mathcal{C}^{(q)}+\mathcal{D}^{(q)}+h.c.\,.
\label{tunneling-op-frac-tun}
\end{equation}
Above, we have not included Klein factors in the tunneling operators. Because of the commutation rules of Eq.~(\ref{commutationrules}), it is straightforward to check that tunneling operators at different QPCs do not commute with each other.

Let us discuss the consequence of a naive calculations of the current-current correlation along the the same lines of the previous sections.
 The lowest nonvanishing contribution to the cross-current correlation would again be of the fourth order in the tunneling amplitudes. A contribution of this form would violate Byers-Yang theorem~\cite{byersYang}: the current-current correlation would show a periodicity of $(1/\nu) \, \Phi_0$ and not $\Phi_0$ as expected from a gauge invariance argument. The approach put forward in Refs.~[\onlinecite{kane03}] and [\onlinecite{feldman}] was to introduce a new set of two-body Klein operators. Such Klein factors would enforce the commutation between tunneling operators at different QPCs to be zero, i.e., restore locality in the tunneling Hamiltonian. A similar approach would be in principle possible here, but is technically hard to follow: unfortunately it turns out that in our case one is soon confronted with expressions which are hardly manageable (see Appendix \ref{twevlth_order_break}).

Here, we follow an alternative route: a quasiparticle in a quantum Hall liquid at Laughlin filling factor $\nu$ can be described as a composite object, consisting of a point charge $q=\nu e$ with a single magnetic flux quantum solenoid, $\Phi_0$, attached to it. When a quasiparticle encircles another quasiparticle, it will pick up an AB-phase $\theta= 2 \pi \nu$ which accounts for their mutual fractional statistics~\cite{Arovas84}. When a quasiparticle tunnels from one of the external edges to the internal ones, its flux is \emph{trapped} inside the interferometer. The magnetic flux enclosed in the active area of the interferometer is then
$\Phi_{\rm tot}=\Phi_{\rm AB}+\Phi_{stat}$, where $\Phi_{stat}$ is the \emph{statistical flux}~\cite{kane03,feldman,feldman2007,feldman2006,law2008} and is  given by $\Phi_0$ times the number $n$ of quasiparticles that have tunneled from the external to the internal edges. This is sketched in Fig. \ref{trappedflux}.  The dynamics of quasiparticles moving along the edges of the interferometer is then entirely determined by $n$ $mod(1/\nu)$. For a given value of
 $\Phi_{\rm AB}$ the system can be found in $1/\nu$ possible \emph{states} characterized by $n=0,1,...,(1/\nu-1)$.
We modify the tunneling operator at QPC $\mathcal{A}$ to include the statistical flux, we have $\mathcal{A}^{(q)} \rightarrow \mathcal{A}_n^{(q)}=(\Gamma_\mathcal{A}/l_c) e^{2\pi i \nu (\Phi_{\rm AB}+n \Phi_0)/\Phi_0 } e^{i\sqrt{\nu}(\phi_1(0)-\phi_3(0))} $.
In the next sections, we study the kinetic equation which describes the Markovian evolution of the statistical flux and calculate transition rates between the three possible flux states using the microscopic model defined above. This will allow us to derive an expression for $S_{\Phi}$ for the regime considered here.

%Again, going to the interaction representation  we have
%\begin{fleqn}[15pt]
%\begin{multline}
%\mathcal{A}_n^{(q)}(t)=\Gamma_\mathcal{A} e^{i e \nu Vt}e^{2\pi i \nu (\Phi_{\rm AB}+n \Phi_0)/\Phi_0 } e^{i\sqrt{\nu}(\phi_1(0,t)-\phi_3(0,t))} ,\\
%\\
%\mathcal{B}^{(q)}(t)=\Gamma_\mathcal{B} e^{i e \nu Vt}e^{-i e \nu V L_4/v}   e^{i\sqrt{\nu}(\phi_4(L_4,t)-\phi_3(L_3,t))}, \\
%\\
%\mathcal{C}^{(q)}(t)=\Gamma_\mathcal{C}e^{i e \nu Vt}   e^{i\sqrt{\nu}(\phi_4(0,t)-\phi_2(0,t))}, \\
%\\
%\mathcal{D}^{(q)}(t)=\Gamma_\mathcal{D}e^{i e \nu Vt} e^{-i e \nu V L_1/v}    e^{i\sqrt{\nu}(\phi_1(L_1,t)-5\phi_2(L_2,t))} . \\
%\end{multline}
%\end{fleqn}
%Notice that in the Aharonov-Bohm we have also included the contribution due to the statistical phase.

\section{Current-current correlation: electron tunneling case}
\label{sec_elec_over_frac}
Here, we consider the part of the current-current correlation that depends on the magnetic flux, $S_{\Phi}$.
For the calculation of such a quantity in the case of tunneling of electrons, we show that it is possible to obtain an analytic expression for the fractional filling factor $\nu=1/(2n+1)$ using the results previously obtained for the integer case $\nu=1$.
Once more,  as in Sec.~\ref{integerbosonic}, we introduce tunneling currents and calculate their correlations. In this case, it follows from Eq.~(\ref{fractionaldensity}) that the charge operator on edge $i$ is $N_i=-(\sqrt{\nu}e/2\pi)\int dx \,\partial_x \phi_i$.
The tunneling currents $I^{(e)}_{T2}$ and $I^{(e)}_{T3}$ are obtained by the equation of motion for the total charge on edges 2 and 3. Here, we have
\begin{equation}
\hat{I}^{(e)}_{T2} = \frac{d}{dt}{\hat Q_2}=-\frac{ie}{\hbar}\left(\mathcal{C}^{(e)}+\mathcal{D}^{(e)}-h.c.\right)\,, \end{equation}
\begin{equation}
\hat{I}^{(e)}_{T3}=  \frac{d}{dt}{\hat Q_3}=-\frac{ie}{\hbar}\left(\mathcal{A}^{(e)}+\mathcal{B}^{(e)}-h.c.\right)\,.
\end{equation}
As previously done in Sec.~\ref{integerbosonic}, we can express the zero-frequency current-current correlations of tunneling current
$I^{(e)}_{T2}$ and $I^{(e)}_{T3}$ as
\begin{equation}\label{stunnel-fractional}
\resizebox{.85\hsize}{!}{$\displaystyle S^{(e)}_{23}(0)=\frac{1}{2}\int_{-\infty}^{\infty}dt\; \sum_{\eta} \langle \langle T_K I^{(e)}_{T2}(0_\eta) I^{(e)}_{T3}(t_{-\eta}) \mathcal{S}_K\rangle \rangle $}\,,
\end{equation}
where $\mathcal{S}_K$ was defined in Eq.~\eqref{keldysh_action}.
In this case $H_0$ is given by Eq.~(\ref{fractionalhamiltonian}) and the tunneling operators by Eq.~(\ref{tunneling-op-frac-e}).
To the lowest non-trivial order in the tunneling amplitudes we have
\begin{multline}\label{s23-nu-e}
S^{(e)}_{23}(0)=\frac{(-i)^2}{4\,\hbar^2}\int_{-\infty}^{\infty} dt \int_K d\tau_1 d\tau_2
\times \\  \langle \langle T_K  I^{(e)}_{T2}(0_\eta) I^{(e)}_{T3}(t_{-\eta})H_T(\tau_1)H_T(\tau_2)
\rangle \rangle \, .
\end{multline}
The analysis of Eq.~(\ref{s23-nu-e}) is formally identical to the analysis performed for Eq.~(\ref{stunnel}) (cf. Appendix \ref{app_integer_bosonic}) but with a different expression for the Keldysh Green's function. In this case we have
\begin{widetext}
\begin{align}
G_{i}^{\eta_1\eta_2}(x_1,t_1;x_2,t_2) &=
%\\ &\resizebox{.9\hsize}{!}{$\displaystyle
 (1/l_c)\langle T_K F_i(t_{1,\eta_1})F_i^\dag(t_{2,\eta_2})      \rangle
\langle T_K e^{-i\phi_i(x_1,t_{1,\eta_1})/\sqrt{\nu}} e^{i\phi_i(x_2,t_{2,\eta_2})/\sqrt{\nu}}        \rangle
%$}
\nonumber
\\
%&\resizebox{.9\hsize}{!}{$\displaystyle
&=
\frac{ l_c^{(1/\nu-1)} e^{i \mu_i[(x_1-x_2)/v-(t_1-t_2)]/\hbar }}
{\left(\frac{\hbar \beta  v}{\pi} \sin\left\{\frac{\pi}{\hbar \beta v}\left(i[(t_1-t_2)v-(x_1-x_2)]+\chi_{\eta_1,\eta_2}(t_1-t_2)\, l_c\right)\right\}\right)^{1/\nu}}
%$}
\,,
\label{green-e}
\end{align}
\end{widetext}

Here too we focus on the flux-dependent part of the current-current correlation and collect
terms proportional to $\Gamma_\mathcal{A} \Gamma_\mathcal{B}^* \Gamma_\mathcal{C} \Gamma_\mathcal{D}^*$ of Eq.~(\ref{s23-nu-e}). In this case we call this quantity $S^{(e)}_{\Phi}$.
As we show in Appendix \ref{app_equiv_bos_ferm}, to compute $S^{(e)}_{\Phi}$ does not require much effort, it can be derived from the expression of $S_{\Phi}$ [Eq.~(\ref{sphi})] in a straightforward  way: The original $S_{\Phi}$ [Eq.~(\ref{sphi})] has been calculated for the geometry depicted in Fig.~\ref{scheme}(a). In the present case, we rely on the geometry of Fig.~\ref{scheme}(b), which involves substitutions, $L_1\leftrightarrow  L_2$, $L_3\leftrightarrow  L_4$, $\Gamma_\mathcal{A}\leftrightarrow  \Gamma_\mathcal{C}$. Following this substitution, by taking derivatives with respect to the interferometer arm lengths, we obtain for any $\nu=1/(2n+1)$
\begin{multline}\label{derivatives1}
S^{(e)}_{\Phi}=l_c^{4(1/\nu-1)}\\
\times \prod_{i=1}^4 \prod_{j=1}^{(1/\nu-1)/2}\frac{1}{j(j+1)}\left[\frac{\partial^2}{\partial L_i^2}-\left(\frac{j\pi}{\hbar \beta v}\right)^2\right]S_{\Phi}  \,.
\end{multline}

\begin{figure*}
\includegraphics[width=\textwidth]{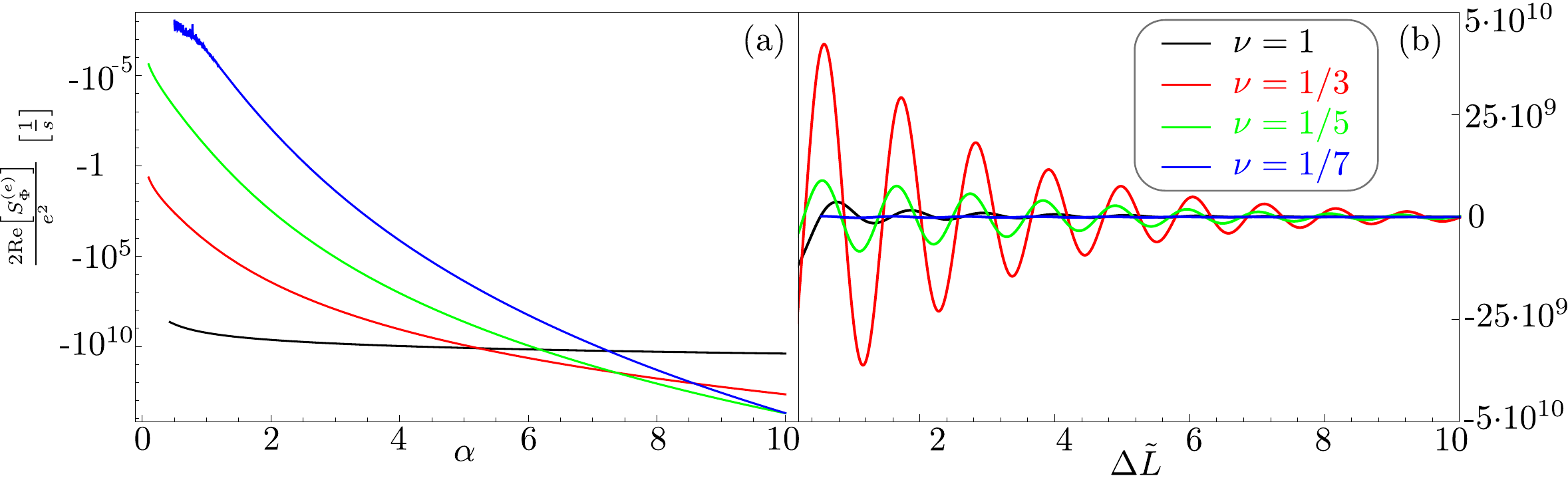}
\caption{\textit{Electronic cross-current correlations}. Plots of the cross-current correlations $2{\rm Re}\left\{S_{\Phi}^{(e)}\right\}$ at $\Phi_{\rm AB}=0$ [cf.~Eqs.~\eqref{sphi} and \eqref{derivatives1}] as a function of (a) voltage [$\alpha=e V \beta/(2\pi)$] for interferometer arms mismatch $\Delta \tilde{L}=\pi(L_1+L_4-L_2-L_3)/(\hbar \beta v)\rightarrow 0$, and (b) $\Delta \tilde{L}$ for $\alpha=6$. Different curves are for filling fractions $\nu=1$, $\nu=1/3$, $\nu=1/5$, and $\nu=1/7$ (black, red, green, and blue, respectively). We assume the following experimental values: temperature~\cite{neder2007} $T=10mK$, and edge veolcity~\cite{McClure:2009} $v\sim 1.5\cdot 10^5 m/s$. For presentation reasons we took the ratio of cutoff length over thermal length to be $\pi l_c/(\hbar \beta v)=0.18$.}
\label{plote}
\end{figure*}

In Fig.~\ref{plote}, we plot the dependence of $S_{\Phi}$ on voltage and for different values of $\nu$ using realistic experimental values. Notice that the function is initially negative -- a sign of fermionic statistics of the interfering electrons.
%%%%%%%%%%%%%%%%%%%%%%%%%%%%%%%%%%%%%%%%%%%%%%%%%%%%%
%%%%%%%%%%%%%%%%%%%%%%%%%%%%%%%%%%%%%%%%%%%%%%%%%%%%%

%%%%%%%%%%%%%%%%%%%%%%%%%%%%%%%%%%%%%%%%%%%%%%%%%%%%
\section{Current-current correlation: quasiparticle tunneling case}
\label{fractionalqp}
In this section, we address the calculation of current-current correlations for the quasiparticle tunneling case. Initially, one may introduce
a set of Klein factors in the tunneling operators and use a Keldysh nonequilibrium approach. This analysis leads, for example, for $\nu=1/3$ to a lowest nonvanishing contribution of $12^{\rm th}$ order in the tunneling perturbation. In Appendix \ref{twevlth_order_break}, we show that, following realistic assumptions, a $12^{\rm th}$-order Keldysh perturbation theory may be dramatically simplified. The main observation underlined in Appendix \ref{twevlth_order_break} is that complex processes along a Keldysh contour may  be decoupled into three consecutive $4^{\rm th}$ order processes, each corresponding to a two $e^\star=1/3$ anyon scattering. The time interval between two consecutive two-anyon processes may be ``dressed'' by single anyon scattering processes. This picture constitutes the basis to the rate equation analysis employed here.

We study the Markovian evolution of the statistical-flux trapped inside the interferometer. Thus, we derive an expression for current-current correlations in terms of quasiparticle transfer rates. In Section \ref{kinetic}, we illustrate the general framework for such a calculation, and in Section \ref{calculationrates} we derive explicit expressions for the quasiparticle transfer rates using the microscopic model introduced in Section \ref{modelquasiparticle}. The analysis presented here for anyonic HBT interferometry applies to all simple odd fraction $\nu$ (Laughlin states). To facilitate the discussion, below we first present our analysis for $\nu=1/3$. Then, consequently, we present expressions for general Laughlin fractions $\nu$.

%%%%%%%%%%%%%%%%%%%%%%%%%%%%%%%%%%%%%%%%%%%%%%%%%%%%%%%%%%%%%%
\subsection{Kinetic approach}\label{kinetic}
In order to calculate the current-current correlations using a master-equation formalism, we introduce a general formalism following Refs.~[\onlinecite{Korotkov94,Koch06}] and adapt it to the problem considered here. The evolution of our system at $\nu=1/3$ is among three possible values of the statistical flux. It is governed by a standard master equation
\begin{equation}
	\resizebox{.85\hsize}{!}{$\displaystyle\frac{d}{dt}P(f,t|i)=\sum_{k =0,1,2 }\left[P(k,t|i)W_{kf}-P(f,t|i)W_{fk}\right]\, .$}
	\label{master}
\end{equation}
Here,  $P(f,t|i)$ denotes the conditional probability to find the system in the statistical flux state $\ket{f}$ at time $t$ given that at time $t=0$  the system was in the state $\ket{i}$, i.e., $P(f,t=0|i)=\delta_{if}$.
$W_{if}$ is the transition rate from the statistical flux state $\ket{i}$ to state $\ket{f}$. Several processes can contribute to a specific $W_{if}$.
Hence, we write these rates as
\begin{align}
	W_{if}=\sum_{\zeta}W_{if}^{(\zeta)}\, ,
	\label{rate_resum}
\end{align}
where $\zeta$ labels the \emph{elementary} processes $\zeta$ contributing to the \emph{total} rate $W_{if}$.

\begin{figure}[!ht]
\includegraphics[width=0.5\columnwidth]{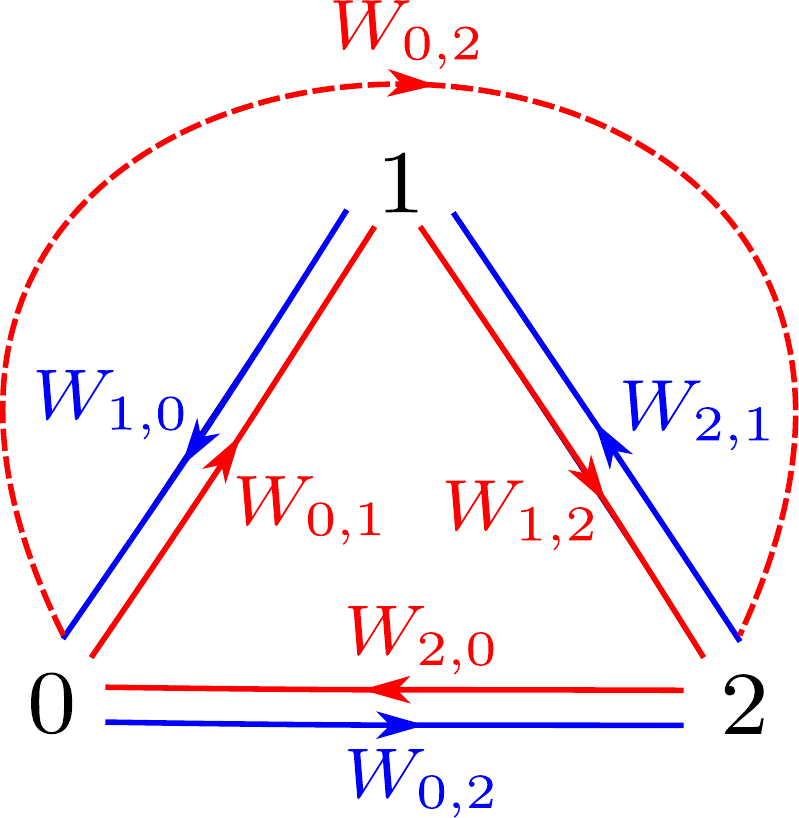}
\caption{Illustrated are the three possible values of the topological flux. The arrows represents transitions
between different states. Clockwise transitions (red solid arrows) represent processes which increase the statistical flux by one (red solid arrows) or two (red dashed arrows). In the high-voltage limit, considered here, we disregard counter-clockwise processes (blue solid arrows) which reduce the statistical flux by taking quasiparticles against the voltage gradient. Similarly, we disregard processes which do not change the topological flux, e.g.,~a process in which a quasiparticle tunnels from edge 1 to edge 3 and finally to edge 4 [cf.~Fig.~\ref{scheme}(a)].}
\label{rates}
\end{figure}

Let us consider the case of $W_{i,i+1}$, i.e.,~processes that increase the statistical flux by one. Note that here and in the following, the indices appear $\mod\, (3)$. A tunneling of a quasiparticle across any of the four QPCs contributes to $W_{i,i+1}$. However, what charge is transferred between which source and drain depends on the specific process. As an example, tunneling of a quasiparticle across QPC $\mathcal{A}$  corresponds to a charge
transfer from edge 1 to edge 3. Consequently, we measure a charge $-q$ in drain $D_1$ and a charge $+q$ in drain $D_3$.
This corresponds to tunneling currents, i.e., currents are measured with respect to the background currents at the drains when the tunneling Hamiltonian is absent.

As in the previous sections, we are interested in the magnetic-flux-modulated component of the cross-current correlation [cf.~Eqs.~\eqref{sphi} and \eqref{derivatives1}]. Due to the geometry of our interferometer, processes involving tunneling of only one quasiparticle, as in the above example, are not sensitive to such a flux. Hence, in order to see flux-dependence in the cross-current correlation, we need to consider AB-dependent processes where tunneling of two or more quasiparticles takes place. Here we add two-quasiparticle processes to our analysis. Note that (i) these processes are calculated using the generalized Fermi golden rule in Sec.~\ref{calculationrates}, (ii) in the following we shall see that the lowest nonvanishing AB-dependent contribution corresponds to a sequence of three two-quasiparticle processes. Note that as far as the leading flux sensitive terms are concerned, this is the most important contribution; due to phase-space arguments, we do not need to include four- and six-quasiparticle processes into our analysis.

Let us define the quantities to be used in the following; $\mean{ I_a} $ is the average current due to tunneling events measured in drain $a$, ($a={1,2,3,4}$). The zero-frequency current-current correlation between two drains $a$ and $b$ is:
\begin{equation}
S_{a,b}=\int_{-\infty}^{\infty} dt \left[  \mean{ I_a(t)I_b(0)}- \mean{I_{a}(t) } \mean{ I_{b}(0)} \right]\,.
\label{zerofreq}
\end{equation}
%In the following part of the article we consider only the case $\nu=1/3$, since the tunneling of quasiparticles inside the interferometer is sensitive to the total magnetic flux  $\Phi_{\rm AB}+\Phi_t$, the system's dynamic can be described by only three "states".{\bf Explain why we introduce these quantities}
%The transition  rate $W_{if}$, refers {\em only to} the change in the number of trapped fluxes.  Several processes contribute to
%the total transition rate between two states. For instance let us consider the case of tunneling of one quasiparticle from the outer to the inner edges of the interferometer---the tunneling can occur  through any of the four QPC's and it will give rise to a  clockwise transition (see Fig. \ref{rates}). On the other hand each of these elementary process is characterized by a different charge transfer between the  terminals. In order to distinguish all the possible processes contributing the total transition rate between two states it is convenient to  define:
The master equation can be written in a more compact form by introducing the following matrix:
\begin{align}
	\resizebox{.85\hsize}{!}{$\displaystyle \mathbb{W}=\left(
\begin{array}{c c c}
	-\sum_{k\neq 0}W_{0k} & W_{10} & W_{20}\\
	W_{01} & -\sum_{k\neq 1}W_{1k} & W_{21}\\
	W_{02} & W_{12} & -\sum_{k\neq 2}W_{2k}
\end{array}\right)\, .$}
\label{rate_matrix}
\end{align}
Thus, Eq.~(\ref{master}) can be rewritten in matrix form as
\begin{align}
	\frac{d}{dt}\mathbf{p}_i(t)=\mathbb{W}\mathbf{p}_i(t)\, ,
\end{align}
with the probability vector
\begin{align}
	\mathbf{p}_i(t)=(P(0,t|i), P(1,t|i),P(2,t|i))\, .
\end{align}
Taking the initial condition $\mathbf{p}_i(t=0)=\hat{e}_i$, the master equation has a formal solution
\begin{align}
	\mathbf{p}_i(t)=e^{\mathbb{W}t}\hat{e}_i\, .
\end{align}

We are interested in the normalized stationary distribution, $\mathbf{P}$, which is the solution of
\begin{align}\label{stationarydistribution}
	\mathbb{W} \, \mathbf{P}=0\, .
\end{align}
Let us introduce a matrix $E$ with all entries set to $1$, one can show~\cite{Korotkov94,Koch06} that $\mathbb{W} + E$ is invertible and that applying its inverse to the vector $\mathbf{e} = (1, 1, 1)$ yields the stationary probability distribution
\begin{align}
	\mathbf{P}=\left(\mathbb{W}+E\right)^{-1}\mathbf{e}\, .
\end{align}

 Here we report general formulas to obtain currents and current-current correlations from the stationary distribution function appearing in Eq.~(\ref{stationarydistribution}). Details can be found in Refs.~[\onlinecite{Korotkov94,Koch06}]. The stationary current in a given drain $a$ can be written as
\begin{align}\label{me-current}
	\mean{I_a}=q\sum_{i,f}\sum_{\nu}s_{if,\zeta}^a P_i W_{if}^{(\zeta)}\, .
\end{align}
Here the coefficents $\{ s_{if,\zeta}^a\}$ specify the charge transfer from a given source to a given drain when a process $\zeta$ takes place (cf. Table \ref{table1}).
The zero-frequency current-current correlation reads as~\cite{Korotkov94,Koch06}
\begin{align}\label{curr-curr-general}	S_{ab}=2q^2\Bigg[
\underbrace{
\text{tr}\left\{\mathbf{u}_{ab}\right\}
}_{\mbox{auto}}
-\underbrace{
\mathbf{w}_{b}\mathbb{W}^{-1}\bar{\mathbf{y}}_{a}-\mathbf{w}_{a}\mathbb{W}^{-1}\bar{\mathbf{y}}_{b}}_{\mbox{cross}}
\Bigg]\, ,
\end{align}
with ``auto'' terms corresponding to processes that affect both drains $a$ and $b$ inherently, and ``cross'' terms corresponding to correlations between two different processes that affect $S_{ab}$. They are expressed by
\begin{align}
	(u_{ab})_i=&\sum_{f}\sum_{\zeta}s_{if,\nu}^a s_{if,\nu}^b P_i W_{if}^{(\zeta)}\, ,\\
	(y_{a})_j=&\sum_{i}\sum_{\zeta}s_{ij,\nu}^a P_i W_{ij}^{(\zeta)}\, ,\\
	(w_{b})_k=&\sum_{f}\sum_{\zeta}s_{kf,\nu}^b W_{kf}^{(\zeta)}\, ,\\
	(\bar{y}_{a})_j=&(y_{a})_j-\frac{\mean{I_a}P_j}{q}\, .
\end{align}

%%%%%%%%%%%%%%%%%%%%%%%%%%%%%%%%%%%%%%%%%%%%%%%%%%%%%
\begin{figure}
\includegraphics[width=\columnwidth]{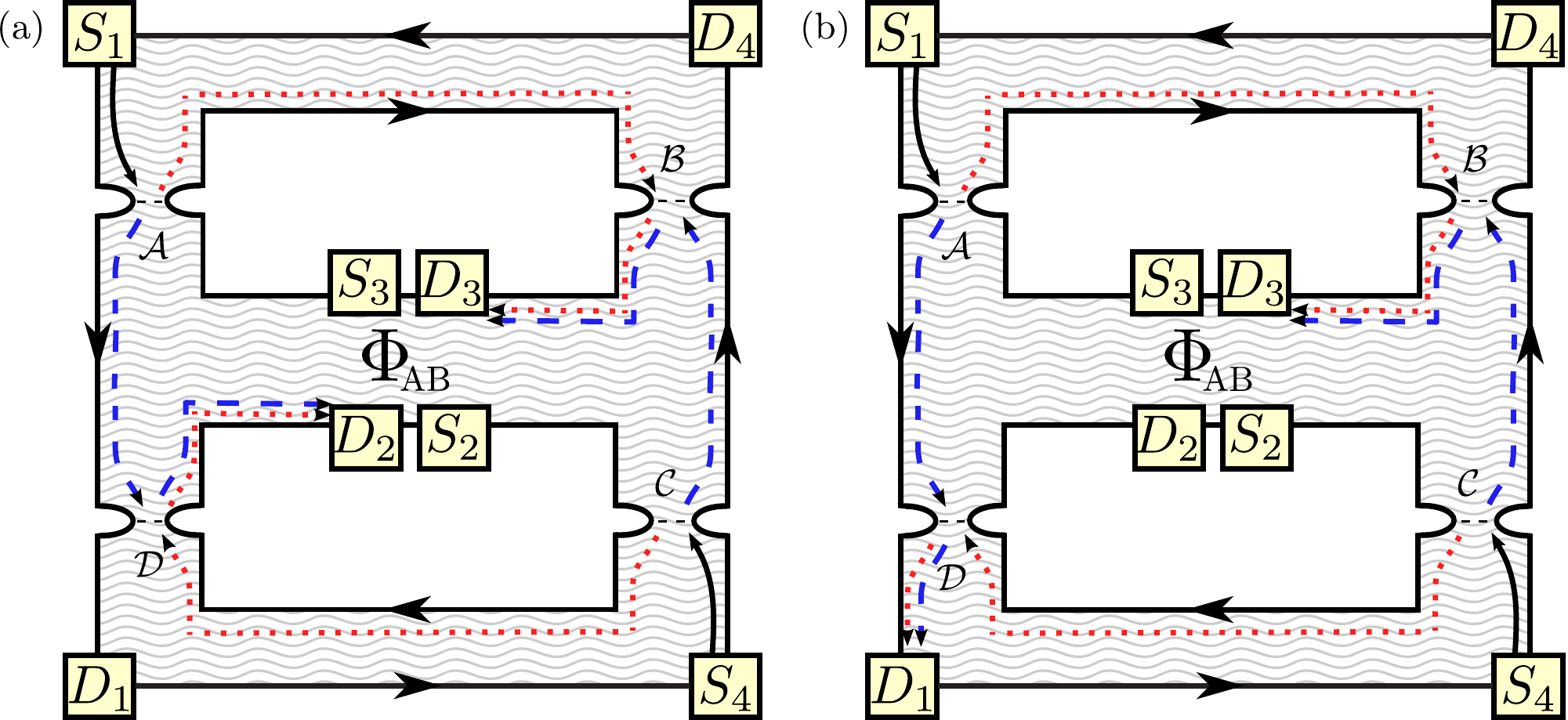}
\caption{Illustration of two-particle processes. (a) In process $(2,\mathcal{A}\mathcal{B}\mathcal{C}\mathcal{D},\Phi_{\text{tot}}(j))$ two QPs are transferred from edges 1 and 4 to edges 2 and 3, the process is AB-sensitive due to the interference between two amplitudes $A_1$ and $A_2$. In $A_1$ a QP tunnels from edge 1 to edge 3 and a second QP tunnels from edge 4 to edge 2 (red dotted line). In $A_2$ a QP tunnels from edge 1 to edge 2 and a second QP tunnels from edge 4 to edge 3 (blue dashed line). This process changes the statistical flux by two. (b) Process $(1,\mathcal{A}\mathcal{B}\mathcal{C}\mathcal{D},\Phi_{\text{tot}}(j))_1$ (and similarly process $(1,\mathcal{A}\mathcal{B}\mathcal{C}\mathcal{D},\Phi_{\text{tot}}(j))_2$) is also AB-sensitive but in this case only one QP is trapped inside the interferometer changing the statistical flux by one.}
\label{processes}
\end{figure}
%%%%%%%%%%%%%%%%%%%%%%%%%%%%%%%%%%%%%%%%%%%%%%%%%%%%%

We consider processes of second and fourth order in the tunneling amplitudes $\Gamma_{\mathcal{A}}, \Gamma_{\mathcal{B}}, \Gamma_{\mathcal{C}}$, and $\Gamma_{\mathcal{D}}$. We limit our analysis to the case of large-bias voltage $V$ compared to the thermal energy $k_B T$. This assumption allows us to disregard processes that result in transfer of quasiparticles from the inner edges (2 and 3) to the outer ones (1 and 4). The probability of such events is thus exponentially suppressed in the ratio $\nu eV/ (k_B T)$ (see Sec.~\ref{calculationrates}).

%%%%%%%%%%%%%%%%%%%%%%%%%%%%%%%%%%%%%%%%%%%%%%%%%%%%%%%%
\begin{table}[!ht]
\begin{tabular}{|l|c|c|r|c|c|r|}
 \hline
 \multicolumn{7}{|c|}{Elementary processes} \\
 \hline
  Process $\zeta$          & Order    & $(j,f)$ & $D_1$    &   $D_2$  &  $D_3$   & $D_4$  \\
  \hline
  $(1,\mathcal{A},0)$  & $\Gamma^2$ & $(j,j+1)$ & $-1$ & $0$ & $1$ &  $0$ \\
  $(1,\mathcal{B},0)$  & $\Gamma^2$ & $(j,j+1)$ &  $0$ & $0$ & $1$ & $-1$ \\
  $(1,\mathcal{C},0)$  & $\Gamma^2$ & $(j,j+1)$ &  $0$ & $1$ & $0$ & $-1$ \\
  $(1,\mathcal{D},0)$  & $\Gamma^2$ & $(j,j+1)$ & $-1$ & $1$ & $0$ &  $0$ \\
  $(2,\mathcal{A},0)$  & $\Gamma^4$ & $(j,j+2)$ & $-2$ & $0$ & $2$ &  $0$ \\
  $(2,\mathcal{B},0)$  & $\Gamma^4$ & $(j,j+2)$ &  $0$ & $0$ & $2$ & $-2$ \\
  $(2,\mathcal{C},0)$  & $\Gamma^4$ & $(j,j+2)$ &  $0$ & $2$ & $0$ & $-2$ \\
  $(2,\mathcal{D},0)$  & $\Gamma^4$ & $(j,j+2)$ & $-2$ & $2$ & $0$ &  $0$ \\
  $(2,\mathcal{A}$$\mathcal{B},0)$ & $\Gamma^4$ & $(j,j+2)$ & $-1$ & $0$ & $2$ & $-1$ \\
  $(2,\mathcal{C}$$\mathcal{D},0)$ & $\Gamma^4$ & $(j,j+2)$ & $-1$ & $2$ & $0$ & $-1$ \\
  $(2,\mathcal{A}$$\mathcal{D},0)$ & $\Gamma^4$ & $(j,j+2)$ & $-2$ & $1$ & $1$ &  $0$ \\
  $(2,\mathcal{B}$$\mathcal{C},0)$ & $\Gamma^4$ & $(j,j+2)$ &  $0$ & $1$ & $1$ & $-2$ \\
  $(2,\mathcal{A}\mathcal{B}\mathcal{C}\mathcal{D},\Phi_{\text{tot}}(j))$        & $\Gamma^4$ & $(j,j+2)$ & $-1$ & $1$ & $1$ & $-1$ \\
  $(1,\mathcal{A}\mathcal{B}\mathcal{C}\mathcal{D},\Phi_{\text{tot}}(j))_{1}$    & $\Gamma^4$ & $(j,j+1)$ &  $0$ & $0$ & $1$ & $-1$ \\
  $(1,\mathcal{A}\mathcal{B}\mathcal{C}\mathcal{D},\Phi_{\text{tot}}(j))_{2}$    & $\Gamma^4$ & $(j,j+1)$ & $-1$ & $1$ & $0$ &  $0$ \\
  \hline
\end{tabular}
\caption{Elementary QP transfer processes. Each process, $(\zeta)=(m,N,\phi)$, is characterized according to the change, $m$, in the number of QPs trapped in the interferometer; $N$ the QPCs at which QP tunneling takes place; and the flux, $\phi$, entering the flux factor $\kappa_j^{(m,N,\phi)}=\cos[2\pi \phi/(3\Phi_0)]$. Note that $\phi=0$ depicts a flux-independent process and $\phi=\Phi_{\text{tot}}(j)=\Phi_{\text{AB}}+j\cdot\Phi_0$ a process that depends on the total trapped flux. The order of the process (second or fourth in the tunneling amplitude $\Gamma$), the initial and final fluxon states [$(j,f)$, where $f-j$ is the added number of statistical fluxons], and the charge added at each drain ($+1$ refers to the absorption of one QP or charge $q=-(1/3)|e|$ at the drain), are indicated. For example (cf. Fig.~\ref{processes}), the process $(\zeta)=(1,\mathcal{A},0)$ corresponds to the emission of a QP from source $S_1$, its tunneling across QPC $\mathcal{A}$, and its trapping at $D_3$. Following the tunneling event a quasi-hole is created at edge $\overline{S_1 D_1}$ and a charge $-q$ is consequently absorbed in $D_1$. The flux dependent processes [the two-QPs trapping process $(2,\mathcal{A}\mathcal{B}\mathcal{C}\mathcal{D},\Phi_{\text{tot}}(j))$ and the single-QP trapping $(1,\mathcal{A}\mathcal{B}\mathcal{C}\mathcal{D},\Phi_{\text{tot}}(j))_{1}$] are illustrated in Fig.~\ref{processes}.
} \label{table1}
\end{table}
%%%%%%%%%%%%%%%%%%%%%%%%%%%%%%%%%%%%%%%%%%%%%%%%%%%%%%%%

Let us summarize the processes that we take into account (cf. Table \ref{table1}). We have four contributions of second order in the tunneling amplitude corresponding to tunneling of a single quasiparticle across any of the four QPCs. These processes do not depend on the flux of the magnetic field, and increase the statistical flux by one (clockwise transition in Fig.~\ref{rates}). For instance, the rate relative to tunneling through QPC $\mathcal{A}$ can be written as
\begin{equation}\label{1-flux-2rate}
W^{(1,\mathcal{A},0)}_{i,i+1}=|\tilde{\Gamma}_{\mathcal{A}}|^2 \gamma(V,T,\nu)\,,
\end{equation}
where $\tilde{\Gamma}_{\mathcal{A}}=\Gamma_{\mathcal{A}}/(\hbar v)$, and $\gamma(V,T,\nu)$ is calculated in Sec.~\ref{calculationrates} and expresses the microscopic details of our model. Similar expressions are obtained for the second-order processes at the other QPCs.

Next, we consider fourth-order processes. Here, the classification of processes is more interesting. We have processes that change the statistical flux by one or by two. They can be either flux-dependent or flux-independent. For instance, flux-independent processes may involve coherent transfer of two quasiparticles  across QPC $\mathcal{A}$ (corresponding to a process of order $|\Gamma_{\mathcal{A}}|^4$ )
or a coherent transfer of two quasiparticles across QPCs $\mathcal{A}$ and $\mathcal{D}$ (corresponding to a process of order $|\Gamma_{\mathcal{A}}|^2|\Gamma_{\mathcal{D}}|^2$). Such processes (e.g.,~$W_{02}$, cf.~Fig.~\ref{rates}) correspond to two-step clockwise transitions. Note that at finite temperature there are processes that correspond to a quasiparticle tunneling \emph{against} the voltage gradient (counter-clockwise transitions in Fig.~\ref{rates}). Assuming that the bias voltage is larger than the temperature, $eV\gg T$, these processes are ignored here. We do not calculate the flux-independent fourth-order processes explicitly as they would not enter the lowest order flux-sensitive contribution to the cross-current correlation.
We just denote the general structure of these flux-independent terms. For instance,
the process involving tunneling across QPCs $\mathcal{A}$ and $\mathcal{D}$ will read as $W_{i,i+2}^{(2,\mathcal{A}\mathcal{D},0)}=\Omega_0(V,T,\nu)|\tilde{\Gamma}_{\mathcal{A}}|^2|\tilde{\Gamma}_{\mathcal{D}}|^2$.

Most interesting for us are fourth-order flux-dependent processes, i.e., sensitive to the applied magnetic field \emph{and} to the number of trapped statistical fluxes. In the large-bias-voltage $V$ limit, considered here, there are three processes of this kind: (i) a process denoted by $(2,\mathcal{A}\mathcal{B}\mathcal{C}\mathcal{D},\Phi_{\text{tot}}(j))$ where two quasiparticles are transferred from the external edges 1 and 4 to the internal edges 2 and 3 via and interfering process of two quasiparticles  [cf.~Fig.~\ref{processes}(a)]. Such a process changes the statistical flux by two, i.e.~it corresponds to a counter-clockwise transition in Fig.~\ref{rates}. The corresponding rate can be written as
\begin{multline}\label{2-flux-4rate}
W_{i,i+2}^{(2,\mathcal{A}\mathcal{B}\mathcal{C}\mathcal{D},\Phi_{\text{tot}}(j))}=|\tilde{\Gamma}_{\mathcal{A}}\tilde{\Gamma}_{\mathcal{B}}\tilde{\Gamma}_{\mathcal{C}}\tilde{\Gamma}_{\mathcal{D}}|
\Omega(V,T,\nu,\Delta L)\\\cos \left[  2\pi \nu (\Phi_{\rm AB}+j \, \Phi_0)/\Phi_0 \right]\,,
\end{multline}
where $\Delta L=(L_1+L_4-L_2-L_3)$ is the length-asymmetry of the interfering paths, and $\Omega(V,T,\nu,\Delta L)$ is calculated in Section \ref{calculationrates} and expresses the microscopic details of our model.

The remaining two flux-dependent processes correspond to events where a single quasiparticle is transferred to an inner edge of the interferometer. Thus, these processes change the statistical flux by one, i.e.,~they correspond to clockwise transitions in Fig.~\ref{rates}. In Fig.~\ref{processes}(b), we illustrate the amplitudes constructing such an interfering process. The corresponding rates can be written as
\begin{multline}\label{1-flux-4rate}
W_{i,i+1}^{(1,\mathcal{A}\mathcal{B}\mathcal{C}\mathcal{D},\Phi_{\text{tot}}(j))_{k}}=|\tilde{\Gamma}_{\mathcal{A}}\tilde{\Gamma}_{\mathcal{B}}\tilde{\Gamma}_{\mathcal{C}}\tilde{\Gamma}_{\mathcal{D}}|
\tilde{\Omega}(V,T,\nu,\Delta L)\\\cos \left[ 2\pi\nu (\Phi_{\rm AB}+j \, \Phi_0)/\Phi_0 \right]\,,
\end{multline}
with $k=1,2$ and $\tilde{\Omega}(V,T,\nu,\Delta L)$ is calculated in Section \ref{calculationrates} corresponding to the microscopic details of our model.

Note that the functions $\Omega(V,T,\nu,\Delta L)$ and $\tilde{\Omega}(V,T,\nu,\Delta L)$ are not independent. Even without a microscopic derivation of their expressions, we can show their relationship. Let us, for example, consider the
average current in drain $D_3$ and in particular its component which might appear to be flux-dependent. Let us write the matrix $\mathbb{W}$ of Eq.~(\ref{rate_matrix}) as
\begin{equation}
\mathbb{W}=\mathbb{W}^{(2)}+\mathbb{W}_0^{(4)}+\mathbb{W}_\Phi^{(4)} \, ,
\end{equation}
where $\mathbb{W}^{(2)}$ includes all contributions from second-order processes, and $\mathbb{W}_0^{(4)}$ and $\mathbb{W}_\Phi^{(4)}$ include fourth-order flux-independent and flux-dependent processes, respectively.
In order to calculate the ``flux-dependent'' part of  $I_3$, denoted by $\langle I_3 \rangle_\Phi$, we can expand the stationary distribution $\mathbf{P}$ in powers of fourth-order processes as
\begin{multline}
\resizebox{.85\hsize}{!}{$\displaystyle\mathbf{P}\simeq\frac{1}{\mathbb{W}^{(2)}+E}\mathbf{e}-
\frac{1}{\mathbb{W}^{(2)}+E}(\mathbb{W}_0^{(4)}+\mathbb{W}_\Phi^{(4)})\frac{1}{\mathbb{W}^{(2)}+E}\mathbf{e} + $}\\
\resizebox{.85\hsize}{!}{$\displaystyle \frac{1}{\mathbb{W}^{(2)}+E}(\mathbb{W}_0^{(4)}+\mathbb{W}_\Phi^{(4)})
\frac{1}{\mathbb{W}^{(2)}+E}(\mathbb{W}_0^{(4)}+\mathbb{W}_\Phi^{(4)})\frac{1}{\mathbb{W}^{(2)}+E}\mathbf{e}\,.$}
\label{dist_expand}
\end{multline}
Substituting Eqs.~(\ref{1-flux-2rate}), (\ref{2-flux-4rate}), (\ref{1-flux-4rate}), and \eqref{dist_expand} into Eq.~(\ref{me-current}), we obtain
\begin{widetext}
\begin{multline}
\langle I_3 \rangle_\Phi=\frac{q|\tilde{\Gamma}_{\mathcal{A}}\tilde{\Gamma}_{\mathcal{B}}\tilde{\Gamma}_{\mathcal{C}}\tilde{\Gamma}_{\mathcal{D}}|^3 }{4(|\tilde{\Gamma}_{\mathcal{A}}|^2+|\tilde{\Gamma}_{\mathcal{B}}|^2+|\tilde{\Gamma}_{\mathcal{C}}|^2+|\tilde{\Gamma}_{\mathcal{D}}|^2)^2\gamma^2(V,T,1/3)}
\
[\Omega(V,T,1/3,\Delta L)+\tilde{\Omega}(V,T,1/3,\Delta L) ]
\\ \times  [4\,  \tilde{\Omega}^2(V,T,1/3,\Delta L)+2 \, \tilde{\Omega}(V,T,1/3,\Delta L) \Omega(V,T,1/3,\Delta L)
+\Omega^2(V,T,1/3,\Delta L) ]\ \cos(2\pi \Phi_{\rm AB}/\Phi_0)\,.
\end{multline}
\end{widetext}

On the other hand, the following argument shows that $\langle I_3 \rangle_{\Phi}$ must vanish: consider, for instance, the current at drain $D_3$. Owing to chiral propagation along the edges, this tunneling current does not depend on the scattering at QPC $\mathcal{D}$. A gauge transformation can then ascribe the total magnetic flux to this QPC. This implies that the current in $D_3$  is independent of the magnetic flux \cite{campagnano2012}. A similar argument holds for tunneling currents collected at other drains. Hence, we have to conclude that  $\tilde{\Omega}(V,T,\nu,\Delta L)=- \Omega(V,T,\nu,\Delta L)$. This is verified by the detailed derivation of these quantities in Sec.~\ref{calculationrates}.

% An important remark here is that even though a QP that tunnels inside the interferometer leaves a quantum of flux behind we assume that its charge will be nevertheless detected at the corresponding drain.

Having established a relationship between $\Omega(V,T,\nu,\Delta L)$ and $\tilde{\Omega}(V,T,\nu,\Delta L)$, we move to the calculation of the flux-sensitive zero-frequency cross-current correlator to $S_{14}$, namely, $S^{(q)}_{\Phi}$ (a calculation of $S_{23}$ leads to an identical expression). We begin by showing that no contribution comes from the cross terms in Eq.~(\ref{curr-curr-general}):
Because of the relation $\Omega=-\tilde{\Omega}$, $\mathbf{w}_4$  is proportional to the unity vector, $\mathbf{w}_4\propto \mathbf{e}$. Moreover, one can show that the matrix $\mathbb{W}$ is invertible in the subspace of traceless vectors such as $\bar{\mathbf{y}}_{1}$, which is traceless by construction \cite{Korotkov94,Koch06}. Hence, $\mathbf{w}_4\mathbb{W}^{-1}\bar{\mathbf{y}}_{1}=0$ and likewise $\mathbf{w}_1\mathbb{W}^{-1}\bar{\mathbf{y}}_{4}=0$.

We remain with cross-current contributions coming from auto-terms,
\begin{equation}\label{14autocorr}
S_{14} = 2 q^2 \sum_{i f \zeta} s_{if,\zeta}^1 s_{if,\zeta}^4 P_i  W_{if}^{(\zeta)}\,.
\end{equation}
In Eq.~\eqref{14autocorr}, only processes $(2,\mathcal{A}\mathcal{B}\mathcal{C}\mathcal{D},\Phi_{\text{tot}}(j))$, $(2,\mathcal{A}\mathcal{D},0)$, and $(2,\mathcal{B}\mathcal{C},0)$ give a non-zero contribution (cf.~Table \ref{table1}). For brevity, let us rewrite Eq.~\eqref{dist_expand} as
\begin{equation} \label{stat-exp}
\mathbf{P}=\mathbf{P}_0+\delta\mathbf{P}\,,
\end{equation}
with
\begin{equation}
\mathbf{P}_0=(\mathbb{W}^{(2)}+E)^{-1}\mathbf{e}=\frac{1}{3}\mathbf{e}\,,
\end{equation}
and
\begin{multline}
\delta \mathbf{P}=-(\mathbb{W}^{(2)}+E)^{-1}   (\mathbb{W}^{(4)}_0+\mathbb{W}^{(4)}_\Phi)  (\mathbb{W}^{(2)}+E)^{-1}\mathbf{e} \\+\cdots
\end{multline}

Notice that $\mbox{Tr} \, \delta\mathbf{P}=0$. As the rates $W_{if}^{(2,\mathcal{A}\mathcal{D},0)}$ and $W_{if}^{(2,\mathcal{B}\mathcal{C},0)}$ are flux-independent, when multiplied by $\delta\mathbf{P}$ the result is zero. This is, once more, an outcome of a scalar product of a traceless vector with a vector proportional to $\mathbf{e}$.

Hence, the only possible flux-dependent contribution may come from $\sum_i P_i W_{i,i+2}^{(2,\mathcal{A}\mathcal{B}\mathcal{C}\mathcal{D},\Phi_{\text{tot}}(j))}$. Going order by order in the expansion of $\mathbf{P}$, we see that the contributions of order $|\Gamma|^4$ vanish. Indeed, they are be proportional to
\begin{equation}
\sum_{if}\cos\left[\frac{2\pi}{3 \Phi_0}(\Phi_{\rm AB}+i \Phi_0)\right]\equiv 0\,.
\end{equation}
In a similar fashion, the $|\Gamma|^8$ is vanishing. The first non-vanishing contribution is
\begin{widetext}
\begin{align}\label{mainres1}
S^{(q)}_{\Phi}= \frac{e^2|\tilde{\Gamma}_\mathcal{A}\tilde{\Gamma}_\mathcal{B}\tilde{\Gamma}_\mathcal{C}\tilde{\Gamma}_\mathcal{D}|^3\Omega^3(V,T,1/3,\Delta L) \cos[2\pi(\Phi_{\text{AB}}/\Phi_0)]}{6 (|\tilde{\Gamma}_\mathcal{A}|^2+|\tilde{\Gamma}_\mathcal{B}|^2+|\tilde{\Gamma}_\mathcal{C}|^2+|\tilde{\Gamma}_\mathcal{D}|^2)^2\gamma^2(V,T,1/3)} \,.
\end{align}
We repeated the kinetic analysis for additional Laughlin fractions. We obtain that, in general, the lowest non-vanishing contribution is
\begin{align}\label{mainres1gen}
S^{(q)}_{\Phi}= \frac{\nu e^2|\tilde{\Gamma}_\mathcal{A}\tilde{\Gamma}_\mathcal{B}\tilde{\Gamma}_\mathcal{C}\tilde{\Gamma}_\mathcal{D}|^{1/\nu}\Omega^{1/\nu}(V,T,\nu,\Delta L) \cos[2\pi(\Phi_{\text{AB}}/\Phi_0)]}{2^{(1/\nu-2)} (|\tilde{\Gamma}_\mathcal{A}|^2+|\tilde{\Gamma}_\mathcal{B}|^2+|\tilde{\Gamma}_\mathcal{C}|^2+|\tilde{\Gamma}_\mathcal{D}|^2)^{(1/\nu-1)}\gamma^{(1/\nu-1)}(V,T,\nu)} \,.
\end{align}
\end{widetext}

\begin{figure}
\includegraphics[width=\columnwidth]{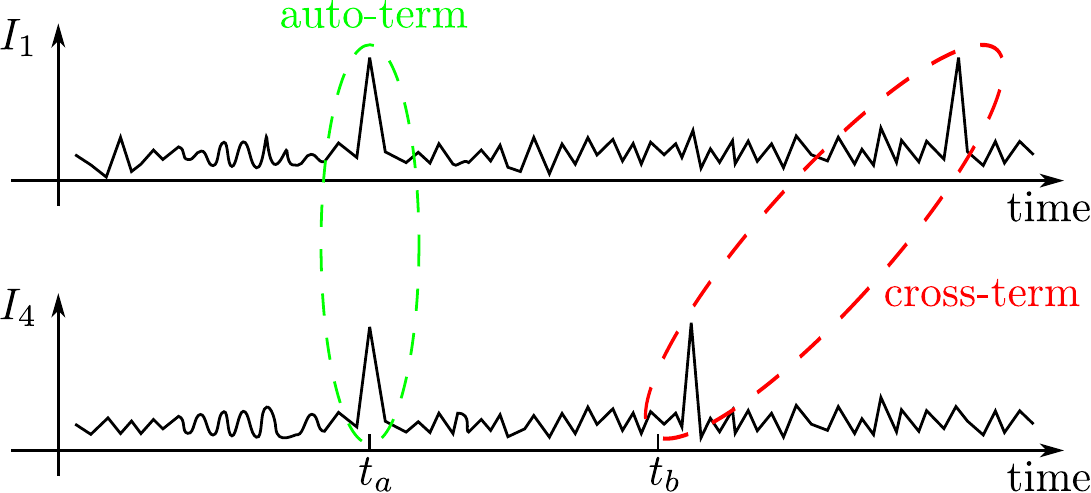}
\caption{Illustration of the currents and their cross-current correlations. Indicated are auto (green) and cross (red) terms.}
\label{autoterm}
\end{figure}

The fact that $S^{(q)}_{\Phi}$, has nonvanishing contributions only from auto terms, could correspond to an averaging out of cross-terms in the  zero-frequency limit. However, cross terms vanish also in the finite-frequency regime. In order to see this, we rewrite~\cite{Koch06} Eqs.~\eqref{zerofreq} and \eqref{curr-curr-general} as
\begin{align}
S_{a,b}(\omega)&=\int_{-\infty}^{\infty} dt e^{i\omega t}\left[  \mean{ I_a(t)I_b(0)}- \mean{I_{a}(t) } \mean{ I_{b}(0)} \right]\nonumber\\
&=2q^2\Bigg[
\underbrace{
\text{tr}\left\{\mathbf{u}_{ab}\right\}
}_{\mbox{auto}}\label{finfreq}
\\
&-\underbrace{
\mathbf{w}_{b}(\mathbb{W}+i\omega \mathds{1})^{-1}\bar{\mathbf{y}}_{a}-\mathbf{w}_{a}(\mathbb{W}+i\omega \mathds{1})^{-1}\bar{\mathbf{y}}_{b}}_{\mbox{cross}}
\Bigg]\, .\nonumber
\end{align}
Here, too, as $\mathbf{w}_1$ and $\mathbf{w}_4$  are proportional to the unity vector, and $\bar{\mathbf{y}}_{1}$ and $\bar{\mathbf{y}}_{4}$ are traceless, $\mathbf{w}_4(\mathbb{W}+i\omega \mathds{1})^{-1}\bar{\mathbf{y}}_{1}=0$ and $\mathbf{w}_1(\mathbb{W}+i\omega \mathds{1})^{-1}\bar{\mathbf{y}}_{4}=0$. Hence, the flux-dependent cross-current correlation has nonvanishing contributions only from auto-terms, as illustrated in Fig.~\ref{autoterm}.

%%%%%%%%%%%%%%%%%%%%%%%%%%%%%%%%%%%%%%%%%%%%%%%%%%%%%
%%%%%%%%%%%%%%%%%%%%%%%%%%%%%%%%%%%%%%%%%%%%%%%%%%%%%

Equation (\ref{mainres1gen}) is one of the main results of this work. Some of its significance and ramifications are discussed in Sec.~\ref{main_results}. Nonetheless, let us discuss here in more detail the implications of Eq.~\eqref{mainres1gen}. We notice, first, that its sign is positive. This is similar to the sign obtained for two-particle boson interference (cf.~Ref.~[\onlinecite{samuelsson04}]) and is contrary to the negative sign obtained for electrons [cf.~Eqs.~\eqref{sphi} and \eqref{derivatives1}]. Similarly to the results of Ref.~[\onlinecite{vish}] it suggests a bunching effect rather than antibunching.

Equally interesting is the non-analytic structure of Eq.~(\ref{mainres1}) in terms of the tunneling amplitudes $\Gamma$'s. As we have shown in Secs.~\ref{integerfermionic} and \ref{integerbosonic}, for an HBT interferometer operating at integer filling fraction,  the AB-dependent signal of the cross-current correlations is proportional to $|\Gamma|^4$. Here, in the case of filling fraction $\nu$, a naive expectation based on gauge invariance arguments would suggest a signal proportional to $|\Gamma|^{4/\nu}$ (cf.~Appendix \ref{twevlth_order_break}). In Eq.~\eqref{mainres1gen}, instead, we find a contribution of order $|\Gamma|^{2/\nu+2}$. This dependence can be understood as follows: we take the lowest-order gauge-invariant contribution corresponding to $|\Gamma_\mathcal{A}\Gamma_\mathcal{B}\Gamma_\mathcal{C}\Gamma_\mathcal{D}|^{1/\nu}$. Such a contribution can be thought of, in the high-voltage-bias limit considered here, as a coherent sequence of $1/\nu$ two-quasiparticle processes akin to $(2,\mathcal{A}\mathcal{B}\mathcal{C}\mathcal{D},\Phi_{\text{tot}}(j))$. However, such a contribution, analyzed within a non-equilibrium Keldysh scheme, in the same spirit of Refs.~[\onlinecite{kane03,feldman}], would be divergent if calculated naively. This results from allowing arbitrarily long-time intervals between two subsequent two-quasiparticle processes. Nonetheless, such time intervals are limited by single-particle processes, which introduce a new time scale $\tau_{s}\simeq 1/(W_{i,i+1}^{(1,\mathcal{A},0)}+W_{i,i+1}^{(1,\mathcal{B},0)}+W_{i,i+1}^{(1,\mathcal{C},0)}+W_{i,i+1}^{(1,\mathcal{D},0)})$. Having $1/\nu-1$, such intervals between the $1/\nu$ $(2,\mathcal{A}\mathcal{B}\mathcal{C}\mathcal{D},\Phi_{\text{tot}}(j))$ processes result in a factor $\sim\Gamma^{4/\nu-2 (1/ \nu -1)}$, and explain the structure of Eq.~(\ref{mainres1}).

%%%%%%%%%%%%%%%%%%%%%%%%%%%%%%%%%%%%%%%%%%%%%%%%%%%%%%%%%%%%%%%%%%%%%
\subsection{Calculation of the rates}\label{calculationrates}
%\begin{figure}
%\includegraphics[scale=.6]{HBT2flux-AB}
%\caption{}
%\label{2fluxAB}
%\end{figure}
%\begin{figure}
%\includegraphics[scale=.6]{HBT1flux-AB}
%\caption{}
%\label{1fluxAB}
%\end{figure}
Here, we calculate explicitly the rates introduced above. As mentioned before, the system can be found in three possible statistical flux states. We focus here on processes that change the statistical flux by one or two flux quanta, limiting ourselves to fourth-order processes in the tunneling amplitudes $\Gamma$s. The Hamiltonian describing the system has been introduced in Sec.~\ref{modelquasiparticle}. In order for us to calculate transition rates, we assume that the system is in a given statistical flux state, reflected in the tunneling operator $\mathcal{A}^{(q)}_j$, and by means of Fermi's golden rule we calculate the rate of transferring  one or two quasiparticles from external to internal edges.
Let $| \hat{\psi}_i \rangle$ and $|\hat{\psi}_f \rangle$ be two many-body eigenstates of the system  in absence of tunneling (the tunneling Hamiltonian is $H_T$). Very generally, the transition rate between them
due to the tunneling Hamiltonian  can be written as
\begin{equation}\label{generalfermigr}
\frac{2\pi}{\hbar}  |\langle\hat{\psi}_i| \tilde{T}|\hat{\psi}_f\rangle|^2 \delta(E_f-E_i) \, ,
\end{equation}
where $\tilde{T}$ is the scattering operator given by
\begin{equation}
\tilde{T}=H_T+H_T\frac{1}{E_i-H_0-i 0^+}H_T+\cdots \, .
\end{equation}
Let us first consider the case of one-particle rate. For the sake of concreteness, we consider here tunneling through QPC
$\mathcal{A}$, all the other single-particle rates being similar. In this case $|\hat{\psi}_f \rangle$ is
obtained by removing a QP from edge 1 and transferring it to edge 3. Since we are interested in the \emph{total} transition rate, we sum over all possible initial and final states. Notice that each  edge is kept at a finite chemical potential $\mu_i$ ($i=\{1,2,3,4\}$) and that the initial states  are weighted  by $w_i=Z^{-1}\langle \hat{\psi}_i|\exp{[-\beta(H_0-\sum_i\mu_i N_i)}]|\hat{\psi}_i\rangle$, with $Z=\mbox{Tr}\exp{[-\beta(H_0-\sum_i\mu_i N_i)]}$.
To the lowest order in the tunneling amplitude the transition rate $W^{(1,\mathcal{A})}_{j,j+1}$ is given by
\begin{equation}
\label{singletransfer}
W_{j,j+1}^{(1,\mathcal{A})}=\frac{2 \pi}{\hbar} \sum_{i,f} w_i \langle \hat{\psi}_i | \ta |\hat{\psi}_f \rangle \langle \hat{\psi}_f |\tad | \hat{\psi}_i \rangle \delta(E_f-E_i) \,.
\end{equation}
Here the operator $\tad$ annihilates a quasiparticle on edge 1 and creates it on edge 3.  Expressing the tunneling operators in the interaction representation (with respect to $H_0$), Eq.~(\ref{singletransfer}) can be  rewritten as
\begin{equation}\label{single-qp-transf2}
W_{j,j+1}^{(1,\mathcal{A})}= \sum_{if}   \frac{w_i}{\hbar^2} \int_{-\infty}^{\infty}dt  \langle \hat{\psi}_i | \ta(0)|\hat{\psi}_f \rangle \langle \hat{\psi}_f | \tad(t) | \hat{\psi}_i \rangle \,.
\end{equation}
Notice that in Eq.~(\ref{single-qp-transf2})  we can extend the sum over final states to a sum over a complete set of states and obtain
\begin{widetext}
\begin{align}
&W_{j,j+1}^{(1,\mathcal{A})}=\frac{1}{\hbar^2}\int_{-\infty}^{\infty}dt
 \langle   \ta(0) \tad(t)  \rangle
=\frac{|\Gamma_\mathcal{A}|^2}{\hbar^2 l_c^2} l_{c}^{2\nu}\int_{-\infty}^{\infty}dt\, e^{-i\nu e V t/\hbar}
\left\{ \frac{\hbar \beta v}{\pi}
\sin \left[ \frac{\pi}{\hbar \beta v} (-i v t+l_c)   \right] \right\}^{-2\nu}
\nonumber\\
&=\frac{|\Gamma_\mathcal{A}|^2}{\hbar^2 v^2 } \frac{eV}{2\hbar}\left(\frac{\hbar\beta v}{2\pi l_{c}}\right)^{2-2\nu}
\frac{(2\pi)^2 e^{\nu \pi \alpha}}{ \alpha\Gamma \left[2\nu\right]\left|\Gamma \left[1-\nu+i\nu\alpha\right]\right|^2}
\frac{1}{\left[\cosh(2\pi\nu\alpha)-\cos(2\pi\nu)\right]}
\equiv\frac{|\Gamma_\mathcal{A}|^2}{\hbar^2 v^2}  \gamma(V,T,\nu) \,,\label{single-qp-transf3}
\end{align}
\end{widetext}
where $\alpha=e V \beta/(2\pi)$. In order to obtain Eq.~(\ref{single-qp-transf3}) we have used that the two point  correlation function for an edge kept at finite temperature and finite chemical potential is given by
\begin{small}
\begin{align}
&\langle e^{i \sqrt{\nu}\phi(x,t)} e^{-i \sqrt{\nu}\phi(0,0)} \rangle_\mu
\label{finite-mu-corr} \\
&=   e^{i\mu \nu (t-x/v)/\hbar}
\langle e^{i \sqrt{\nu}\phi(x,t)} e^{-i \sqrt{\nu}\phi(0,0)} \rangle_{\mu=0}
\nonumber\\
&=e^{i\mu \nu (t-x/v)/\hbar} l_c^{\nu} \left\{\frac{\hbar \beta v }{\pi}\sin\left[ \frac{\pi}{\hbar \beta v}(i(vt-x)+l_c)\right]\right\}^{-\nu}
\,.\nonumber
\end{align}
\end{small}
We now consider the \emph{total} rate of transferring two quasiparticles from the external to the internal edges. Since there are no contributions to such a rate from second  and third order terms in the tunneling amplitudes $\Gamma$s, we need to consider the fourth order, we thus have
\begin{multline}\label{total-2qp-rate}
W_{j,j+2}=\frac{2\pi}{\hbar} \sum_{if}w_i  \bra{\hat{\psi}_i} H_T \frac{1}{E_i-H_0-i 0^+} H_T \ket{\hat{\psi}_f}
\\ \times
\bra{\hat{\psi}_f} H_T \frac{1}{E_i-H_0+i 0^+} H_T \ket{\hat{\psi}_i}\delta(E_f-E_i)\, .
\end{multline}
Notice that in this case the many-body eigenstate $\ket{\hat{\psi}_f}$ is obtained from $\ket{\hat{\psi}_i}$ by transferring two quasiparticles. Being interested only  in the lowest contribution to the current-current correlation modulated by the magnetic flux, we consider the contributions proportional to $|\Gamma_\mathcal{A}\Gamma_\mathcal{B}\Gamma_\mathcal{C}\Gamma_\mathcal{D}|$. We have
\begin{widetext}
\begin{multline}
W^{(2,\phi)}_{j,j+2}=\frac{2\pi}{\hbar} \sum_{if} w_i \left\{
 \langle \hat{\psi}_i| \tb \frac{1}{E_i-H_0-i 0^+} \td| \hat{\psi}_f \rangle +
  \langle \hat{\psi}_i| \td  \frac{1}{E_i-H_0-i 0^+} \tb| \hat{\psi}_f \rangle
 \right\}  \\
\times \left\{
 \langle \hat{\psi}_f |\tad \frac{1}{E_i-H_0+i 0^+} \tcd | \hat{\psi}_i\rangle +
 \langle \hat{\psi}_f |\tcd  \frac{1}{E_i-H_0+i 0^+} \tad | \hat{\psi}_i\rangle \right\} \ \delta(E_f-E_i)+c.c..
 \label{2phi-2qp-rate}
\end{multline}
\end{widetext}
The above contribution corresponds to the rate $(2,\phi)$ of Table \ref{table1}, the corresponding amplitudes are represented in Fig.~\ref{processes}. Indeed the operator $\tad$ ($\tcd$)  annihilates a quasiparticle on edge 1 (on edge 4) and then creates it on edge 2 (3) respectively; similar statements apply to the operators  $\tbd$ and  $\tdd$.
Notice that we do not take into account other contributions such as, for instance, $W_{j,j+2}^{(2, \mathcal{A}\mathcal{D})}$ (proportional to $|\Gamma_\mathcal{A}|^2 |\Gamma_\mathcal{D}|^2$)  and   $W_{j,j+2}^{(2, \mathcal{B}\mathcal{C})}$ (proportional to $|\Gamma_\mathcal{B}|^2 |\Gamma_\mathcal{C}|^2$).

Let us consider now one of the four contributions proportional to $\Gamma_\mathcal{A}^* \Gamma_\mathcal{B} \Gamma_\mathcal{C}^*\Gamma_\mathcal{D}$ we obtain from Eq.~(\ref{2phi-2qp-rate}),
\begin{multline}
\mbox{I}=\frac{2\pi}{\hbar} \sum_{if} w_i \langle \hat{\psi}_i | \tb  \frac{1}{E_i-H_0-i 0^+} \td|  \hat{\psi}_f \rangle \\
\times
 \langle  \hat{\psi}_f |\tad  \frac{1}{E_i-H_0+i 0^+} \tcd|  \hat{\psi}_i\rangle \delta(E_f-E_i)\,.
\end{multline}
Once again, moving to the interaction representation one can  rewrite the previous expression as
\begin{multline}
\mbox{I}=\sum_{if}\frac{w_i}{\hbar^2}\int_{-\infty}^{+\infty}dt  \langle  \hat{\psi}_i| \tb(0)  \frac{1}{E_i-H_0-i 0^+} \td(0)|  \hat{\psi}_f \rangle
\\ \times
\langle  \hat{\psi}_f |\tad(t)  \frac{1}{E_i-H_0+i 0^+} \tcd(t) |  \hat{\psi}_i\rangle\,.
\end{multline}
The sum over the final states may be changed to a sum over a complete set of states; we can rewrite the expression as
\begin{multline}
\mbox{I}=\sum_{i}\frac{w_i}{\hbar^2}\int_{-\infty}^{+\infty}dt
 \langle \hat{\psi}_i| \tb(0)  \frac{1}{E_i-H_0-i 0^+} \td(0) \\ \times\tad(t)  \frac{1}{E_i-H_0+i 0^+} \tcd(t) |\hat{\psi}_i\rangle\,.
\end{multline}
This may be rewritten as
\begin{multline}\label{contribution-I}
\mbox{I}=\frac{1}{\hbar^4}\int_{-\infty}^{+\infty}dt \int_{-\infty}^{0}dt_1\int_{-\infty}^{0} dt_2
\\ \times
\langle \tb(t_1)\td(0)\tad(t)\tcd(t+t_2)\rangle\,.
\end{multline}
The other three contributions proportional to $\Gamma_\mathcal{A}^* \Gamma_\mathcal{B} \Gamma_\mathcal{C}^*\Gamma_\mathcal{D}$
 are
\begin{multline}
\mbox{II}=\frac{1}{\hbar^4}\int_{-\infty}^{+\infty}dt \int_{-\infty}^{0}dt_1\int_{-\infty}^{0} dt_2
\\ \times
\langle \tb(t_1)\td(0)\tcd(t)\tad(t+t_2)\rangle \, ,
\end{multline}
\begin{multline}
\mbox{III}=\frac{1}{\hbar^4}\int_{-\infty}^{+\infty}dt \int_{-\infty}^{0}dt_1\int_{-\infty}^{0} dt_2
\\ \times
\langle \td(t_1)\tb(0)\tad(t)\tcd(t+t_2)\rangle \, ,
\end{multline}
\begin{multline}
\mbox{IV}=\frac{1}{\hbar^4}\int_{-\infty}^{+\infty}dt \int_{-\infty}^{0}dt_1\int_{-\infty}^{0} dt_2
\\ \times
\langle \td(t_1)\tb(0)\tcd(t)\tad(t+t_2)\rangle \, .
\end{multline}
We thus obtain
\begin{equation}
W_{j,j+2}^{(2,\mathcal{A}\mathcal{B}\mathcal{C}\mathcal{D},\Phi_{\text{tot}}(j))}=(\mbox{I}+\mbox{II}+\mbox{III}+\mbox{IV})+c.c.
\end{equation}
%We proceed now to the calculation of Eq.\ref{contribution-I},  using results from reference \cite{vondelft} we have:
%\begin{multline}
%\langle    e^{\pm i\sqrt{\nu} \phi(x_1,t_1)} e^{\mp i\sqrt{\nu} \phi(x_2,t_2)}\rangle_\beta =
%\\
%\tau_c^\nu\left\{ i \frac{\beta}{\pi} \sinh[ \frac{\pi}{\beta} (t_1-t_2-x_1+x_2-i \epsilon)]\right\}^{-\nu}
%\end{multline}
Using Eq.~(\ref{finite-mu-corr}), we can write Eq.~(\ref{contribution-I}) as
\begin{multline}\label{rate}
\mbox{I}=\frac{\Gamma^*_\mathcal{A} \Gamma_\mathcal{B} \Gamma^*_\mathcal{C} \Gamma_\mathcal{D}}{\hbar^4 l_c^4} e^{-2\pi i \nu(\Phi_{\rm AB}+j\,\Phi)/\Phi_0 } e^{-i \nu e  V (L_4+L_1)/(\hbar v) } \\ \times
\left(\frac{\pi l_c}{ \hbar \beta v}\right)^{4\nu}
\int_{-\infty}^{+\infty}dt\int_{-\infty}^{0}dt_1\int_{-\infty}^{0} dt_2 e^{-2i \nu e V t/\hbar} \\ \times
\sin^{-\nu} \left\{ \frac{\pi}{\hbar \beta v}[i v (-t+\frac{t_1}{2}-\frac{t_2}{2})-i L_4+ l_c ]\right\}
 \\ \times
 \sin^{-\nu} \left\{ \frac{\pi}{\hbar \beta v}[i v (-t+\frac{t_1}{2}+\frac{t_2}{2})-i L_3+ l_c ]\right\}
  \\ \times
\sin^{-\nu} \left\{ \frac{\pi}{\hbar \beta v}[i v (-t-\frac{t_1}{2}-\frac{t_2}{2})-i L_2+ l_c ]\right\}
  \\ \times
  \sin^{-\nu} \left\{ \frac{\pi}{\hbar \beta v}[i v (-t-\frac{t_1}{2}+\frac{t_2}{2})-i L_1+ l_c ]\right\}\,,
\end{multline}
where $t$ has been shifted by $(t_1-t_2)/2$.

%%%%%%%%%%%%%%%%%%%%%%%%%%%%%%%%%%%%%%%%%%%%%%%%%%%%%
\begin{figure*}
\includegraphics[width=\textwidth]{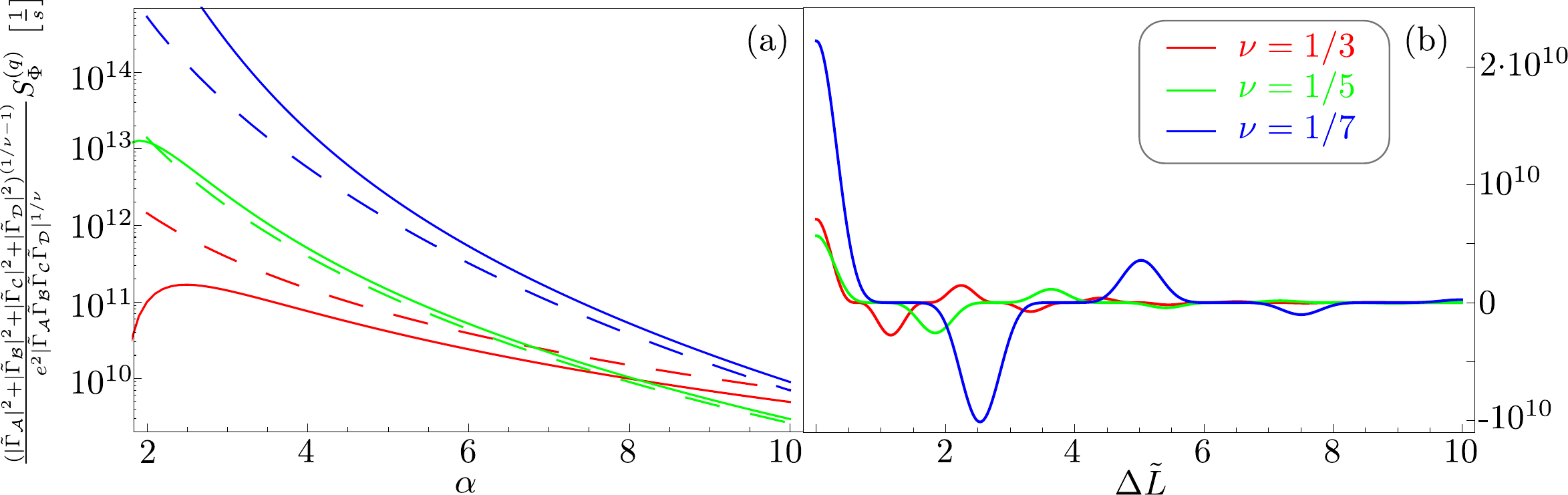}
\caption{\textit{Quasiparticle cross-current correlations}. Plots of the cross-current correlations $S_{\Phi}^{(q)}$ at $\Phi_{\rm AB}=0$ [cf.~Eqs.~\eqref{mainres1gen}, \eqref{single-qp-transf3}, \eqref{2-phi-rate-bis}, and \eqref{trueRes}] as a function of (a) voltage [$\alpha=e V \beta/(2\pi)$] for interferometer arms mismatch $\Delta \tilde{L}=\pi(L_1+L_4-L_2-L_3)/(\hbar \beta v)\rightarrow 0$, and (b) $\Delta \tilde{L}$ for $\alpha=9$. Different curves are for filling fractions $\nu=1/3$, $\nu=1/5$, and $\nu=1/7$ (red, green, and blue, respectively). Dashed lines correspond to using $\Omega(V,T,\nu)$ from Eq.~\eqref{2-phi-rate-bis} and full lines to Eq.~\eqref{trueRes}. We assume the following experimental values: temperature~\cite{neder2007} $T=10mK$, edge veolcity~\cite{McClure:2009} $v\sim 1.5\cdot 10^5 m/s$. We rescaled the tunneling amplitudes such that we took $2[\hbar \beta v/(2\pi l_c)]^{1-1/v}=1$.}
\label{plotqp}
\end{figure*}
%%%%%%%%%%%%%%%%%%%%%%%%%%%%%%%%%%%%%%%%%%%%%%%%%%%%%

Remarkably, changing variables in the terms II, III and IV, yields exactly the missing  sectors in the  $t_1$ and $t_2$ integrals of contribution I.
Hence, we can combine the four contributions into a single expression
\begin{widetext}
\begin{multline}\label{eq_rates}
\mbox{I+II+III+IV}= \frac{\Gamma^*_\mathcal{A} \Gamma_\mathcal{B} \Gamma^*_\mathcal{C} \Gamma_\mathcal{D}}{\hbar^4 l_c^4} e^{-2\pi i \nu(\Phi_{\rm AB}+j \,\Phi)/\Phi_0 }e^{-i \nu e  V (L_4+L_1)/( \hbar v)}
\left(\frac{\pi l_c}{ \hbar \beta v}\right)^{4\nu}
\int_{-\infty}^{+\infty}dt\int_{-\infty}^{+\infty}dt_1\int_{-\infty}^{+\infty} dt_2 e^{-2i \nu e V t/\hbar}
\\ \times
\sin^{-\nu} \left\{ \frac{\pi}{\hbar \beta v}[i v (-t+\frac{t_1}{2}-\frac{t_2}{2})-i L_4+ l_c ]\right\}
  \sin^{-\nu} \left\{ \frac{\pi}{\hbar \beta v}[i v (-t+\frac{t_1}{2}+\frac{t_2}{2})-i L_3+ l_c ]\right\}
  \\ \times
  \sin^{-\nu} \left\{ \frac{\pi}{\hbar \beta v}[i v (-t-\frac{t_1}{2}-\frac{t_2}{2})-i L_2+ l_c ]\right\}
    \sin^{-\nu} \left\{ \frac{\pi}{\hbar \beta v}[i v (-t-\frac{t_1}{2}+\frac{t_2}{2})-i L_1+ l_c ]\right\}\,.
\end{multline}
This integral can be evaluated explicitly (cf.~Appendix \ref{app_rate2}). In the limit of $\nu e V \beta >>1$ and $\Delta L<<\hbar v/(\nu e V)$, we obtain
a presentable expression
%\begin{align}\label{2-phi-rate-bis}
%&W_{j,j+2}^{(2,\phi)}= 2 \frac{\left| \Gamma_\mathcal{A} \Gamma_\mathcal{B} \Gamma_\mathcal{C} \Gamma_\mathcal{D} \right|}{\hbar^4 v^4 \hbar\beta} \cos \left[ 2\pi\nu(\Phi_\Gamma+\Phi_{\rm AB}+j \,\Phi_0)/\Phi_0 \right] \nonumber\\
%&\quad\quad\quad\times\left(\frac{\hbar \beta v}{l_c}\right)^{4-4\nu}\frac{16 \pi^{1/3}}{2^{\nu}\Gamma(1-\nu)}\\
%&=  \Omega(V,T)\frac{\left| \Gamma_\mathcal{A} \Gamma_\mathcal{B} \Gamma_\mathcal{C} \Gamma_\mathcal{D}  \right|}{\hbar^4 v^4}  \cos \left[ 2\pi\nu(\Phi_\Gamma+\Phi_{\rm AB}+j\,\Phi_0)/\Phi_0 \right]
%   \, ,\nonumber
%\end{align}

\begin{align}\label{2-phi-rate-bis}
W_{j,j+2}^{(2,\mathcal{A}\mathcal{B}\mathcal{C}\mathcal{D},\Phi_{\text{tot}}(j))} &=\left(\mbox{I+II+III+IV}\right)+c.c.
%\\
%&=
%2\frac{ \left| \Gamma_\mathcal{A} \Gamma_\mathcal{B} \Gamma_\mathcal{C} \Gamma_\mathcal{D} \right|}{\hbar^4 v^4}
%\frac{\pi^2}{\hbar\beta}
%\left(\frac{\hbar \beta v}{2\pi l_c}\right)^{4-4\nu} e^{\pi \alpha} 2^{7 - 2 \nu} \alpha^{-2 + 2 \nu}\Gamma[1 - 2\nu]\sin(\pi\nu)\cos \left[ %2\pi\nu(\Phi_\Gamma+\Phi_{ab}+j \,\Phi_0)/\Phi_0 \right]\nonumber\\
%&\equiv
=
2\frac{\left| \Gamma_\mathcal{A} \Gamma_\mathcal{B} \Gamma_\mathcal{C} \Gamma_\mathcal{D}  \right|}{\hbar^4 v^4} \Omega(V,T,\nu) \cos \left[ 2\pi\nu(\Phi_\Gamma+\Phi_{\rm AB}+j\,\Phi_0)/\Phi_0 \right]\,,\nonumber
\end{align}
\end{widetext}
where we used
\begin{align}
\Gamma^*_\mathcal{A} \Gamma_\mathcal{B} \Gamma^*_\mathcal{C} \Gamma_\mathcal{D} &= \left| \Gamma_\mathcal{A} \Gamma_\mathcal{B} \Gamma_\mathcal{C} \Gamma_\mathcal{D} \right|
\exp(-i 2\pi \nu \Phi_\Gamma/\Phi_0) \, ,
\\
\Omega(V,T,\nu)&=\frac{\nu eV}{\hbar}
\left(\frac{\hbar \beta v}{2\pi l_c}\right)^{4-4\nu}
\nonumber
\\&\times \pi 2^{4} (\nu\alpha)^{-3 + 2 \nu}\Gamma[1 - 2\nu]\sin(\pi\nu)
.
\label{Omega}
\end{align}
The expressions for the rate at finite $\Delta L$ and the corresponding function $\Omega(V,T,\nu,\Delta L)$ are presented in Appendix  \ref{app_rate2}. Importantly, in contrast to the single-particle scattering rate, the two-particle rate is not cut off by voltage only. As a result, even in the
high-voltage regime $eV\gg T$, the two-particle rate scales with temperature. This implies that the scaling of the two-particle rate can not
be simply expressed through a product of four renormalized tunneling amplitudes. Hence, in the limit of zero temperature, the correlation function (\ref{Omega}), diverges; in this case, the cutoff is provided by the single-particle scattering rate.

As for $W_{j,j+1}^{(1,\mathcal{A}\mathcal{B}\mathcal{C}\mathcal{D},\Phi_{\text{tot}}(j))_1}$, the calculation is more involved and we report here only the results in terms of Green's function; the details are presented in Appendix \ref{app_rate1}.
We find
\begin{widetext}
\begin{align}
W_{j,j+1}^{(1,\mathcal{A}\mathcal{B}\mathcal{C}\mathcal{D},\Phi_{\text{tot}}(j))_1}&=
\left| \Gamma_\mathcal{A} \Gamma_\mathcal{B} \Gamma_\mathcal{C} \Gamma_\mathcal{D} \right| \cos \left[ 2\pi\nu(\Phi_\Gamma+\Phi_{\rm AB}+j \Phi_0)/\Phi_0 \right] \nonumber \\
&\times
\int \frac{d\epsilon}{2 \pi}
\big[
G_{4}^{<}(\epsilon,-L4)G_{3}^{>}(\epsilon+\nu eV,L_3)
G_{2}^{t}(\epsilon+\nu eV,L_2)G_{1}^{t}(\epsilon,-L_1)\nonumber \\
&+G_{4}^{t}(\epsilon,-L4)G_{3}^{>}(\epsilon+\nu eV,L_3)
G_{2}^{\bar t}(\epsilon+\nu eV,L_2)G_{1}^{<}(\epsilon,-L_1)
\big]\nonumber\\
&= \bar{\Omega}(V,T,\nu,\Delta L) \left| \Gamma_\mathcal{A} \Gamma_\mathcal{B} \Gamma_\mathcal{C} \Gamma_\mathcal{D} \right| \cos \left[ 2\pi\nu(\Phi_\Gamma+\Phi_{\rm AB}+j \Phi_0)/\Phi_0 \right] \, ,
\end{align}
\end{widetext}
where, as explained in the previous Section, in the limit considered here $\bar\Omega\simeq-\Omega$.

In Fig.~\ref{plotqp}, we plot the dependence of $S_{\Phi}$ on voltage and for different values of $\nu$ using realistic experimental values. Notice that the function is initially positive -- a sign of bosonic statistics of the interfering anyons.

\section{Summary and outlook}
\label{summary}

In this work, we have presented an extensive theoretical study of the magnetic flux-dependent cross-current correlations in a Hanbury Brown and Twiss (HBT) interferometer realized with edge states of a quantum Hall system. This work substantiates and details some of the results reported in Ref.~[\onlinecite{campagnano2012}], and extends the analysis of electronic (and anyonic) HBT interferometry to new setups and further filling fractions. Our analysis applies to a two-dimensional electron gas in both an integer ($\nu=1$) filling fraction, and a fractional (Laughlin) $\nu=1/(2n+1)$ filling fraction. 

There are two obvious generalizations of our model. One involves the introduction of interactions in both the integer, $\nu=1$, and the fractional, $\nu=1/(2n+1)$, cases. As for the second, we note that the same approach used here for Laughlin quasiparticles (analysis of master equation) could be also employed to study HBT interferometry for other filling fractions, e.g., $\nu=5/2$, where exotic quasiparticles, non-Abelian anyons, emerge.

\section{ACKNOWLEDGMENTS}
We thank A. Carmi for interesting discussions. We are particularly grateful to A.~C.~Potter and D.~E.~Feldman, our co-authors on a previous work~\cite{campagnano2012}, for useful discussion.  We acknowledge financial support by BSF under Grant No. 2006371, GIF, ISF, Minerva Foundation of the DFG, DFG-SPP 1666, DFG-CFN, BMBF, SNF, MIUR-FIRB 2012 project, Grant No. RBFR1236VV, and MIUR-PRIN 2009 project, Grant No. 2009HS2F7N.

%%%%%%%%%%%%%%%%%%%%%%%%%%%%%%%%%%%%%%%%%%%%%%%%
\appendix

%%%%%%%%%%%%%%%%%%%%%%%%%%%%%%%%%%%%%%%%%%%%%%%%%%%%
\section{Scattering Approach for $\nu=1$}
\label{app_scatter}
In this appendix, we derive the results of Sec.~\ref{integerfermionic} using a Landauer-B\"{u}ttiker scattering approach (cf. Ref.~[\onlinecite{samuelsson04}] and Fig.~\ref{scheme}).
We can picture the two-particle interferometer as a multi-channel scatterer.
In principle, in such an analysis one should take into account incoming and outgoing channels from all reservoirs $\{S_j\}$ and $\{D_j\}$, $j=1,2,3,4$. This would lead to an $8\times 8$ scattering matrix. However, for our electronic HBT with chemical potentials $\mu_1=\mu_4=eV$ and $\mu_2=\mu_3=0$ and chiral edges, it suffices to consider outgoing channels from sources $\{S_j\}$, which come into drains $\{D_j\}$, i.e. scattering between the edges $i=1,2,3,4$. Hence, for each wave number $k$, the outgoing second quantized operators  $\{\hat{b}_{j,k}\}$ are related to those of the incoming states $\{\hat{a}_{j,k}\}$ by  the scattering matrix $\hat s(k)$,
\begin{equation}
\left(
\begin{array}{c}
\hat{b}_{1,k} \\  \hat{b}_{2,k} \\ \hat{b}_{3,k} \\ \hat{b}_{4,k}
\end{array}
\right)
=\hat{s}(k)
\left(
\begin{array}{c}
\hat{a}_{1,k} \\  \hat{a}_{2,k} \\ \hat{a}_{3,k} \\ \hat{a}_{4,k}
\end{array}
\right)\,.
\end{equation}
Thus, the current-current correlator $S_{23}$, in terms of the scattering matrix $\hat s$ reads as~\cite{samuelsson04}
\begin{align}\label{scattering}
S_{23}=&-\frac{e^2}{2 \pi}\int d \omega \left| s_{21}^*(\omega/v)  s_{31}(\omega/v)+s^*_{24}(\omega/v)s_{34}(\omega/v)  \right|^2 \nonumber\\ &\times \left[f(\hbar\omega)-f(\hbar\omega-eV) \right]^2 \, .
\end{align}

In order for us to calculate the scattering matrix elements, we need to solve the simple scattering problem between the edges of our system. The Schr\"{o}dinger equation in first quantization reads as
\begin{widetext}
\begin{equation}\label{firstquantization}
\resizebox{.9\hsize}{!}{$\displaystyle \left(
\begin{array}{cccc}
-i v \partial_x & \frac{\Gamma_D}{\hbar} \delta(x-L_1) e^{(L_2-L_1)\partial_x} & \frac{\Gamma_A}{\hbar} \delta(x) &0\\
\frac{\Gamma_D^*}{\hbar} \delta(x-L_2)e^{(L_1-L_2)\partial_x} & -i v \partial_x & 0 & \frac{\Gamma_C^*}{\hbar} \delta(x) \\
\frac{\Gamma_A^*}{\hbar} \delta(x) & 0 & -i v \partial_x & \frac{\Gamma_B^*}{\hbar} \delta(x-L_3) e^{(L_4-L_3)\partial_x} \\
0 & \frac{\Gamma_C}{\hbar} \delta(x) & \frac{\Gamma_B}{\hbar} \delta(x-L_4) e^{(L_3-L_4)\partial_x} & -i v \partial_x
\end{array}
\right)
\left(
\begin{array}{c}
\varphi_1(x) \\ \varphi_2(x) \\ \varphi_3(x) \\ \varphi_4(x)
\end{array}
\right)
= E
\left(
\begin{array}{c}
\varphi_1(x) \\ \varphi_2(x) \\ \varphi_3(x) \\ \varphi_4(x)
\end{array}
\right)\,,$}
\end{equation}
\end{widetext}

This equation, at energy $E=\hbar v k$, is readily solved by
\begin{equation}
\varphi_j(x)=e^{i k x} \left\{
\begin{array}{ll}
\alpha_j^{I}   & x<0 \\
\alpha_j^{II}  & 0<x<L_i \\
\alpha_j^{III} & x>L_i
\end{array}
\right.\,,
\end{equation}
where the coefficents $\{\alpha_j^{l}\}$ ($l=I,II,III$) are determined by the wavefunctions' matching conditions imposed by the delta potentials. For instance, integrating the first row of Eq.~(\ref{firstquantization}) from $x=0^-$ to $x=0^+$ we obtain
\begin{align}
\resizebox{.85\hsize}{!}{$\displaystyle -iv \left[ \varphi_1(0^+)-\varphi_1(0^-) \right]+\frac{\Gamma_A}{2\hbar} \left[  \varphi_3(0^+)+\varphi_3(0^-)\right] =0 \, , $}
\end{align}
and from the second row,
\begin{align}
\resizebox{.85\hsize}{!}{$\displaystyle -iv \left[ \varphi_3(0^+)-\varphi_3(0^-) \right]+\frac{\Gamma_A^*}{2\hbar} \left[  \varphi_1(0^+)+\varphi_1(0^-)\right] =0 \, ,$}
\end{align}
where we have used $\varphi_j(0)=( \varphi_j(0^+)+\varphi_j(0^-))/2$. Similarly, one obtains the remaining matching conditions
at QPCs $\mathcal{B}, \mathcal{C}$ and $\mathcal{D}$.

To calculate $s_{j1}$, we set  $\alpha_1^I=1$ and $\alpha_2^I=\alpha_3^I=\alpha_4^I=0$. We, then, solve the system of linear equations given by the matching conditions. Similarly one obtains the other  matrix elements of the scattering matrix.  Let us  focus on the contributions to $S_{23}$ which involve tunneling at all four QPCs. This contribution is proportional to $s_{21} s_{31}^*s_{24}^*s_{s34}$. We  obtain
\begin{align}\label{scattering1}
s_{21}(k)=&\exp[ik(L_1-L_2)]\frac{1-|\Gamma_A|^2/(4\hbar^2 v^2)}{1+|\Gamma_A|^2/(4\hbar^2 v^2)} \nonumber\\
&\quad\quad\quad\times \frac{-i(\Gamma^*_D/(\hbar v))}{(1+|\Gamma_B|^2/(4 \hbar^2 v^2))}\,,\\
\label{scattering2}
s_{31}(k)=&\frac{1-|\Gamma_B|^2/(4\hbar^2 v^2)}{1+|\Gamma_B|^2/(4\hbar^2 v^2)}
\frac{-i(\Gamma^*_A/(\hbar v))}{(1+|\Gamma_A|^2/(4\hbar^2 v^2))}\,,\\
\label{scattering3}
s_{24}(k)=&\frac{1-|\Gamma_D|^2/(4\hbar^2 v^2)}{1+|\Gamma_D|^2/(4\hbar^2 v^2)}
\frac{-i(\Gamma^*_C/(\hbar v))}{(1+|\Gamma_C|^2/(4\hbar^2 v^2))}\,,\\
\label{scattering4}
s_{34}(k)=&\exp[ik(L_4-L_3)]\frac{1-|\Gamma_C|^2/(4\hbar^2 v^2)}{1+|\Gamma_C|^2/(4\hbar^2 v^2)} \nonumber\\
&\quad\quad\quad\times
\frac{-i(\Gamma^*_B/(\hbar v))}{(1+|\Gamma_B|^2/(4\hbar^2 v^2))}\,.
\end{align}
Substituting Eqs.~(\ref{scattering1})-(\ref{scattering4}) in Eq.~(\ref{scattering}), and  keeping  terms up to the fourth order in tunneling amplitudes $\Gamma$, we obtain
\begin{multline}
S_{23}=-\frac{e^2}{2\pi \hbar^4{v}^4}\int d\omega \left[ f(\omega)-f(\omega-eV)\right]^2 \\ \times
\Big[\left( \Gamma_A \Gamma_B^* \Gamma_C \Gamma_D^*   \exp \left\{ i \frac{\omega}{v}(L_1+L_4-L_2-L_3)\right\} +c.c. \right) \\
+|\Gamma_A|^2|\Gamma_D|^2+|\Gamma_B|^2|\Gamma_C|^2 \Big] \, ,
\end{multline}
which coincides with our previous  expressions of  Eqs. (\ref{s0}) and (\ref{sphi}).

\section{Perturbative calculation of $S_{23}$ for $\nu=1$ in fermionic langauge}
\label{app_calcs23}
In this Appendix, we derive in detail Eqs.~(\ref{s0}) and (\ref{sphi}). We remind the reader that for our choice of chemical potentials,  $\mu_1=\mu_4=eV$ and $\mu_2=\mu_3=0$, we can write $S_{23}$ as [cf.~Eq.~\eqref{s23_divide_contrib}]
\begin{multline}
S_{23}=\frac{1}{\hbar^4 v^4}\Big[ (|\Gamma_\mathcal{A}|^2|\Gamma_\mathcal{D}|^2+|\Gamma_\mathcal{B}|^2|\Gamma_\mathcal{C}|^2)S_0\\ +(\Gamma_\mathcal{A} \Gamma_\mathcal{B}^* \Gamma_\mathcal{C} \Gamma_\mathcal{D}^*S_{\Phi}+c.c. )\Big]\,.
\end{multline}
From Eq.~(\ref{generalsij}), we have
\begin{widetext}
\begin{align}\label{s23gen}
S_{23}=&\frac{e^2 v^2}{2}\frac{(-i)^4}{4!\hbar^4} \sum_{\eta= \pm1 }
\int_{-\infty}^{\infty} dt_1 \int_K d\tau_1 d\tau_2 d\tau_3 d \tau_4
 \langle \langle  T_K
: \hat{\psi}^\dag_2(x_2,t_{1,\eta}+ \eta \, 0^+) \hat{\psi}_2(x_2,t_{1,\eta}):
 \nonumber\\
 &\times : \hat{\psi}^\dag_3(x_3,t_{2,-\eta}- \eta \, 0^+) \hat{\psi}_3(x_3,t_{2,-\eta}): H_T(\tau_1)H_T(\tau_2)H_T(\tau_3)H_T(\tau_4)
\rangle \rangle \, .
\end{align}
\end{widetext}

\subsection{Calculation of $S_{\Phi}$}
Let us consider first the calculation of $S_{\Phi}$. Collecting the relevant contributions from Eq.~(\ref{s23gen}), we can write
\begin{widetext}
\begin{align}\label{fluxdep}
\frac{\left(\Gamma_\mathcal{A} \Gamma_\mathcal{B}^* \Gamma_\mathcal{C} \Gamma_\mathcal{D}^*S_{\Phi}+c.c.\right)}{\hbar^4 v^4}=&
\frac{e^2 v^2}{2 \hbar^4}\sum_{\eta=\pm1}\int_{-\infty}^{\infty} dt_1
\int_K d\tau_1 d\tau_2 d\tau_3 d \tau_4
\langle \langle  T_K
: \hat{\psi}^\dag_2(x_2,t_{1,\eta}+ \eta \, 0^+) \hat{\psi}_2(x_2,t_{1,\eta}):\nonumber\\
&\times : \hat{\psi}^\dag_3(x_3,t_{2,-\eta}- \eta \, 0^+) \hat{\psi}_3(x_3,t_{2,-\eta}):    H_{T_\mathcal{A}}(\tau_1)H_{T_\mathcal{B}}(\tau_2)H_{T_\mathcal{C}}(\tau_3)H_{T_\mathcal{D}}(\tau_4)\rangle \rangle \, .
\end{align}
\end{widetext}
Notice that in Eq.~(\ref{fluxdep}) we have considered a particular permutation of the tunneling operators
$H_{T_\mathcal{A}}$,$H_{T_\mathcal{B}}$,$H_{T_\mathcal{C}}$ and $H_{T_\mathcal{D}}$.
Indeed, for the integer filling factor case considered here, all these operators commute with each other.
Any other permutation, for example, $H_{T_\mathcal{D}}(\tau_1)H_{T_\mathcal{B}}(\tau_2)H_{T_\mathcal{C}}(\tau_3)H_{T_\mathcal{A}}(\tau_4)$
yields exactly the same contribution and is taken into account in the prefactor of Eq.~(\ref{fluxdep}) (Anyonic tunneling operators, in the geometry discussed here, commute as well. This can be arranged by proper selection of Klein factors~\cite{feldman}).

In the following, we write explicitly a point $\tau_i$ on the Keldysh contour as $s \in (-\infty,\infty)$ and its branch index $\eta=\{-1,+1\}$.
Although the calculation presented here is not very involved, we nevertheless give the reader some
detail. Let us, as example, consider the term proportional to $\Gamma_\mathcal{A} \Gamma_\mathcal{B}^* \Gamma_\mathcal{C} \Gamma_\mathcal{D}^*$ in the $\langle \langle ...\rangle \rangle$ average of Eq.~(\ref{fluxdep}).
For such a contribution the $\langle \langle ...\rangle \rangle$ average coincides with the $ \langle ...\rangle $ average. Indeed, there are no terms proportional to $\Gamma_\mathcal{A} \Gamma_\mathcal{B}^* \Gamma_\mathcal{C} \Gamma_\mathcal{D}^*$ in $\langle I_2\rangle \langle I_3\rangle$ due to the chiral propagation along the edge channels. We therefore have
\begin{widetext}
\begin{multline}\label{fluxdep2}
\Gamma_\mathcal{A} \Gamma_\mathcal{B}^* \Gamma_\mathcal{C} \Gamma_\mathcal{D}^*\langle T_K ... \rangle=
 \langle  T_K  : \hat{\psi}^\dag_2(x_2,t_{1,\eta}+ \eta \, 0^+) \hat{\psi}_2(x_2,t_{1,\eta}):
 : \hat{\psi}^\dag_3(x_3,t_{2,-\eta}- \eta \, 0^+) \hat{\psi}_3(x_3,t_{2,-\eta}):    \\ \times
 \Gamma_\mathcal{A} \hat{\psi}^\dag_1(0,s_{1,\eta_1}) \hat{\psi}_3(0,s_{1,\eta_1})
 \Gamma_\mathcal{B}^* \hat{\psi}^\dag_3(L3,s_{2,\eta_2}) \hat{\psi}_4(L_4,s_{2,\eta_2})
 \Gamma_\mathcal{C} \hat{\psi}^\dag_4(0,s_{3,\eta_3}) \hat{\psi}_2(0,s_{3,\eta_3})
 \Gamma_\mathcal{D}^* \hat{\psi}^\dag_2(L_2,s_{4,\eta_4}) \hat{\psi}_1(L_1,s_{4,\eta_4})
 \rangle\,.
\end{multline}
\end{widetext}
We can now write explicitly $:\hat{\psi}_j^\dag \hat{\psi}_j:=\hat{\psi}_j^\dag \hat{\psi}_j-\langle vac |\hat{\psi}_j^\dag \hat{\psi}_j|vac\rangle$, where $|vac\rangle$ has been introduced in section \ref{integerfermionic} as $\prod_{j=1,..,4; k_j<0} c_{k,j}^\dag |0\rangle$. This yields four contributions in Eq.~(\ref{fluxdep2}). Each of these contributions can be decomposed using Wick's theorem. Notice that only the term with the  contraction of $\hat{\psi}^\dag_2(x_2,t_{1,\eta}+ \eta \, 0^+)$ with $\hat{\psi}_2(0,s_{3,\eta_3})$ and of $\hat{\psi}^\dag_3(x_3,t_{2,-\eta}- \eta \, 0^+)$
with $\hat{\psi}_3(0,s_{1,\eta_1})$ has non-vanishing contribution. The other terms are either unconnected contractions, which are identically zero in Keldysh formalism, or terms that cancel one  another
due to the normal ordering. This gives us
\begin{widetext}
\begin{align} \label{sphiinter}
S_{\Phi}=& \frac{e^2 v^6}{2}\sum_{\eta,\eta_1,...,\eta_4=\pm1}\eta_1\eta_2\eta_3\eta_4
\int_{-\infty}^{\infty} dt_1 ds_1 ds_2 ds_3 d s_4
 {G}^{\eta_3 \eta}_{2}(-x_2,s_3-t_1)  {G}^{\eta \eta_4}_{2}(x_2-L_2,t_1-s_4) \nonumber\\
 &\times{G}^{\eta_1 -\eta}_{3}(-x_3,s_1-t_2)
{G}^{-\eta \eta_2}_{3}(x_3-L_3,t_2-s_2)  {G}^{\eta_4 -\eta_1}_{1}(L_1,s_4-s_1)   {G}^{\eta_2 -\eta_3}_{4}(L_4,s_2-s_3)\,.
\end{align}
In Eq.~(\ref{sphiinter}) we can rewrite the Green's function using a mixed  space-energy representation,
\begin{align}
S_{\Phi}= &\frac{e^2 v^6}{2(2\pi)^6 }\sum_{\eta,\eta_1,...,\eta_4=\pm1}\eta_1\eta_2\eta_3\eta_4
\int dt_1 ds_1 ds_2 ds_3 d s_4
\int d \omega_1d \omega_2d \omega_3d \omega_4d \omega_5 d \omega_6 \nonumber\\
&\times {G}^{\eta_3 \eta}_{2}(-x_2,\omega_1)  {G}^{\eta \eta_4}_{2}(x_2-L_2,\omega_2) {G}^{\eta_1 -\eta}_{3}(-x_3,\omega_3)
{G}^{-\eta \eta_2}_{3}(x_3-L_3,\omega_4)   {G}^{\eta_4 -\eta_1}_{1}(L_1,\omega_5)   {G}^{\eta_2 -\eta_3}_{4}(L_4,\omega_6)\nonumber\\
& \times e^{-i \omega_1(s_3-t_1)-i\omega_2(t_1-s_4)-i\omega_3(s_1-t_2)-i\omega_4(t_2-s_2)-i\omega_5(s_4-s_1)-i\omega_6(s_2-s_3)} \, .
\end{align}
The integrations over the times $t_1,s_1,...,s_4$ are trivial, yielding delta functions with arguments $\omega_1, ... ,\omega_6$. The remaining integrals are also straightforward, we obtain
\begin{multline}
\label{b7}
S_{\Phi}= \frac{e^2 v^6}{4\pi}\sum_{\eta,\eta_1,...,\eta_4=\pm1}\eta_1\eta_2\eta_3\eta_4
\int d \omega \,
{G}^{\eta_3 \eta}_{2}(-x_2,\omega)  {G}^{\eta \eta_4}_{2}(x_2-L_2,\omega) {G}^{\eta_1 -\eta}_{3}(-x_3,\omega)
{G}^{-\eta \eta_2}_{3}(x_3-L_3,\omega)  \\ \times  {G}^{\eta_4 -\eta_1}_{1}(L_1,\omega)   {G}^{\eta_2 -\eta_3}_{4}(L_4,\omega) \,.
\end{multline}
\end{widetext}
We can now substitute in Eq.~\eqref{b7} the mixed space-energy representation Green's functions from Eqs.~(\ref{FT1}-\ref{FT4}). Performing the sum over the Keldysh indices $\eta,\eta_1,..,\eta_4$, we obtain
\begin{multline}
S_{\Phi}=-\frac{e^2}{2\pi }\int d\omega \left[ f(\hbar \omega)-f(\hbar \omega-eV)\right]^2
\\ \times
 \exp \left\{ i \frac{\omega}{v}(L_1+L_4-L_2-L_3)\right\} \, .
\end{multline}

\subsection{Calculation of $S_0$}
Let us now turn to the calculation of $S_0$, i.e. the component of $S_{23}$ not modulated by the flux of the magnetic field. From Eq.~(\ref{s23gen}) we collect terms proportional to $|\Gamma_\mathcal{A}|^2|\Gamma_\mathcal{D}|^2$
to find
\begin{small}
\begin{multline}
\frac{|\Gamma_\mathcal{A}|^2|\Gamma_\mathcal{D}|^2}{\hbar^4 v^4} S_0=\frac{e^2 v^2}{8\,\hbar^4}\sum_{\eta=\pm1}\int_{-\infty}^{+\infty} dt_1
\int_K d\tau_1 d\tau_2 d\tau_3 d \tau_4
 \times \\
\langle \langle  T_K
: \hat{\psi}^\dag_2(x_2,t_{1,\eta}+ \eta \, 0^+) \hat{\psi}_2(x_2,t_{1,\eta}):\\
 \times: \hat{\psi}^\dag_3(x_3,t_{2,-\eta}- \eta \, 0^+) \hat{\psi}_3(x_3,t_{2,-\eta}):
 \\ \times
   H_{T_\mathcal{A}}(\tau_1)H_{T_\mathcal{A}}(\tau_2)H_{T_\mathcal{D}}(\tau_3)H_{T_\mathcal{D}}(\tau_4)\rangle \rangle \, .
\end{multline}
\end{small}
In this case the calculation is slightly more involved compared to the previous one. Indeed, in $\langle I_2\rangle\langle I_3\rangle$ there are contributions proportional to $|\Gamma_\mathcal{A}|^2|\Gamma_\mathcal{D}|^2$.
Taking care of the normal ordering, and subtracting terms coming from $\langle I_2\rangle \langle I_3\rangle$, we obtain
\begin{widetext}
\begin{small}
\begin{align}\label{fluxind}
&S_0 =\frac{e^2 v^6}{8}\sum_{\eta,\eta_1,...,\eta_4=\pm1}\eta_1\eta_2\eta_3\eta_4
\int dt_1 ds_1 ds_2 ds_3 d s_4
 \\   &\Big[ {G}^{\eta_1 \eta}_{3}(-x_3,s_1-t_1)  {G}^{\eta \eta_2}_{3}(x_3,t_1-s_2) {G}^{\eta_3 -\eta}_{2}(L_2-x_2,s_3-t_2)
{G}^{-\eta \eta_4}_{2}(x_2-L_2,t_2-s_4)    {G}^{\eta_4 -\eta_1}_{1}(L_1,s_4-s_1)   {G}^{\eta_2 -\eta_3}_{1}(-L_1,s_2-s_3)+ \nonumber\\
& {G}^{\eta_1 \eta}_{3}(-x_3,s_1-t_1)  {G}^{\eta \eta_2}_{3}(x_3,t_1-s_2) {G}^{\eta_4 -\eta}_{2}(L_2-x_2,s_4-t_2)
{G}^{-\eta \eta_3}_{2}(x_2-L_2,t_2-s_3)    {G}^{\eta_3 -\eta_1}_{1}(L_1,s_3-s_1)   {G}^{\eta_2 -\eta_4}_{1}(-L_1,s_2-s_4)+ \nonumber\\
& {G}^{\eta_2 \eta}_{3}(-x_3,s_2-t_1)  {G}^{\eta \eta_1}_{3}(x_3,t_1-s_1) {G}^{\eta_3 -\eta}_{2}(L_2-x_2,s_3-t_2)
{G}^{-\eta \eta_4}_{2}(x_2-L_2,t_2-s_4)   {G}^{\eta_4 -\eta_2}_{1}(L_1,s_4-s_2)   {G}^{\eta_1 -\eta_3}_{1}(-L_1,s_1-s_3)+ \nonumber\\
&{G}^{\eta_2 \eta}_{3}(-x_3,s_2-t_1)  {G}^{\eta \eta_1}_{3}(x_3,t_1-s_1) {G}^{\eta_4 -\eta}_{2}(L_2-x_2,s_4-t_2)
{G}^{-\eta \eta_3}_{2}(x_2-L_2,t_2-s_3)   {G}^{\eta_1 -\eta_4}_{1}(L_1,s_2-s_4)   {G}^{\eta_3 -\eta_2}_{1}(-L_1,s_3-s_2)
\Big]\,.\nonumber
\end{align}
\end{small}
\end{widetext}
As in the calculation of $S_{\Phi}$, we express the Green's function in the previous expression via their mixed energy-space representation. Performing the straightforward integrals and the sum over the Keldysh indices we find
\begin{equation}
S_0=-\frac{e^2}{2\pi }\int d\omega \left[ f(\hbar \omega)-f(\hbar \omega-eV)\right]^2\,.
\end{equation}

%%%%%%%%%%%%%%%%%%%%%%%%%%%%%%%%%%%%%%%%%%%%%%%%%%%
\section{Perturbative calculation of $S_{\Phi}$ for $\nu=1$ in bosonic language}
\label{app_integer_bosonic}
In this Appendix we present the calculation of $S_{\Phi}$ obtained from the tunneling currents introduced in Section \ref{integerbosonic}, Eqs. (\ref{tunncurr1}) and (\ref{tunncurr2}). From Eq. (\ref{s23t}), we write explicitly the integrals over the Keldysh contour and obtain
\begin{align}
&S^{(T)}_{23}(0)=\frac{(-i)^2}{4\hbar^2}\int_{-\infty}^{\infty}\int_{-\infty}^{\infty}\int_{-\infty}^{\infty}dt\;dt_1\;dt_2\times\\ &\sum_{\eta,\eta_1,\eta_2}\eta_1 \eta_2 \langle \langle T_K  I_{T2}(0_\eta) I_{T3}(t_{-\eta})H_T(t_{1,\eta_1})H_T(t_{2,\eta_2})\rangle \rangle \,.\nonumber
\end{align}

Let us collect terms proportional to $\Gamma_{\mathcal{A}}\Gamma_{\mathcal{B}}^\star\Gamma_{\mathcal{C}}\Gamma_{\mathcal{D}}^\star$ from the previous expression. We have
\begin{widetext}
\begin{multline}
\frac{\Gamma_{\mathcal{A}}\Gamma_{\mathcal{B}}^\star\Gamma_{\mathcal{C}}\Gamma_{\mathcal{D}}^\star}{\hbar^4 v^4} S_{\Phi}^{(T)}=\frac{e^2}{4\hbar^4}
\int_{-\infty}^{\infty} dt \int_{-\infty}^{\infty} dt_1\int_{-\infty}^{\infty}\;dt_2 \sum_{\eta,\eta_1,\eta_2}\eta_1 \eta_2  \langle T_K
\Big\{
\mathcal{C}(0_\eta)\mathcal{A}(t_{-\eta})\left[\mathcal{B}^\dagger(t_{1,\eta_1})\mathcal{D}^\dagger(t_{2,\eta_2})+\mathcal{D}^\dagger(t_{1,\eta_1})\mathcal{B}^\dagger(t_{2,\eta_2})\right]  \\
 -\mathcal{C}(0_\eta)\mathcal{B}^\dagger(t_{-\eta})\left[\mathcal{A}(t_{1,\eta_1})\mathcal{D}^\dagger(t_{2,\eta_2})+\mathcal{D}^\dagger(t_{1,\eta_1})\mathcal{A}(t_{2,\eta_2})\right]
 -\mathcal{D}^\dagger(0_\eta)\mathcal{A}(t_{-\eta})\left[\mathcal{B}^\dagger(t_{1,\eta_1})\mathcal{C}(t_{2,\eta_2})+\mathcal{C}(t_{1,\eta_1})\mathcal{B}^\dagger(t_{2,\eta_2})\right] \\
+ \mathcal{D}^\dagger(0_\eta)\mathcal{B}^\dagger(t_{-\eta})\left[\mathcal{A}(t_{1,\eta_1})\mathcal{C}(t_{2,\eta_2})+\mathcal{C}(t_{1,\eta_1})\mathcal{A}(t_{2,\eta_2})\right] \Big\} \rangle\,.
\label{app_integer_bosonic_c2}
\end{multline}
Recall that for the above contribution $\langle \langle ...\rangle \rangle$-averaging coincides with $\langle ... \rangle$-averaging. Collecting the non-vanishing contractions, we obtain
\begin{small}
\begin{multline}\label{stphig}
S_{\Phi}^{(T)}=-\frac{e^2 v^4}{2}
\int_{-\infty}^{\infty} dt \int_{-\infty}^{\infty} dt_1\int_{-\infty}^{\infty}\;dt_2 \sum_{\eta,\eta_1,\eta_2}\eta_1 \eta_2
\Big\{
G_{1}^{\eta_2 -\eta}(L_1,t_2-t)G_{2}^{\eta \eta_2}(-L_2,-t_2)G_{3}^{-\eta \eta_1}(-L_3,t-t_1)G_{4}^{\eta_1 \eta}(L_4,t_1)
\\
-G_{1}^{\eta_2 \eta_1}(L_1,t_2-t_1)G_{2}^{\eta \eta_2}(-L_2,-t_2)G_{3}^{\eta_1 -\eta}(-L_3,t_1-t)G_{4}^{-\eta \eta}(L_4,t)
\\
-G_{1}^{\eta -\eta}(L_1,-t)G_{2}^{\eta_2 \eta}(-L_2,t_2)G_{3}^{-\eta \eta_1}(-L_3,t-t_1) G_{4}^{\eta_1 \eta_2}(L_4,t_1-t_2)
\\
+G_{1}^{\eta \eta_1}(L_1,-t_1)G_{2}^{\eta_2 \eta}(-L_2,t_2)G_{3}^{\eta_1 -\eta}(-L_3,t_1-t)G_{4}^{-\eta \eta_2}(L_4,t-t_2)\,.
\Big\}
\end{multline}
\end{small}
\end{widetext}

We proceed  here with a slightly different way than what was done in Appendix \ref{app_calcs23}. Such an approach
will turn out to be useful in the case of a fractional filling factor.
Due to the finite lengths of the interferometer arms ($L_i>0$), we notice that in the expressions of the bosonic Green's functions [cf. Eq.~(\ref{eq:GtoS})] we can substitute $\chi_{\eta_1 \eta_2}(t_1-t_2)\rightarrow \chi_{\eta_1 \eta_2}(x_1-x_2)$.

After this substitution, we can perform the sums over the Keldysh indices in Eq.~(\ref{stphig}).
Remarkably, the first terms in Eq.~(\ref{stphig}) sum up to zero. We are left with the following expression:
\begin{widetext}
\begin{small}
\begin{align}\label{stphig2}
&S_{\Phi}^{(T)}=\frac{e^2 v^4}{2(2\pi)^4}\sum_{\eta, \eta_1,\eta_2} \eta_1 \eta_2
\int dt dt_1 dt_2
 e^{i\mu(L_1/v+L_4/v+t_1+t_2-t)/\hbar}
\frac{1}{\frac{\hbar\beta v}{\pi}\sinh[\frac{\pi}{\hbar\beta v}(-v\, t1-L_1-i l_c   \chi_{\eta \eta_1}(L_1) )]}\times
\\ &
\frac{1}{\frac{\hbar\beta v}{\pi}\sinh[\frac{\pi}{\hbar\beta v}(v\, t_2+L_2-i l_c \chi_{\eta_2 \eta}(-L_2))]}
\frac{1}{\frac{\hbar\beta v}{\pi}\sinh[\frac{\pi}{\hbar\beta v}(v\, (t_1-t)+L_3-i l_c  \chi_{\eta_1-\eta}(-L_3))]}
\frac{1}{\frac{\hbar\beta v}{\pi}\sinh[\frac{\pi}{\hbar\beta v}(v\, (t-t_2)-L_4-i l_c  \chi_{-\eta \eta_2}(L_4))]}\,.\nonumber
\end{align}
\end{small}
\end{widetext}
Notice that  $\chi_{\eta \eta_1}(L_1)=\eta_1$, and similarly for the other $\chi$-factors. Substituting this into Eq.~(\ref{stphig2}), we obtain Eq.~\eqref{electrons_in_integer}.

%%%%%%%%%%%%%%%%%%%%%%%%%%%%%%%%%%%%%%%%%%%%%%%%%%%%%%%%%%%%%
\section{Perturbative calculation of $S_{\Phi}$ for electron hopping over $\nu=1/(2n+1)$ edges}
\label{app_equiv_bos_ferm}
In this appendix, we prove explicitly Eq.~(\ref{derivatives1}), i.e., the relation between $S^{(e)}_{\Phi}$ and $S_{\Phi}$.
As previously done for $S_{\Phi}$ in Appendix \ref{app_integer_bosonic}, we collect contributions proportional
to $\Gamma_\mathcal{A} \Gamma_\mathcal{B}^* \Gamma_\mathcal{C} \Gamma_\mathcal{D}^*$ in Eq.~(\ref{s23-nu-e}) in order to calculate the magnetic-flux-modulated part of the cross-current correlations. We obtain a contribution similar to Eq.~\eqref{app_integer_bosonic_c2}:
\begin{widetext}
\begin{multline}\label{1-3-allterms}
\frac{\Gamma_{\mathcal{A}}\Gamma_{\mathcal{B}}^\star\Gamma_{\mathcal{C}}\Gamma_{\mathcal{D}}^\star}{\hbar^4 v^4} S_{\Phi}^{(e)}=\frac{e^2}{4\hbar^4}
\int_{-\infty}^{\infty} dt \int_{-\infty}^{\infty} dt_1\int_{-\infty}^{\infty}\;dt_2 \sum_{\eta,\eta_1,\eta_2}\eta_1 \eta_2  \\\Big\langle T_K
\Big\{
\mathcal{C}^{(e)}(0_\eta)\mathcal{A}^{(e)}(t_{-\eta})\left[\mathcal{B}^{(e)\dagger}(t_{1,\eta_1})\mathcal{D}^{(e)\dagger}(t_{2,\eta_2})+\mathcal{D}^{(e)\dagger}(t_{1,\eta_1})\mathcal{B}^{(e)\dagger}(t_{2,\eta_2})\right]  \\
 -\mathcal{C}^{(e)}(0_\eta)\mathcal{B}^{(e)\dagger}(t_{-\eta})\left[\mathcal{A}^{(e)}(t_{1,\eta_1})\mathcal{D}^{(e)\dagger}(t_{2,\eta_2})+\mathcal{D}^{(e)\dagger}(t_{1,\eta_1})\mathcal{A}^{(e)}(t_{2,\eta_2})\right]\\
 -\mathcal{D}^{(e)\dagger}(0_\eta)\mathcal{A}^{(e)}(t_{-\eta})\left[\mathcal{B}^{(e)\dagger}(t_{1,\eta_1})\mathcal{C}^{(e)}(t_{2,\eta_2})+\mathcal{C}^{(e)}(t_{1,\eta_1})\mathcal{B}^{(e)\dagger}(t_{2,\eta_2})\right] \\
+ \mathcal{D}^{(e)\dagger}(0_\eta)\mathcal{B}^{(e)\dagger}(t_{-\eta})\left[\mathcal{A}^{(e)}(t_{1,\eta_1})\mathcal{C}^{(e)}(t_{2,\eta_2})+\mathcal{C}^{(e)}(t_{1,\eta_1})\mathcal{A}^{(e)}(t_{2,\eta_2})\right] \Big\} \Big\rangle\,.
\end{multline}
%\end{widetext}
In this case, in contrast to Appendices \ref{app_calcs23} and \ref{app_integer_bosonic} (tunneling of electrons in integer quantum Hall systems), the tunneling operators $\mathcal{A}^{(e)}, \mathcal{B}^{(e)}, \mathcal{C}^{(e)}$, and $\mathcal{D}^{(e)}$ describe tunneling of electrons between the  edge channels of a fractional Hall liquid.

Let us consider, for example, the following contribution from the above expression.
\begin{equation}
\frac{e^2}{4\hbar^4} \sum_{\eta \eta_1 \eta_2}\eta_1 \eta_2 \int dt \, dt_1 \, dt_2\langle T_K
\mathcal{D}^{(e)\dagger}(0_\eta)\mathcal{B}^{(e)\dagger}(t_{-\eta}) \mathcal{A}^{(e)}(t_{1,\eta_1})\mathcal{C}^{(e)}(t_{2,\eta_2})
\rangle \, .
\end{equation}
Making use of Eq.~(\ref{green-e}) we obtain
%\begin{widetext}
\begin{small}
\begin{align}\label{electron-in-1/3}
&\frac{e^2}{2\hbar^4} \frac{\Gamma_{\mathcal{A}}\Gamma_{\mathcal{B}}^\star\Gamma_{\mathcal{C}}\Gamma_{\mathcal{D}}^\star}{l_c^4} l_c^{4/\nu} \sum_{ \eta_1,\eta_2} \eta_1 \eta_2
\int dt \, dt_1 \, dt_2
 e^{i\mu(L_1/v+L_4/v+t_1+t_2-t)/\hbar}
\frac{1}{\left\{\frac{\hbar\beta v}{\pi}\sinh[\frac{\pi}{\hbar\beta v}(-v\, t1-L_1-i \epsilon   \eta_1 )]\right\}^{1/\nu}}
\\ &\times
\frac{1}{\left\{\frac{\hbar\beta v}{\pi}\sinh[\frac{\pi}{\hbar\beta v}(v\, t_2+L_2+i \epsilon \eta_2)]\right\}^{1/\nu}}
\frac{1}{\left\{\frac{\hbar\beta v}{\pi}\sinh[\frac{\pi}{\hbar\beta v}(v\, (t_1-t)+L_3+i \epsilon \eta_1 )]\right\}^{1/\nu}}
\frac{1}{\left\{\frac{\hbar\beta v}{\pi}\sinh[\frac{\pi}{\hbar\beta v}(v\, (t-t_2)-L_4-i \epsilon  \eta_2)]\right\}^{1/\nu}}\,.\nonumber
\end{align}
\end{small}
\end{widetext}
It is now straightforward to evaluate Eq.~\eqref{electron-in-1/3} by taking derivatives of the calculated Eq.~\eqref{electrons_in_integer}. For example, for $\nu=1/3$ we can use the relation
\begin{equation}
\frac{1}{2}\left(\frac{\partial^2}{\partial L^2}-b^2\right)\sinh(a \pm b L)^{-1}=\sinh(a\pm b L)^{-3}\,,
\end{equation}
while similar relations apply to other fractions $\nu=1/(2n+1)$. Hence, by taking the derivatives of  Eq.~\eqref{electrons_in_integer} with respect to $L_1,..,L_4$, we obtain Eq.~(\ref{electron-in-1/3}). The same considerations apply to all contributions in Eq.~(\ref{1-3-allterms}). Thus, eventually we obtain Eq.~\eqref{derivatives1}.

%%%%%%%%%%%%%%%%%%%%%%%%%%%%%%%%%%%%%%%%%%%%%%%%%%%%%%%%%%%%%
\section{Quasiparticle Tunneling--Perturbation approach.}
\label{twevlth_order_break}
In this appendix, we employ the Keldysh perturbation theory for the calculation of the current-current correlation function
in the case of quasiparticle tunneling. In contrast to the electron tunneling, a straightforward perturbative calculation,
(apart from being much more cumbersome than the kinetic approach) turns out to be insufficient in the
quasiparticle case in view of arising divergencies. These divergencies are similar to those encountered in the
treatment of the quasiparticle Mach-Zehnder interference \cite{feldman}; in both cases, they are intimately related
to the Byers-Yang theorem~\cite{byersYang}.

The purpose of this appendix is to demonstrate how the Keldysh perturbation theory works in the quasiparticle case.
This consideration allows us to establish a bridge between this approach and the kinetic framework adopted in the main text (cf. Sec.~\ref{kinetic}).
In particular, we discuss how the perturbative treatment should be modified to avoid divergencies, and show how this
regularization is related to the kinetic approach of the main text.
For simplicity, in this appendix, we will focus on the case of $\nu=1/3$.

We are going to calculate the lowest-order contribution to the current-current correlation function
with the correct periodicity in $\Phi_{\rm AB}$.
In view of the Byers-Yang theorem, this turns out to be proportional
to $\Gamma_\mathcal{A}^3 \Gamma_\mathcal{B}^{*3}\Gamma_\mathcal{C}^3 \Gamma_\mathcal{D}^{*3}$, i.e., it arises only at the $12^{\rm th}$ order in the
expansion of the current correlation function
in the tunneling amplitude. All lower-order contributions have wrong periodicity
(for example, the term proportional to $\Gamma_\mathcal{A} \Gamma_\mathcal{B} \Gamma_\mathcal{C} \Gamma_\mathcal{D}$ is $3\Phi_0$ periodic).
Within the bosonization framework, the terms with wrong periodicity vanish automatically by the proper attachment of Klein factors to the tunneling operators~\cite{feldman}.

Here, we restrict ourselves to considering a typical term of this $12^{\rm th}$-order expansion (we remind the reader that
our actual calculation relies on the rate equations). Since we are interested only in the general structure
of the perturbative expressions, we do not write all the prefactors and use the proportionality sign.
The formal expression for the total $12^{\rm th}$-order contribution to the current-current correlation involves 11 time integrations:
\begin{align}
&S_{23} \propto \sum_{\eta_0,\eta_1,...,\eta_{10}=\pm 1} \eta_1\dots \eta_{10} \int _{-\infty}^{+\infty} dt_0 dt_1\dots dt_{10}\nonumber\\
 &\langle T_k I_2(0_{\eta_0})   I_3(t_{0,-{\eta_0}}) H_T(t_{1,\eta_1})    \dots H_T(t_{10,\eta_{10}})  \rangle,
\end{align}
where $H_T(t_{i,\eta_i})$ is defined in Eqs.~\eqref{tunneling-op-frac} and \eqref{tunneling-op-frac-tun}.
As in the integer case, here too we have four types of contributions in the integrand:
\begin{widetext}
\begin{align}
\label{four-contr}
&\langle T_k I_2(0_{\eta_0})   I_3(t_{0,-{\eta_0}}) H_T(t_{1,\eta_1})    \dots H_T(t_{10,\eta_{10}}) \rangle\propto \big\{ \langle T_k   \mathcal{C}(0_{\eta_0}) \mathcal{A}(t_{0,-{\eta_0}})\times\\
&[ \mathcal{C}(t_{1,\eta_1})  \mathcal{C}(t_{2,\eta_2})  \mathcal{A}(t_{3,\eta_3})  \mathcal{A}(t_{4,\eta_4})  \mathcal{D}^\dag(t_{5,\eta_5})
 \mathcal{D}^\dag(t_{6,\eta_6})    \mathcal{D}^\dag(t_{7,\eta_7})    \mathcal{B}^\dag(t_{8,\eta_8})
 \mathcal{B}^\dag(t_{9,\eta_9})    \mathcal{B}^\dag(t_{10,\eta_{10}}) + \text{permutations}]   \rangle \nonumber\\
& -  \langle T_k   \mathcal{C}(0_{\eta_0}) \mathcal{B}^\dag(t_{0,-{\eta_0}})[ \dots]   \rangle
 + \langle T_k \mathcal{D}^\dag(0_{\eta_0}) \mathcal{A}(t_{0,-{\eta_0}})[\dots] \rangle
 - \langle T_k \mathcal{D}^\dag(0_{\eta_0}) \mathcal{B}^\dag(t_{0,-{\eta_0}})[\dots] \rangle \big\}\,.\nonumber
 \end{align}
 \end{widetext}
Here, however, the time-dependent tunneling operators are given by Eq.~\eqref{tunneling-op-frac} with time-dependent fields $\phi_i(x,t)$.
For brevity, in this Appendix we suppress the superscript ``($q$)'' in the quasiparticle tunneling operators.
Furthermore, here we do not include the Klein factors, focusing only on the structure of quasiparticle propagators
(we will briefly comment on the role of the Klein factors at the end of this appendix).

\begin{figure}
\includegraphics[width=\columnwidth]{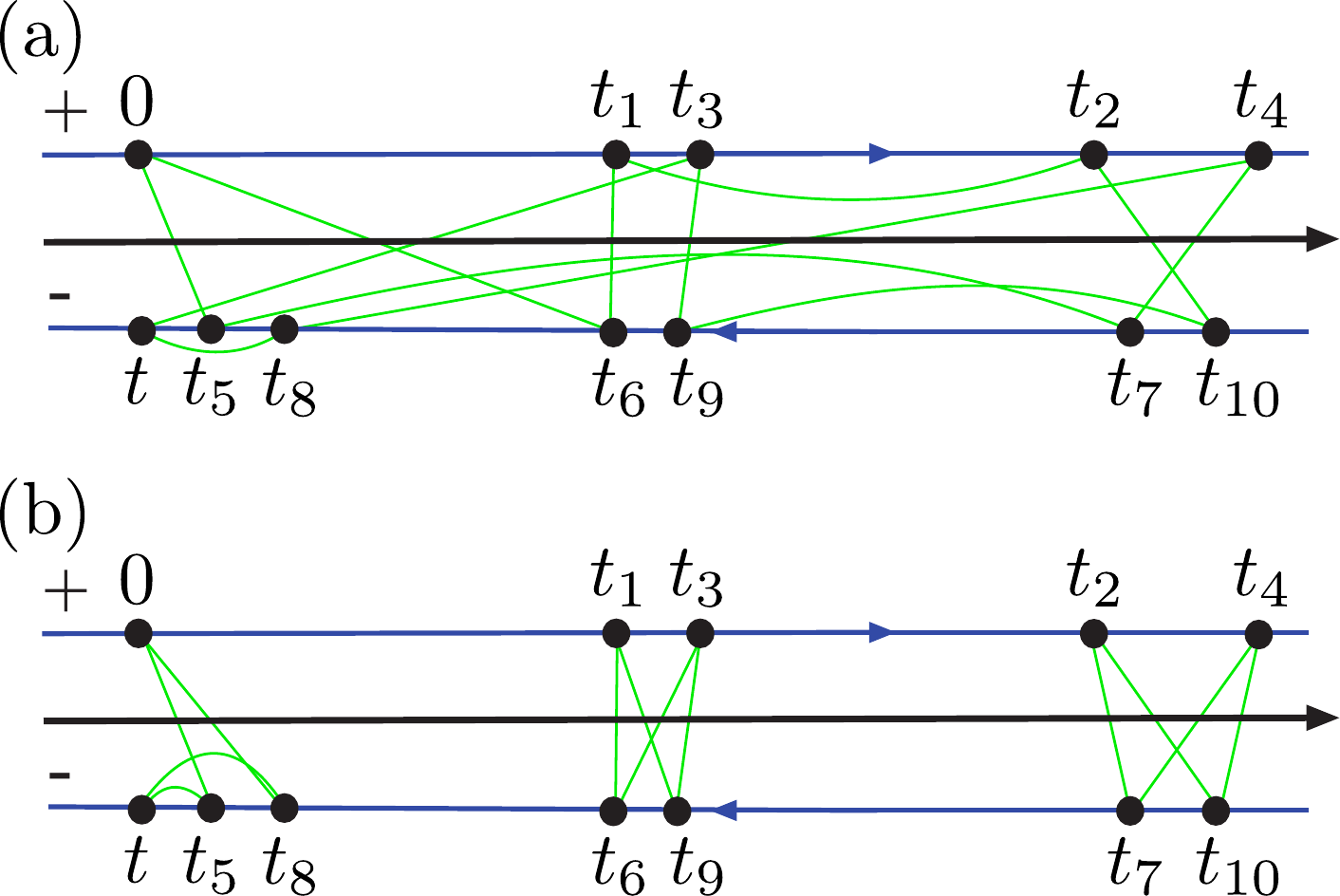}
\caption{(a) A typical Keldysh diagram for the correlator $S_{23}$. (b) Possible groupings of times forming loops.
The particular choice, shown here, of placing the times on the two Keldysh branches does not affect the separation into three loops. The separation in time between consecutive loops is assumed to be much larger than $\hbar \beta=L_T/v$.}
\label{Fig:grouping}
\end{figure}

Let us now focus on the first (the one shown explicitly) term in Eq.~(\ref{four-contr}).
Representing the tunneling Hamiltonian in the bosonized form, we write the corresponding contribution to the current correlation
function as:
\begin{widetext}
\begin{multline}
\label{S23-bosonized}
S_{23}^{\mathcal{CA}} \propto \sum_{\eta_0,\eta_1,...,\eta_{10}=\pm 1} \eta_1\dots \eta_{10} \int _{-\infty}^{+\infty} dt_0 dt_1\dots dt_{10}
e^{-i \nu e V [t_0+t_1+t_2+t_3+t_4-t_5-t_6-t_7-t_8-t_9-t_{10}]}  \\
\times \langle T_k e^{i\sqrt{\nu}  \phi_1(0,t_{0,-{\eta_0}})} e^{i\sqrt{\nu} \phi_1(0,t_{3,\eta_3})}e^{i\sqrt{\nu} \phi_1(0,t_{4,\eta_4})}e^{-i\sqrt{\nu} \phi_1(L_1,t_{5,\eta_5})}e^{-i\sqrt{\nu} \phi_1(L_1,t_{6,\eta_6})}e^{-i\sqrt{\nu} \phi_1(L_1,t_{7,\eta_7})}\rangle \\
\times
\langle T_k e^{-i\sqrt{\nu} \phi_2(0,0_{\eta_0})}   e^{-i\sqrt{\nu} \phi_2(0,t_{1,\eta_1})}e^{-i\sqrt{\nu} \phi_2(0,t_{2,\eta_2})}e^{i\sqrt{\nu} \phi_2(L_2,t_{5,\eta_5})}e^{i\sqrt{\nu} \phi_2(L_2,t_{6,\eta_6})}e^{i\sqrt{\nu} \phi_2(L_2,t_{7,\eta_7}) }\rangle \\
\times
\langle T_k e^{-i\sqrt{\nu} \phi_3(0,t_{0,-{\eta_0}})}e^{-i\sqrt{\nu} \phi_3(0,t_{3,\eta_3})} e^{-i\sqrt{\nu} \phi_3(0,t_{4,\eta_4})}e^{i\sqrt{\nu} \phi_3(L_3,t_{8,\eta_8})}e^{i\sqrt{\nu} \phi_3(L_3,t_{9,\eta_9})}e^{i\sqrt{\nu} \phi_3(L_3,t_{10,\eta_{10} }) } \rangle \\
\times
\langle T_k e^{i\sqrt{\nu} \phi_4(0,0_{\eta_0})}e^{i\sqrt{\nu} \phi_4(0,t_{1,\eta_1})} e^{i\sqrt{\nu} \phi_4(0,t_{2,\eta_2})}e^{-i\sqrt{\nu} \phi_4(L_4,t_{8,\eta_8})}
e^{-i\sqrt{\nu} \phi_4(L_4,t_{9,\eta_9})}e^{-i\sqrt{\nu} \phi_4(L_4,t_{10,\eta_{10} }) } \rangle\,.
\end{multline}
\end{widetext}
The integrand here is a product of four Keldysh traces, each corresponding to one of the edges.
Each trace contains three vertex operators (exponentials of bosonic fields)
with $+i \sqrt{\nu} \phi$ and three exponentials with $-i \sqrt{\nu} \phi$
that correspond to the tunneling of three quasiparticles into the edge and from the edge.
As usual, averaging of the product of exponential operators with the free (quadratic) action
generates all possible contractions between the points on the Keldysh contour.
For averages involving same (opposite) signs in front of bosonic fields, one gets (see, e.g., Ref.~[\onlinecite{vondelft}]):
\begin{align}
\left\langle e^{\pm i\sqrt{\nu} \phi(x,t_{m,\eta_m})} e^{\mp i\sqrt{\nu} \phi(0,t_{n,\eta_n})}\right\rangle &\propto& \left[\frac{1}{s_{\eta_m,\eta_n}^{(m,n)}(x)}\right]^\nu\,,
\end{align}
where we use the short-hand notation
\begin{equation}
s_{\eta_m,\eta_n}^{(m,n)}(x)=\sinh \left\{
\frac{\pi}{\hbar \beta v}\,   [x-v(t_n-t_m) -i \chi_{\eta_m,\eta_n}^{(m,n)} l_c] \right\}\,,
\label{sinh}
\end{equation}
with
\begin{equation}
\chi_{\eta_m,\eta_n}^{(m,n)}=\frac{\eta_m+\eta_n}{2}\text{sgn}(t_m-t_n)-\frac{\eta_m-\eta_n}{2}.
\end{equation}
Performing all the contractions, we arrive at
\begin{widetext}
\begin{multline}\label{twelve}
S_{23}^{\mathcal{CA}} \propto \sum_{\eta_0,\eta_1,...,\eta_{10}=\pm 1} \eta_1\dots \eta_{10} \int _{-\infty}^{+\infty} dt_0 dt_1\dots dt_{10}
e^{-i \nu e V (t_0+t_1+t_2+t_3+t_4-t_5-t_6-t_7-t_8-t_9-t_{10})}  \\
\times \left[
\frac{s_{-\eta_0,\eta_3}^{(0,3)}(0)     s_{-\eta_0,\eta_4}^{(0,4)}(0)    s_{\eta_3,\eta_4}^{(3,4)}(0) s_{\eta_5,\eta_6}^{(5,6)}(0)     s_{\eta_5,\eta_7}^{(5,7)}(0)    s_{\eta_6,\eta_7}^{(6,7)}(0)}
     {s_{-\eta_0,\eta_5}^{(0,5)}(L_1)    s_{-\eta_0,\eta_6}^{(0,6)}(L_1)   s_{-\eta_0,\eta_7}^{(0,7)}(L_1) s_{\eta_3,\eta_5}^{(3,5)}(L_1)    s_{\eta_3,\eta_6}^{(3,6)}(L_1)   s_{\eta_3,\eta_7}^{(3,7)}(L_1) s_{\eta_4,\eta_5}^{(4,5)}(L_1)    s_{\eta_4,\eta_6}^{(4,6)}(L_1)   s_{\eta_4,\eta_7}^{(4,7)}(L_1) }
\right]^{1/3}
\\
\times
\left[
\frac{s_{\eta_0,\eta_1}^{(i,1)}(0)     s_{\eta_0,\eta_2}^{(i,2)}(0)    s_{\eta_1,\eta_2}^{(1,2)}(0)
      s_{\eta_5,\eta_6}^{(5,6)}(0)    s_{\eta_5,\eta_7}^{(5,7)}(0)    s_{\eta_6,\eta_7}^{(6,7)}(0)    }
{s_{\eta_0,\eta_5}^{(i,5)}(L_2)    s_{\eta_0,\eta_6}^{(i,6)}(L_2)   s_{\eta_0,\eta_7}^{(i,7)}(L_2)
s_{\eta_1,\eta_5}^{(1,5)}(L_2)    s_{\eta_1,\eta_6}^{(1,6)}(L_2)   s_{\eta_1,\eta_7}^{(1,7)}(L_2)
s_{\eta_2,\eta_5}^{(2,5)}(L_2)    s_{\eta_2,\eta_6}^{(2,6)}(L_2)   s_{\eta_2,\eta_7}^{(2,7)}(L_2)}
\right]^{1/3}
\\
\times\left[
\frac{s_{-\eta_0,\eta_3}^{(0,3)}(0)     s_{-\eta_0,\eta_4}^{(0,4)}(0)    s_{\eta_3,\eta_4}^{(3,4)}(0)
s_{\eta_8,\eta_9}^{(8,9)}(0)     s_{\eta_8,\eta_{10}}^{(8,10)}(0)    s_{\eta_9,\eta_{10}}^{(9,10)}(0)  }
{s_{-\eta_0,\eta_8}^{(0,8)}(L_3)    s_{-\eta_0,\eta_9}^{(0,9)}(L_3)   s_{-\eta_0,\eta_{10}}^{(0,10)}(L_3)
s_{\eta_3,\eta_8}^{(3,8)}(L_3)    s_{\eta_3,\eta_9}^{(3,9)}(L_3)   s_{\eta_3,\eta_{10}}^{(3,10)}(L_3)
s_{\eta_4,\eta_8}^{(4,8)}(L_3)    s_{\eta_4,\eta_9}^{(4,9)}(L_3)   s_{\eta_4,\eta_{10}}^{(4,10)}(L_3)}
\right]^{1/3}
\\
\times
\left[
\frac{s_{\eta_0,\eta_1}^{(i,1)}(0)     s_{\eta_0,\eta_2}^{(i,2)}(0)    s_{\eta_1,\eta_2}^{(1,2)}(0)
s_{\eta_8,\eta_9}^{(8,9)}(0)     s_{\eta_8,\eta_{10}}^{(8,10)}(0)    s_{\eta_9,\eta_{10}}^{(9,10)}(0)  }
{s_{\eta_0,\eta_8}^{(i,8)}(L_4)    s_{\eta_0,\eta_9}^{(i,9)}(L_4)   s_{\eta_0,\eta_{10}}^{(i,10)}(L_4)
s_{\eta_1,\eta_8}^{(1,8)}(L_4)    s_{\eta_1,\eta_9}^{(1,9)}(L_4)   s_{\eta_1,\eta_{10}}^{(1,10)}(L_4)
s_{\eta_2,\eta_8}^{(2,8)}(L_4)    s_{\eta_2,\eta_9}^{(2,9)}(L_4)   s_{\eta_2,\eta_{10}}^{(2,10)}(L_4)
}
\right]^{1/3},
\end{multline}
\end{widetext}
where we have introduced $t_i=0$.
The expression in Eq.~(\ref{twelve}) contains all possible correlations between the quasiparticles
traveling along the edges of the interferometer. The diagram, Fig.~\ref{Fig:grouping}(a), depicts one of these contributions. The integrand in Eq.~(\ref{twelve}) is the product of four factors that correspond to the Keldysh traces along the four edges
with lengths $L_1,\ldots,L_4$.
Importantly, the same time arguments appear
in the different edge-blocks in this equation, since these blocks are connected by instantaneous
tunneling events.

%%%%%%%%%%%%%%%%%%%%%%%%%%%%%%%%%%%%%%%%%%%%%%%%%%%%%%%%%%%%%
%%%%%%%%%%%%%%%%%%%%%%%%%%%%%%%%%%%%%%%%%%%%%%%%%%%%%%%%%%%%

In each of the four factors the numerator contains the functions $s(x)$
taken at zero arguments $x=0$. These functions dress the tunneling amplitudes [note that each such
function appears twice in Eq.~(\ref{twelve}) since each tunneling contact connects two edges].
All the denominators contain the corresponding lengths of the edges between the tunneling contacts.
Semiclassically, these denominators describe the propagation of quasiparticles along the edges as well as correlations between
the tunneling events. Without such correlations, there would be only $3\times 4 = 12$ functions $s$ in the denominators that form
three closed loops consisting of four edges.
All other $24$ functions $s$ in the denominators describe the correlations between these loops.

We now show that in the high temperature limit
\begin{equation}
T\gg \hbar v/L_i,
\label{highT-condition}
\end{equation}
it is, however, possible to disentangle the quasiparticle loops.
The condition \eqref{highT-condition} implies that the thermal length defined by $\beta \hbar v$ is shorter than the length of each of the interferometer's arms. In this limit, we group the integrals into three separate blocks, and
make use of the exponential behavior of functions $s$ away from the ``light cone'':
\begin{align}
&\sinh \left\{
\frac{\pi}{\hbar \beta v}\,   [x-vt -i \chi l_c] \right\}
\simeq \nonumber\\
&\exp\left(
\frac{\pi}{\hbar \beta v}\,   | x-vt | \right), \qquad |x-vt|\gg \hbar \beta v.
\label{sinh-exp}
\end{align}
There are several possible choices of loops in Eq.~(\ref{twelve}).
In the following, we will show only one representative possibility.

The general procedure is as follows: we first notice that in each edge block in Eq.~(\ref{twelve}),
the denominator consists of three groups of three functions $s^{(m,n)}(L_1)$ with the same first-time argument $m$.
In particular, the first block in Eq.~(\ref{twelve}) contains three $s$ functions with $m=0$, three with $m=3$, and three with
$m=4$. Similarly, these functions can be grouped into triples with regard to the second-time argument: $n=5,6,7$
(this reflects the fact that three quasiparticles tunnel through each junction in the $12^{\rm th}$-order process).
Take one function $s^{(m,n)}(L_1)$ characterized by the
times $t_m$ and $t_n$ in the denominator of the first edge block.
In the second and third blocks, we find the functions $s^{(m',n)}(L_2)$ and $s^{(m,n')}(L_3)$, correspondingly, that share one time
argument with our choice in the first edge. Next, we find the function $s^{(m',n')}(L_4)$ in the fourth edge block that matches the
remaining time arguments $t_{m'}$ and $t_{n'}$ of the functions in edges 2 and 3. These four functions are connected by common
tunneling events and form a closed quasiparticle loop.
Repeating this procedure two more times for the not yet used time arguments (clearly, there are $3\times 2\times 1=6$
distinct possibilities of doing that), we single out three closed quasiparticle loops.
This allows us to rewrite Eq.~(\ref{twelve}) in a way highlighting the semiclassical dynamics of quasiparticles.
In particular, the grouping with
\begin{align}
&m_1=0,\ n_1=5,\ m_1^\prime=i,\ n_1^\prime=8; \nonumber\\
& m_2=3,\ n_2=6,\ m_2^\prime=1,\ n_2^\prime=9; \\
& m_3=4,\ n_3=7,\ m_3^\prime=2,\ n_3^\prime=10\,,\nonumber
\end{align}
yields the following representation of Eq.~(\ref{twelve}):
\begin{widetext}
\begin{align}\label{twelve-grouping}
S_{23}^{\mathcal{CA}}& \propto \sum_{\eta_0,\eta_1,...,\eta_{10}=\pm 1} \eta_1\dots \eta_{10} \int _{-\infty}^{+\infty} dt_0 dt_1\dots dt_{10}
e^{-i \nu e V (t_0+t_1+t_2+t_3+t_4-t_5-t_6-t_7-t_8-t_9-t_{10})}
\nonumber
\\
&\times
\left\{
\left[\frac{1}{s_{-\eta_0,\eta_5}^{(0,5)} s_{\eta_0,\eta_5}^{(i,5)} s_{-\eta_0,\eta_8}^{(0,8)} s_{\eta_0,\eta_8}^{(i,8)} }
\right]_{1,2,3,4}
\left[
\frac{1}{s_{\eta_3,\eta_6}^{(3,6)} s_{\eta_1,\eta_6}^{(1,6)} s_{\eta_3,\eta_9}^{(3,9)} s_{\eta_1,\eta_9}^{(1,9)} }
\right]_{1,2,3,4}
\left[
\frac{1}{ s_{\eta_4,\eta_7}^{(4,7)} s_{\eta_2,\eta_7}^{(2,7)} s_{\eta_4,\eta_{10}}^{(4,10)}   s_{\eta_2,\eta_{10}}^{(2,10)}}
\right]_{1,2,3,4}
\right\}^{1/3}\nonumber
\\
&\times
\left\{
\left[
\frac{s_{-\eta_0,\eta_3}^{(0,3)}     s_{-\eta_0,\eta_4}^{(0,4)}    s_{\eta_3,\eta_4}^{(3,4)} s_{\eta_5,\eta_6}^{(5,6)}     s_{\eta_5,\eta_7}^{(5,7)}    s_{\eta_6,\eta_7}^{(6,7)} }
     {    s_{-\eta_0,\eta_6}^{(0,6)}   s_{-\eta_0,\eta_7}^{(0,7)} s_{\eta_3,\eta_5}^{(3,5)}      s_{\eta_3,\eta_7}^{(3,7)} s_{\eta_4,\eta_5}^{(4,5)}    s_{\eta_4,\eta_6}^{(4,6)}   }\right]_{1}
\right.
%\\
%&\times
\left[\frac{
s_{\eta_0,\eta_1}^{(i,1)}    s_{\eta_0,\eta_2}^{(i,2)}    s_{\eta_1,\eta_2}^{(1,2)}
      s_{\eta_5,\eta_6}^{(5,6)}    s_{\eta_5,\eta_7}^{(5,7)}    s_{\eta_6,\eta_7}^{(6,7)}
      }
{
s_{\eta_0,\eta_6}^{(i,6)}   s_{\eta_0,\eta_7}^{(i,7)}
s_{\eta_1,\eta_5}^{(1,5)}     s_{\eta_1,\eta_7}^{(1,7)}
s_{\eta_2,\eta_5}^{(2,5)}    s_{\eta_2,\eta_6}^{(2,6)}
}
\right]_{2}
\nonumber
\\
&\times
\left[
\frac{
s_{-\eta_0,\eta_3}^{(0,3)}     s_{-\eta_0,\eta_4}^{(0,4)}    s_{\eta_3,\eta_4}^{(3,4)}
s_{\eta_8,\eta_9}^{(8,9)}     s_{\eta_8,\eta_{10}}^{(8,10)}    s_{\eta_9,\eta_{10}}^{(9,10)}
}
{
s_{-\eta_0,\eta_9}^{(0,9)}  s_{-\eta_0,\eta_{10}}^{(0,10)}
s_{\eta_3,\eta_8}^{(3,8)}       s_{\eta_3,\eta_{10}}^{(3,10)}
s_{\eta_4,\eta_8}^{(4,8)}    s_{\eta_4,\eta_9}^{(4,9)}
}
\right]_{3}
\left.
\left[
\frac{s_{\eta_0,\eta_1}^{(i,1)}     s_{\eta_0,\eta_2}^{(i,2)}    s_{\eta_1,\eta_2}^{(1,2)}
s_{\eta_8,\eta_9}^{(8,9)}     s_{\eta_8,\eta_{10}}^{(8,10)}    s_{\eta_9,\eta_{10}}^{(9,10)}  }
{    s_{\eta_0,\eta_9}^{(i,9)}   s_{\eta_0,\eta_{10}}^{(i,10)}
s_{\eta_1,\eta_8}^{(1,8)}       s_{\eta_1,\eta_{10}}^{(1,10)}
s_{\eta_2,\eta_8}^{(2,8)}    s_{\eta_2,\eta_9}^{(2,9)}
}
\right]_{4}
\right\}^{1/3}\,,
\end{align}

Here the three fractions in the first curly brackets correspond to the chosen quasiparticle loops.
The remaining terms in the second curly brackets describe the correlations between the loops.
For the sake of compactness, we have suppressed the arguments of $s$ functions here.
In each quasiparticle loop, the arguments of the four subsequent $s$-functions are $L_1$, $L_2$, $L_3$, and $L_4$, respectively,
as indicated by the corresponding subscripts. For example,
\begin{equation}
\left[\frac{1}{s_{-\eta_0,\eta_5}^{(0,5)} s_{\eta_0,\eta_5}^{(i,5)} s_{-\eta_0,\eta_8}^{(0,8)} s_{\eta_0,\eta_8}^{(i,8)} }
\right]_{1,2,3,4} \equiv \frac{1}{s_{-\eta_0,\eta_5}^{(0,5)}(L_1) s_{\eta_0,\eta_5}^{(i,5)}(L_2)
s_{-\eta_0,\eta_8}^{(0,8)}(L_3) s_{\eta_0,\eta_8}^{(i,8)} (L_4)}\,.
\end{equation}
The arguments of $s$ functions in the correlation term are all zero in the numerator and are $L_i$ in the denominator for the $i^{\rm th}$ edge,
as indicated by the subscript:
\begin{equation}
\left[
\frac{s_{\eta_0,\eta_1}^{i,1}     s_{\eta_0,\eta_2}^{(i,2)}    s_{\eta_1,\eta_2}^{1,2}
s_{\eta_8,\eta_9}^{(8,9)}     s_{\eta_8,\eta_{10}}^{(8,10)}    s_{\eta_9,\eta_{10}}^{(9,10)}  }
{    s_{\eta_0,\eta_9}^{(i,9)}   s_{\eta_0,\eta_{10}}^{(i,10)}
s_{\eta_1,\eta_8}^{(1,8)}       s_{\eta_1,\eta_{10}}^{(1,10)}
s_{\eta_2,\eta_8}^{(2,8)}    s_{\eta_2,\eta_9}^{(2,9)}
}
\right]_{4}\equiv
\frac{s_{\eta_0,\eta_1}^{i,1}(0)     s_{\eta_0,\eta_2}^{(i,2)}(0)    s_{\eta_1,\eta_2}^{1,2}(0)
s_{\eta_8,\eta_9}^{(8,9)}(0)     s_{\eta_8,\eta_{10}}^{(8,10)}(0)    s_{\eta_9,\eta_{10}}^{(9,10)}(0)  }
{    s_{\eta_0,\eta_9}^{(i,9)}(L_4)   s_{\eta_0,\eta_{10}}^{(i,10)}(L_4)
s_{\eta_1,\eta_8}^{(1,8)}(L_4)       s_{\eta_1,\eta_{10}}^{(1,10)}(L_4)
s_{\eta_2,\eta_8}^{(2,8)}(L_4)    s_{\eta_2,\eta_9}^{(2,9)}(L_4)
}\,.
\end{equation}
\end{widetext}

Let us consider the terms in the first quasiparticle loop (the first factor in the square brackets with $m=0$, $n=5$, $m'=i$, and $n'=8$).
We assume $L_1+L_4=L_2+L_3$, otherwise the result of the calculation
will be exponentially small in $\delta L / L_T$, where $L_T=\hbar \beta v$ is the thermal length and $\delta L$ is the length mismatch between the external and internal arms of the interferometer.
Analyzing this first grouping, we notice that in the high-temperature limit, i.e., $L_T \ll L_1,\ldots,L_4$, the main contribution to
the integrals comes from time arguments such that:
\begin{align} \label{semiclassical}
& v(t_5-t_0) \simeq L_1, \quad v( t_5-t_i) \simeq L_2,\nonumber\\
& v(t_8-t_0)\simeq L_3, \quad v(t_8 -t_i) \simeq L_4\,.
 \end{align}
Clearly, the above conditions indeed require $L_1+L_4=L_2+L_3$.
Notice that here time $t_i=0$ is a fixed (external) point. In the other two groupings, there is no external time and one can always satisfy a condition similar to
Eq.~\eqref{semiclassical} having one free parameter. For the other two groupings with $t_1,t_3,t_6,t_9$  and $t_2,t_4,t_7,t_{10}$, the main contribution to the integrals comes from times such that
\begin{align} \label{semiclassical2}
& v(t_6-t_3)\simeq L_1, \quad v(t_6 -t_1) \simeq L_2,\nonumber\\
& v(t_9-t_3) \simeq L_3, \quad v(t_9-t_1) \simeq L_4\,,
 \end{align}
 and
\begin{align} \label{semiclassical3}
& v(t_7-t_4) \simeq L_1, \quad v(t_7-t_2) \simeq L_2, \nonumber\\
& v(t_{10} -t_4) \simeq L_3, \quad v(t_{10}-t_2)\simeq L_4\,.
 \end{align}
These conditions define quasiclassical trajectories of the quasiparticles.

We next show that the remaining correlation functions (those in the second curly brackets) in Eq.~(\ref{twelve-grouping}) simplify to unity.
We fix one time in each loop (for instance, $t_i=0$ in the first loop, $t_9$ in the second loop and $t_{10}$ in the third loop) and express the remaining times using Eqs.~(\ref{semiclassical}), (\ref{semiclassical2}), and (\ref{semiclassical3}).
 As an example, we consider the function $s_{-\eta,\eta_3}^{(0,3)}(0)$ from the numerator in edge block 1.
 Using the above quasiclassical conditions, we substitute $t_0 \simeq (L_2-L_1)/v$ and $t_3 \simeq t_9-L_3/v$.
 In Fig.~\ref{Fig:grouping}(b), we show one possible arrangement of the times.
Assuming that the time distance between the groups is much larger than $L_T/v$
and with reference to the arrangement of times illustrated in Fig.~\ref{Fig:grouping}(b), and using Eq.~(\ref{sinh-exp}), we have:
\begin{align}
\label{s03}
& s_{-\eta,\eta_3}^{(0,3)}(0) \simeq \exp\left({\frac{\pi}{\hbar \beta}|t_0-t_3|}\right)\\
& = \exp\left({\frac{\pi}{\hbar \beta v}|L_2-L_1-v t_9 + L_3|}\right)=\exp\left({\frac{\pi}{\hbar \beta v}|L_4-v t_9|}\right)\,.\nonumber
\end{align}
On the other hand, in the denominator of the edge block 4 in Eq.~(\ref{twelve-grouping}),
we find the function
\begin{equation}
s_{\eta_0,\eta_9}^{(i,9)}(L_4)\simeq \exp\left({\frac{\pi}{\hbar \beta v}|L_4-v t_9|}\right)\,,
\label{s09}
\end{equation}
which cancels the contribution of $s_{-\eta,\eta_3}^{(0,3)}(0)$.
Using the same exponential asymptotic form for all the remaining $s$ functions connecting times belonging to
different groups, one can see that the correlation factor does indeed simplify to unity.
The thermal fluctuations $\delta t_i \alt \hbar/T$ around the quasiclassical trajectories
yield exponentially small corrections from the correlation part. These fluctuations are,
however, important within each quasiparticle loop: the integration over $\delta t_i$
gives rise to a power-law renormalization of the tunneling amplitudes.
We note in passing that the disentanglement of the quasiclassical trajectories and renormalization effects described above
is similar to the procedure used in Refs.~[\onlinecite{Gornyi:2007,Yashenkin:2008,Schneider:2012}]
for calculating the weak-localization correction in disordered Luttinger liquids.

Thus, we are able to represent the $12^{\rm th}$-order perturbative contribution to the current correlation
function as a sum of all possible triple products corresponding to the three quasiparticle loops:
\begin{widetext}
\begin{align}\label{three loops}
S_{23}^{\mathcal{CA}}& \propto \sum_{\eta_0,\eta_1,...,\eta_{10}=\pm 1} \eta_1\dots \eta_{10}
\\
&\times \Big\{
[(0,5),(i,5),(0,8),(i,8)][(3,6),(1,6),(3,9),(1,9)][(4,7),(2,7),(4,10),(2,10)]
\nonumber\\
&+ [(0,5),(i,5),(0,8),(i,8)][(3,7),(1,7),(3,10),(1,10)][(4,6),(2,6),(4,9),(2,9)]
\nonumber\\
&+
[(0,6),(i,6),(0,9),(i,9)][(3,5),(1,5),(3,8),(1,8)][(4,7),(2,7),(4,10),(2,10)]
\nonumber\\
&+
[(0,6),(i,6),(0,9),(i,9)][(3,7),(1,7),(3,10),(1,10)][(4,5),(2,5),(4,8),(2,8)]
\nonumber\\
&+
[(0,7),(i,7),(0,10),(i,10)][(3,5),(1,5),(3,8),(1,8)][(4,6),(2,6),(4,9),(2,9)]
\nonumber\\
&+
[(0,7),(i,7),(0,10),(i,10)][(3,6),(1,6),(3,9),(1,9)][(4,5),(2,5),(4,8),(2,8)]
\Big\}\,.\nonumber
\end{align}
where we denote the products of quasiparticle loops symbolically as follows:
\begin{align}
\{abc\}&=
\underbrace{[(0,5),(i,5),(0,8),(i,8)]}_\text{loop a}\underbrace{[(3,6),(1,6),(3,9),(1,9)]}_\text{loop b}
\underbrace{[(4,7),(2,7),(4,10),(2,10)]}_\text{loop c}
\propto \Gamma_\mathcal{A}^3 \Gamma_\mathcal{B}^{*3}\Gamma_\mathcal{C}^3 \Gamma_\mathcal{D}^{*3}
\nonumber
\\
&\quad \times
\int _{-\infty}^{+\infty} dt_0 dt_5 dt_8 e^{-i \nu e V (t_i+t_0-t_5-t_8)}
\left[
\frac{1}{s_{-\eta_0,\eta_5}^{(0,5)}(L_1) s_{\eta_0,\eta_5}^{(i,5)}(L_2) s_{-\eta_0,\eta_8}^{(0,8)}(L_3) s_{\eta_0,\eta_8}^{(i,8)} (L_4)}
\right]^{1/3}
\nonumber\\
&\quad \times
\int _{-\infty}^{+\infty} dt_1 dt_3 dt_6 dt_9 e^{-i \nu e V (t_1+t_3-t_6-t_9)}
\left[
\frac{1}{s_{\eta_3,\eta_6}^{(3,6)}(L_1) s_{\eta_1,\eta_6}^{(1,6)}(L_2) s_{\eta_3,\eta_9}^{(3,9)}(L_3) s_{\eta_1,\eta_9}^{(1,9)}(L_4) }
\right]^{1/3}
\nonumber
\\
&\quad \times
\int _{-\infty}^{+\infty} dt_2 dt_4 dt_7 dt_{10} e^{-i \nu e V (t_2+t_4-t_7-t_10)}
\left[
\frac{1}{ s_{\eta_4,\eta_7}^{(4,7)}(L_1) s_{\eta_2,\eta_7}^{(2,7)}(L_2) s_{\eta_4,\eta_{10}}^{(4,10)}(L_3)   s_{\eta_2,\eta_{10}}^{(2,10)}(L_4)}
\right]^{1/3}\,.
\label{loop-integrals}
\end{align}
\end{widetext}
Comparing the structure of integrals in Eq.~(\ref{loop-integrals}) within each loop with Eq.~\eqref{rate},
we observe that each loop here is equivalent to a two-particle rate $W^{(2)}$.

It is convenient to introduce the center-of-mass times characterizing each loop in Eq.~(\ref{loop-integrals}):
\begin{align}
& t_a=(t_i+t_0+t_5+t_8)/4, \qquad t_b=(t_1+t_3+t_6+t_9)/4,\nonumber\\
&t_c=(t_2+t_4+t_7+t_{10})/4\,,
\end{align}
as well as their differences:
\begin{equation}
t_{ab}=t_b-t_a, \qquad t_{ac}=t_c-t_a, \qquad t_{bc}=t_c-t_b.
\end{equation}
Within each loop, one can then introduce the relative time variables.
 For instance, in the second integral (loop $b$) we change the variables to
 \begin{equation}
 \tau_1^{(b)}=t_3-t_9, \qquad \tau_2^{(b)}=t_9-t_1, \qquad
 \tau_3^{(b)}=t_1-t_6.
 \end{equation}
 With such variable change, the contribution (\ref{loop-integrals}) to the current correlation function takes the form:
 \begin{equation}
 \{abc\}=\int_{-\infty}^\infty dt_{ab} dt_{bc} W_a^{(2)} W_b^{(2)} W_c^{(2)},
 \end{equation}
 with $W_{a,b,c}^{(2)}$ being independent of the time distances $t_{ab}$ and $t_{bc}$ between the blocks.
 We see that the integration within each choice of the loops yields a seemingly divergent contribution
 for any given configuration of $\eta_{0,\ldots,10}$.
 A similar divergence was encountered in Ref.~[\onlinecite{feldman}] in the problem of Mach-Zehnder interference
 of anyons.

 This divergence can be cured by incorporating a finite lifetime into the propagators connecting different loops.
 Indeed, if we ``dress'' the propagators associated with the integration over $t_{ab}$ by single-particle processes
 (and similarly for $t_{bc}$),  we end up with $\int  dt_{ab} \exp(-W^{(1})  t_{ab})$ and $\int dt_{bc} \exp(-W^{(1)} t_{bc})$,
 where $W^{(1)}\propto \Gamma^2$ are the single-particle scattering rates.
 This renders the ``diverging integrals'' finite, and will introduce a factor $\Gamma^{-4}$ into the final expression
  for each term in the current correlation function:
  \begin{equation}
  S_{23} \propto \frac{\Gamma^{12}}{\Gamma_4} = \Gamma^8,
  \end{equation}
  in agreement with the kinetic approach.
  Physically, the inclusion of the exponentially decaying factors into the perturbative expression
  corresponds to the probability of not completing the interference loop for the involved quasiparticles,
  as well as the probability of changing the flux state due to the single-particle processes. The latter thus
  leads to the dephasing of the two-quasiparticle interference~\cite{footnote:zero_bias}.

 It is worth recalling at this point that in the above calculation, we have completely ignored the Klein factors.
 However, as was emphasized in Ref.~[\onlinecite{feldman}],
 the summation over all possible configurations of $\eta_n$ on the Keldysh contour, without taking into account the Klein
 factors, leads to a cancellation of the divergencies: the total contribution of all terms is then zero
 instead of infinity.
 As we have seen from the kinetic approach, the inclusion of the Klein factors in the perturbative treatment
 can be replaced by considering the fluctuating statistical flux. Note that in the perturbative approach,
 averaging over the dynamics of Klein's factors should provide both the quasiparticle decay discussed above and
 an additional $\eta$-dependent structure that prevent the full ``Keldysh cancellation''.
 This is efficiently done within the kinetic framework adopted in the main text.
 The consideration of this appendix serves as a justification of the master-equation approach.
 The perturbative treatment of the current correlation function demonstrates,  in particular,
 that the correlations between the two-particle tunneling processes can be neglected, which is
 crucial for the master equation.

 We have assumed above that the temperature is still sufficiently high, so that the thermal length is much
shorter than the lengths of the edges  [Eq.~(\ref{highT-condition})].
In fact, the same procedure of the disentanglement of the correlations
can be employed at zero $T$ (we further comment on the conditions below).
In this case, the correlation functions (\ref{sinh}) should be replaced by power-law functions
\begin{equation}
\tilde{s}_{\eta_m,\eta_n}^{(m,n)}(x)=
\frac{\pi}{\hbar \beta v}\,   [x-v(t_n-t_m) -i \chi_{\eta_m,\eta_n}^{(m,n)} l_c]\,,
\label{stilde}
\end{equation}
and hence do not behave exponentially away from the light cone.
Nevertheless, choosing the quasiparticle loops according to the prescription outlined
above Eq.~(\ref{twelve-grouping}), it is possible to demonstrate the cancellation
of the terms in the correlation part of the integrand in Eq.~(\ref{twelve-grouping})
for the values of times satisfying the quasiclassical conditions (\ref{semiclassical}),
(\ref{semiclassical2}), and (\ref{semiclassical3}). This is done similarly to Eqs. (\ref{s03})
and (\ref{s09}).
Under these conditions, assuming that the separation between the loops
is much larger than the typical size of the loops (set by voltage),
the correlation factor becomes unity.
This property of the zero-$T$ correlation block taken at the quasiclassical trajectory
was discussed, e.g., in Ref.~[\onlinecite{feldman}], where it was linked
to the properties of an equivalent Coulomb-gas model.

The conditions for the ``block decoupling'' at $T=0$ (i.e.~, the possibility to consider the cross-correlation signal as made of pair-wise anyonic correlations) requires that the width (in time) of such a ``block'' (an anyonic pair interference) is smaller than the distance between consecutive blocks, $\delta t_{\rm block}$ ($t_{ab}$ and $t_{bc}$). The size of a block is proportional to $\hbar/W^{(1)}$ ($W^{(1)}$ is the single-particle rate). We could now require that
\begin{align}
\frac{\hbar}{W^{(1)}} \gg \delta t_{\rm block}\,,
\label{blockSize}
\end{align}
for the decoupling to hold. We, note, though that barring additional manipulations $\delta t_{\rm block} \sim \hbar/W^{(1)}$, and the inequality Eq.~\eqref{blockSize} is not satisfied. Then, the three loops become correlated,
which in terms of the kinetic approach necessitates accounting for higher-order scattering processes
(involving three and more particles).

%%%%%%%%%%%%%%%%%%%%%%%%%%%%%%%%%%%%%%%%%%%%%%%%%%%%%%%%%%%%%
\section{Evaluation of AB-dependent tunneling rates where the quasiparticles reach $D_2$ and $D_3$}
\label{app_rate2}
In this Appendix we evaluate explicitly Eq.~(\ref{2-phi-rate-bis}), the magnetic flux-dependent component of the two quasiparticle transferring rate, Eq.~(\ref{2phi-2qp-rate}). The expression to compute reads as [cf.~Eq.~\eqref{eq_rates}]
\begin{multline}\label{rate2}
\mbox{I+II+III+IV}=\\
\frac{\Gamma^*_\mathcal{A} \Gamma_\mathcal{B} \Gamma^*_\mathcal{C} \Gamma_\mathcal{D}}{\hbar^4 l_c^4} e^{-2\pi i \nu(\Phi_{\rm AB}+j \Phi_0)/\Phi_0 } e^{-i e \nu V (L_4+L_1)/{v \hbar}} \\ \times
\left(\frac{\pi l_c}{\hbar \beta v}\right)^{4\nu}
\int_{-\infty}^{+\infty}dt\int_{-\infty}^{\infty}dt_1\int_{-\infty}^{\infty} dt_2 e^{-(2 i\nu  e V t)/\hbar} \\ \times
\frac{1}{\sin [ \frac{i\pi}{\hbar \beta }(-tv+\frac{t_1 v}{2}-\frac{t_2 v}{2}-L_4) + l_c ]^{\nu}} \\ \times
 \frac{1}{\sin [ \frac{i\pi}{\hbar \beta }(-tv+\frac{t_1 v}{2}+\frac{t_2 v}{2}-L_3)+ l_c ]^{\nu}} \\ \times
  \frac{1}{\sin [ \frac{i\pi}{\hbar \beta }(-tv-\frac{t_1 v}{2}-\frac{t_2 v}{2}-L_2)+ l_c ]^{\nu}}  \\ \times
   \frac{1}{ \sin [ \frac{i\pi}{\hbar \beta }(-tv-\frac{t_1 v}{2}+\frac{t_2 v}{2}-L_1)+ l_c ]^{\nu}} \, .
\end{multline}
It is convenient to make the integrals dimensionless and to  introduce the Fourier transform of the correlation function
\begin{equation}
g(\omega)=\int_{-\infty}^{+\infty} dt \exp(-i \omega t) \frac{1}{{\sin(it+ \delta)}^{\nu}}\,,
\end{equation}
which, for $\delta\rightarrow0^+$, reads as
\begin{equation}
g(\epsilon)=\frac{1}{2^{1-\nu}\Gamma(\nu)} \Gamma\left(\frac{1}{2}(\nu-i\omega)\right) \Gamma\left(\frac{1}{2}(\nu+i\omega)\right)
\exp(-\frac{\pi \omega}{2})\,.
\end{equation}
We can now represent the correlation functions using the above Fourier transform. The integrations over $t,t_1,t_2$ give delta functions. After the integration over energies, we obtain
\begin{align}
&\mbox{I+II+III+IV}=\nonumber\\
&\frac{\Gamma^*_\mathcal{A} \Gamma_\mathcal{B} \Gamma^*_\mathcal{C} \Gamma_\mathcal{D}}{\hbar^4 v^4}\frac{\pi}{\hbar\beta} e^{-2\pi i \nu(\Phi_{\rm AB}+j \Phi_0)/\Phi_0 }e^{-i 2\nu\alpha \Delta \tilde{L}}\\
&\times
\left(\frac{\hbar \beta v}{2\pi l_c}\right)^{4-4\nu} \frac{1}{[\Gamma(\nu)]^4}
e^{2\pi \nu\alpha} \nonumber\\
&\times\int_{-\infty}^{\infty}\frac{d\epsilon}{2\pi}e^{-i \epsilon \Delta \tilde{L}}\Big[
\Gamma\left(\frac{1}{2}(\nu-i\epsilon)\right) \Gamma\left(\frac{1}{2}(\nu+i\epsilon)\right)\nonumber\\
&\times \Gamma\left(\frac{1}{2}(\nu-i\epsilon)-i\nu\alpha\right) \Gamma\left(\frac{1}{2}(\nu+i\epsilon)+i\nu\alpha\right)
\Big]^2\,.\nonumber
\end{align}
Where $\alpha=e V \beta/(2\pi) $ and $\Delta \tilde{L}=\pi(L_1+L_4-L_2-L_3)/(\hbar \beta v) $.
We compute the following integral,
\begin{align}\label{integral-reformulated}
&\mathcal{I}=\int_{-\infty}^{\infty}\frac{d\epsilon}{2\pi}e^{-i\epsilon \Delta \tilde{L}}
\Big[
\Gamma\left(\frac{1}{2}(\nu-i\epsilon)\right)\Gamma\left(\frac{1}{2}(\nu+i\epsilon)\right)\nonumber\\
&\Gamma\left(\frac{1}{2}(\nu-i\epsilon)-i\nu\alpha\right) \Gamma\left(\frac{1}{2}(\nu+i\epsilon)+i\nu\alpha\right)
\Big]^2 \,,
\end{align}
which can be performed by the residue method.
Let us consider the poles in the lower half complex plane. The calculation of the residues may be quite cumbersome, as each pole is of second order; it is convenient to first manipulate the above  expression.
One can use the following property of the  Gamma function:
\[
\Gamma(z)\Gamma(1-z)=\frac{\pi}{\sin(\pi z)}\,,
\]
and rewrite the integral in Eq.~(\ref{integral-reformulated}) as
\begin{multline}
\mathcal{I}=\pi^2 \frac{\partial}{\partial a_1} \frac{\partial}{\partial a_2}
\int_{-\infty}^{\infty}\frac{d\epsilon}{2\pi}e^{-i\epsilon \Delta \tilde{L}}\\
 \left[ \frac{\Gamma\left(\frac{1}{2}(\nu+i\epsilon)\right) \Gamma\left(\frac{1}{2}(\nu+i\epsilon)+i\nu\alpha\right) }
{\Gamma\left(1-\frac{1}{2}(\nu-i\epsilon)\right) \Gamma\left(1-\frac{1}{2}(\nu-i\epsilon)-i\nu\alpha\right) }
\right]^2
\\ \times  \left.
\frac{1}{\tan[\pi(a_1-i\frac{\epsilon}{2})]\tan[\pi(a_2-i\frac{\epsilon}{2}+i\nu\alpha)]}\right|_{a_1,a_2=\nu/2} \, .
\end{multline}
Notice that now the poles in the lower half of the complex plane are of first order. This renders the
calculation of the residues much less involved. We obtain the following converging sum
\begin{widetext}
\begin{multline}
\label{fullIntVal}
\mathcal{I}=-\frac{4 \pi^2 e^{-\Delta \tilde{L}\nu}e^{i \nu\alpha\Delta \tilde{L}}}{ \sin^2\left(\pi \nu\alpha\right)}\sum_{n=0}^{
\infty}\Bigg[e^{i \nu\alpha\Delta \tilde{L}}\frac{ e^{-2 \Delta \tilde{L} n}   \Gamma^2\left[n + \nu\right]  \Gamma^2\left[1 + n + i \nu\alpha\right] \Gamma^2\left[n - i \nu\alpha + \nu\right] }{\Gamma^2\left[1 + n\right] \Gamma^2\left[1 + n - i \nu\alpha\right] \Gamma^2\left[1 + n + i \nu\alpha\right]}\\
\times \left(\Delta \tilde{L} + i \pi \coth\left[\pi\nu\alpha\right] + \mathcal{H}\left[n\right] +\mathcal{H}\left[n - i \nu\alpha\right] - \mathcal{H}\left[n - 1 + \nu \right] - \mathcal{H}\left[n - 1 + \nu - i \nu\alpha\right]\right)+c.c. \Bigg]\, ,
\end{multline}
\end{widetext}
where $\mathcal{H}$ is the harmonic number function.

Using the result of Eq.~\eqref{fullIntVal}, the obtained 2-qp rate is
\begin{align}
\label{trueRes}
W_{j,j+2}^{(2,\mathcal{A}\mathcal{B}\mathcal{C}\mathcal{D},\Phi_{\text{tot}}(j))} &=\left(\mbox{I+II+III+IV}\right)+c.c. \\
&=\frac{ \left| \Gamma_\mathcal{A} \Gamma_\mathcal{B} \Gamma_\mathcal{C} \Gamma_\mathcal{D} \right|}{\hbar^4 v^4}\Omega(V,T,\nu,\Delta L)\nonumber\\
&\times\cos \left[ 2\pi\nu\frac{\Phi_\Gamma+\Phi_{ab}+j \,\Phi_0}{\Phi_0} + \nu\alpha\Delta \tilde{L}\right]\,,\nonumber
\end{align}
where we defined $\Omega(V,T,\nu,\Delta L)\equiv\frac{eV}{\hbar} \left(\frac{\hbar \beta v}{2\pi l_c}\right)^{4-4\nu} \frac{e^{2\pi \nu\alpha}}{\alpha} \mathcal{I}/\Gamma^4[\nu]$.

Assuming that $(L_1+L_4-L_2-L_3) \ll \hbar v/(\nu e V)$ such that we can take $\Delta \tilde{L}=0$, we find Eq.~\eqref{2-phi-rate-bis}. In particular, for $\nu=1/3$ the rate has the form
\begin{align}
&W_{j,j+2}^{(2,\mathcal{A}\mathcal{B}\mathcal{C}\mathcal{D},\Phi_{\text{tot}}(j))} =2\frac{ \left| \Gamma_\mathcal{A} \Gamma_\mathcal{B} \Gamma_\mathcal{C} \Gamma_\mathcal{D} \right|}{\hbar^4 v^4}\frac{eV}{3\hbar} \left(\frac{\hbar \beta v}{2\pi l_c}\right)^{8/3} \\
&\times  \pi 2^{3}\sqrt{3} (\alpha/3)^{-7/3}\Gamma[1/3]\cos \left[ \frac{2\pi}{3}\frac{\Phi_\Gamma+\Phi_{ab}+j \,\Phi_0}{\Phi_0} \right]\,.\nonumber
\end{align}
In the opposite limit of $\Delta \tilde{L} \gg \hbar v/(\nu e V)$  the rate is exponentially suppressed:
\begin{multline}
W_{j,j+2}^{(2,\mathcal{A}\mathcal{B}\mathcal{C}\mathcal{D},\Phi_{\text{tot}}(j))} =2\frac{ \left| \Gamma_\mathcal{A} \Gamma_\mathcal{B} \Gamma_\mathcal{C} \Gamma_\mathcal{D} \right|}{\hbar^4 v^4}\frac{\pi}{\hbar\beta} \left(\frac{\hbar \beta v}{2\pi l_c}\right)^{4-4\nu}\\
 \times e^{2\pi \nu\alpha} \frac{2^{4}(\nu\alpha)^{-2}}{\Gamma^2[\nu]}e^{-\Delta \tilde{L}\nu}\\
\times\Bigg[e^{i \nu\alpha\Delta \tilde{L}}\Gamma^2\left[1 + i \nu\alpha\right] \Gamma^2\left[\nu - i \nu\alpha\right]\Big(\mathcal{H}\left[\nu - 1 \right]  \\
 + \mathcal{H}\left[\nu - 1  - i \nu\alpha\right]-\Delta \tilde{L}- i \pi \coth\left[\pi\nu\alpha\right] \\
-\mathcal{H}\left[- i \nu\alpha\right]\Big)+c.c. \Bigg]\cos \left[ 2\pi\nu\frac{\Phi_\Gamma+\Phi_{ab}+j \,\Phi_0}{\Phi_0}+ \nu\alpha\Delta \tilde{L}\right]\, .
\end{multline}

%%%%%%%%%%%%%%%%%%%%%%%%%%%%%%%%%%%%%%%%%%%%%%%%%%%%%%%%%%%%%%
\section{Evalutation of AB-dependent tunneling rates where the quasiparticles reach, for example, $D_2$ and $D_4$}
\label{app_rate1}
In this appendix, we consider processes of type $(1,\mathcal{A}\mathcal{B}\mathcal{C}\mathcal{D},\Phi_{\text{tot}}(j))$, i.e., fourth-order  processes that change the statistical flux by one and are sensitive to the flux of the magnetic field (cf.~Table \ref{table1}).
In Sec.~\ref{kinetic}, we have shown that, because of a gauge invariance argument, currents are magnetic flux insensitive, hence a relation between rates of type $(2,\mathcal{A}\mathcal{B}\mathcal{C}\mathcal{D},\Phi_{\text{tot}}(j))$ and $(1,\mathcal{A}\mathcal{B}\mathcal{C}\mathcal{D},\Phi_{\text{tot}}(j))$ must hold. Here we nevertheless show an explicit expression of rates of type $(1,\mathcal{A}\mathcal{B}\mathcal{C}\mathcal{D},\Phi_{\text{tot}}(j))$.

When considering fourth-order processes, in addition to the rate of transferring two quasiparticles from the external edges to the internal ones, we also need to consider the case of fourth order processes where a single quasiparticle is transferred inside the interferometer. Such a process would not contribute to the current-current correlation signal but it would make sure that no AB sensitive terms are contributing to  the current. We address here only the AB sensitive part of this rate. Starting with Eq.~(\ref{generalfermigr}), we write all the possible contributions corresponding to a single-quasiparticle transfer.
Again looking at terms proportional to $\Gamma_\mathcal{A}^* \Gamma_\mathcal{B} \Gamma_\mathcal{C}^*\Gamma_\mathcal{D}$,  we have two sets of contributions:
The first  set is given by
\begin{multline}\label{ABlike}
\frac{2 \pi}{\hbar} \sum_{if} w_i  \langle \hat{\psi}_i | \tb |\hat{\psi}_f \rangle
 \langle \hat{\psi}_f | \td \frac{1}{E_i-H_0+i 0^+} \tcd  \times \\
  \frac{1}{E_i-H_0+i 0^+}  \tad |\hat{\psi}_i \rangle
 \delta(E_i-E_f)\,,
\end{multline}
plus all permutations of $\td, \tad$ and $\tcd$. Notice that in this case the many-body states $|\hat{\psi}_f\rangle$ differ from $|\hat{\psi}_i$ for the transfer of \emph{one} quasiparticle from the external edges to the inners ones.
The second set reads,
\begin{align}\label{ABlike2}
&\frac{2 \pi}{\hbar} \sum_{if} w_i  \langle \hat{\psi}_i | \tb \frac{1}{E_i-H_0-i 0^+} \td\\
&\times \frac{1}{E_i-H_0-i 0^+} \tcd
|\hat{\psi}_f  \rangle
\langle \hat{\psi}_f | \tad | \hat{\psi}_i \rangle \delta(E_i-E_f)\,,\nonumber
\end{align}
plus all permutations of $\tb,\td$ and $\tcd$.

Let us introduce four Green's functions for each field $\phi_k$:
\begin{align}
G_k^t(t,x)=&\langle T  e^{-i \sqrt{\nu}\phi_k(x,t)} e^{i \sqrt{\nu}\phi_k(0,0)}     \rangle\,,\\
G_k^{\bar{t}}(t,x)=&\langle \bar{T}  e^{-i \sqrt{\nu}\phi_k(x,t)} e^{i \sqrt{\nu}\phi_k(0,0)}     \rangle\,,\\
G_k^{>}(t,x)=&\langle  e^{-i \sqrt{\nu}\phi_k(x,t)} e^{i \sqrt{\nu}\phi_k(0,0)}     \rangle\,,\\
G_k^{<}(t,x)=&\langle  e^{i \sqrt{\nu}\phi_k(0,0)} e^{-i \sqrt{\nu}\phi_k(x,t)}      \rangle\,,
\end{align}
where $T$ and $\bar{T}$ are respectively time ordering and time anti-ordering operators.

\begin{figure}
%\vspace{-5cm}
\includegraphics[scale=1]{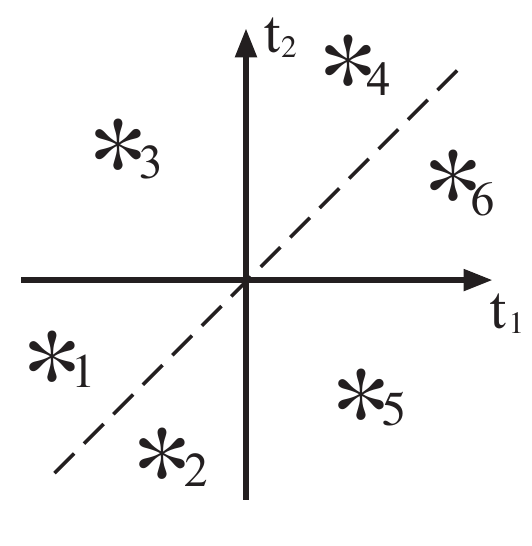}
\caption{The different sectors of the $t_1,\, t_2$ integrals of Eq. \eqref{starcontributions} and the corresponding contributions.}
\label{sectors}
\end{figure}

Repeating the steps of Section \ref{calculationrates}, we change from energy to time representation,
and rewrite Eq.~(\ref{ABlike}) plus all the other permutations as the following contributions
\begin{align}
\star_1&=\int_{-\infty}^{+\infty} dt \int_{-\infty}^0 dt_1\int_{-\infty}^{0}dt_2 e^{i \nu eV t/\hbar} \nonumber\\
&  G_{4}^<(-t-t_1-t_2,-L_4)G_{3}^>(t+t_1,L_3)\nonumber\\
&\times G_{2}^>(-t_1,L_2) G_1^< (t_1+t_2,-L_1) \, ,\\
\star_2&=\int_{-\infty}^{+\infty} dt \int_{-\infty}^0 dt_1\int_{-\infty}^{0}dt_2 e^{i \nu eV t\hbar}   \nonumber\\
& G_{4}^{<}(-t-t_1,-L_4)G_{3}^{>}(t+t_1+t_2,L_3)\nonumber\\
& \times G_{2}^{>}(-t_1-t_2,L_2) G_1^{<}(t_1,-L_1) \, ,\\
\star_3&=\int_{-\infty}^{+\infty} dt \int_{-\infty}^0 dt_1\int_{-\infty}^{0}dt_2 e^{i \nu eV t\hbar}   \nonumber\\
& G_{4}^{<}(-t-t_2,-L_4)G_{3}^{>}(t-t_1,L_3)\nonumber\\
& \times G_{2}^{<}(t_1,L_2) G_1^{<}(t_2,-L_1) \, ,\\
\star_4&=\int_{-\infty}^{+\infty} dt \int_{-\infty}^0 dt_1\int_{-\infty}^{0}dt_2 e^{i \nu eV t\hbar}   \nonumber\\
&  G_{4}^{<}(-t+t_2,-L_4)G_{3}^{>}(t-t_1-t_2,L_3)\nonumber\\
& \times G_{2}^{<}(t_1+t_2,L_2) G_1^{>}(-t_2,-L_1) \, ,\\
\star_5&=\int_{-\infty}^{+\infty} dt \int_{-\infty}^0 dt_1\int_{-\infty}^{0}dt_2 e^{i \nu eV t\hbar}   \nonumber\\
& G_{4}^{<}(-t+t_1,-L_4)G_{3}^{>}(t+t_2,L_3)\nonumber\\
& \times G_{2}^{>}(-t_2,L_2) G_1^{>}(-t_1,-L_1) \, ,\\
\star_6&=\int_{-\infty}^{+\infty} dt \int_{-\infty}^0 dt_1\int_{-\infty}^{0}dt_2 e^{i \nu eV t\hbar}   \nonumber\\
& G_{4}^{<}(-t+t_1+t_2,-L_4)G_{3}^{>}(t-t_2,L_3)\nonumber\\
&\times G_{2}^{<}(t_2,L_2) G_1^{>}(-t_1-t_2,-L_1) \, .
\end{align}
Remarkably all the previous contributions can be cast as a single expression,
\begin{small}
\begin{multline}\label{starcontributions}
\sum_{i=1}^6 \star_i=
\int_{-\infty}^{+\infty} dt \int_{-\infty}^{+\infty} dt_1\int_{-\infty}^{+\infty} dt_2 e^{i \nu eV t\hbar} G_{4}^{<}(-t-t_1,-L_4)  \\ \times
 G_{3}^{>}(t+t_2,L_3)G_{2}^{t}(-t_2,L_2) G_1^{t}(t_1,-L_1)\,.
\end{multline}
\end{small}
One can verify as shown in Fig.~\ref{sectors} that different sectors in the $t_1,\, t_2$ integrals correspond to the six above contributions.
Introducing the Fourier transform of the Green's function, we have
\begin{small}
\begin{align}
\int \frac{d\epsilon}{2 \pi}
G_{4}^{<}(\epsilon,-L4)G_{3}^{>}(\epsilon+\nu eV,L_3)
G_{2}^{t}(\epsilon+\nu eV,L_2)G_{1}^{t}(\epsilon,-L_1)\,.
\end{align}
\end{small}
In a similar fashion one can show that the terms coming from Eq.~(\ref{ABlike2}) give
\begin{small}
\begin{align}
\int \frac{d\epsilon}{2 \pi}
G_{4}^{t}(\epsilon,-L4)G_{3}^{>}(\epsilon+\nu eV,L_3)
G_{2}^{\bar t}(\epsilon+\nu eV,L_2)G_{1}^{<}(\epsilon,-L_1) \,.
\end{align}
\end{small}
Summing up the two contributions, we finally have
\begin{small}
\begin{align}\label{singletransfergreen}
&W_{j,j+1}^{(1,\mathcal{A}\mathcal{B}\mathcal{C}\mathcal{D},\Phi_{\text{tot}}(j))_1}=
\left| \Gamma_\mathcal{A} \Gamma_\mathcal{B} \Gamma_\mathcal{C} \Gamma_\mathcal{D} \right| \cos \left[ 2\pi\nu(\Phi_\Gamma+\Phi_{\rm AB}+j \Phi_0)/\Phi_0 \right]  \\
&\times \int \frac{d\epsilon}{2 \pi} \big[
G_{4}^{<}(\epsilon,-L4)G_{3}^{>}(\epsilon+\nu eV,L_3)
G_{2}^{t}(\epsilon+\nu eV,L_2)G_{1}^{t}(\epsilon,-L_1) \nonumber\\
& + G_{4}^{t}(\epsilon,-L4)G_{3}^{>}(\epsilon+\nu eV,L_3)
G_{2}^{\bar t}(\epsilon+\nu eV,L_2)G_{1}^{<}(\epsilon,-L_1) \big]\,.\nonumber
\end{align}
\end{small}
We do not evaluate explicitly Eq.~(\ref{singletransfergreen}), but notice that if one considers the same expression at zero temperature and in the case of integer filling factor, $\nu=1$, i.e. for transferring of electrons, it is possible to show, using Eq.~(\ref{FT1}) that indeed $W_{j,j+1}^{(1,\mathcal{A}\mathcal{B}\mathcal{C}\mathcal{D},\Phi_{\text{tot}}(j))_1}=-W_{j,j+2}^{(2,\mathcal{A}\mathcal{B}\mathcal{C}\mathcal{D},\Phi_{\text{tot}}(j))}$.

%\bibliographystyle{apsrev}
%\bibliography{HBTPRB}

\begin{thebibliography}{59}
\bibitem[{rem()}]{remarkYuval}
\bibinfo{note}{It is possible to stipulate that the exchange of two anyons
  always takes place in a certain chirality, e.g., clockwise, in which case the
  ambiquity in defining the the statistical factor $\theta$ is removed.}
\bibitem[{foo()}]{footnote:zero_bias}
\bibinfo{note}{A similar source of dephasing (``zero-mode dephasing'') due to
  the change of the number of particles inside the setup was found in a
  single-channel Aharonov-Bohm ring in Ref.~[\onlinecite{gornyiZB}].}
\end{thebibliography}

\end{document}